\documentclass[11pt]{article}
\usepackage{amsmath,mathrsfs,amsthm}
\usepackage{caption}
\usepackage{verbatim}
\usepackage{subcaption}
\usepackage{graphicx,psfrag,epsf}
\usepackage{enumerate}
\usepackage{natbib}
\usepackage{bbm}
\usepackage{bm}
\usepackage{amssymb}
\usepackage[affil-it]{authblk}
\usepackage{verbatim}
\usepackage{extarrows}
\usepackage{framed}
\usepackage[colorlinks=true,citecolor=blue]{hyperref}
\usepackage{algorithm}
\usepackage{algorithmic}
\usepackage{url} 
\usepackage{float}
\usepackage{geometry}
\usepackage{lscape} 
\usepackage{multirow}
\usepackage{booktabs}
\usepackage{pifont}
\usepackage{ragged2e}

\DeclareMathOperator*{\argmin}{arg\,min}

\newtheorem{assumption}{Assumption}[section]

\newtheorem{proposition}{Proposition}[section]
\theoremstyle{remark}
\newtheorem{remark}{Remark}[section]
\newtheorem{lemma}{Lemma}[section]
\newtheorem{corollary}{Corollary}[section]
\newtheorem{condition}{Condition}[section]

\newcommand{\cmark}{\text{\ding{51}}}
\newcommand{\xmark}{\text{\ding{55}}}

\begin{document}
\def\spacingset#1{\renewcommand{\baselinestretch}%
{#1}\small\normalsize} \spacingset{1}

\title{A Scale-free Approach for False Discovery Rate Control in Generalized Linear Models}
\author{Chenguang Dai\footnote{These authors contribute equally to this work.}}
\author{Buyu Lin{$^\ast$}}
\author{Xin Xing}
\author{Jun S. Liu}
\affil{Department of Statistics, Harvard University}
\maketitle

\begin{abstract}
The generalized linear models (GLM) have been widely used in practice to model non-Gaussian response variables. 
When the number of explanatory features is relatively large, scientific researchers are of interest to perform controlled feature selection in order to simplify the downstream analysis. 
This paper introduces a new framework for feature selection in GLMs that can achieve false discovery rate (FDR) control in two asymptotic regimes. 
The key step is to construct a \textit{mirror statistic} to measure the importance of each feature, which is based upon two (asymptotically) independent estimates of the corresponding true coefficient obtained via either the data-splitting method or the Gaussian mirror method. 
The FDR control is achieved by taking advantage of the mirror statistic’s property that, for any null feature, its sampling distribution is (asymptotically) symmetric about 0. 
In the moderate-dimensional setting in which the ratio between the dimension (number of features) $p$ and the sample size $n$ converges to a fixed value, i.e., $p/n\to \kappa \in (0,1)$, we construct the mirror statistic based on the maximum likelihood estimation.
In the high-dimensional setting where $p \gg n$, we use the debiased Lasso to build the mirror statistic.
Compared to the Benjamini-Hochberg procedure, which crucially relies on the asymptotic normality of the $Z$ statistic, the proposed methodology is scale free as it only hinges on the symmetric property, thus is expected to be more robust in  finite-sample cases.
Both simulation results  and a real data application show that the proposed methods are capable of controlling the FDR, and are often more powerful than existing methods including the Benjamini-Hochberg procedure and the knockoff filter.
\end{abstract}

\section{Introduction}
The generalized linear model (GLM) is a useful tool for modeling non-Gaussian response variables such as categorical data and count data. In the current big data era, researchers are often capable of collecting a large number of explanatory features $X_1, \cdots, X_p$ for a given response variable $y$, in which the number of features $p$ is potentially comparable to or larger than the sample size $n$. As the response variable $y$ most likely only depends on a small subset of features, it is of primary interest to identify those relevant features in order to enhance the computability of the analysis as well as the interpretability of the results.

A desired feature selection procedure is expected to control the quality of the selection, which can be mathematically calibrated by the false discovery rate (FDR) \citep{benjamini1995controlling} defined as:
\begin{equation}
\text{FDR} = \mathbbm{E}[\text{FDP}],\ \ \ \text{FDP}\footnote{We assume $\text{FDP} = 0$ if $\#\{j \in \widehat{S}\} = 0$.} = \frac{\#\{j: j\notin S_1,\ j \in \widehat{S}\}}{\#\{j \in \widehat{S}\}},
\end{equation}
where $S_1$ denotes the index set of relevant features, $\widehat{S}$ denotes the index set of selected features, and FDP refers to the false discovery proportion. The expectation is taken with respect to  the randomness both in the data and in the selection procedure if it is not deterministic. Existing FDR control methods that can be applied to GLMs include the Benjamini-Hochberg (BHq) procedure (\citet{ma2020global} specifically consider the logistic regression model) and the knockoff filter \citep{candes2018panning, lu2018deeppink, huang2019relaxing}. A brief discussion on these existing methods as well as their comparisons to our proposed strategy are given in Section \ref{subsec:GLM-comparison-high-dimension}.

In this paper, we develop new methodologies for exercising controlled feature selection in GLMs based upon the recently developed FDR control framework in \citet{xing2019controlling} and \citet{Dai2020DS}. Under the guiding principle of data perturbation, we construct a mirror statistic for each feature to measure its relative importance, based upon two estimates of the corresponding true coefficient obtained via either the Gaussian mirror method \citep{xing2019controlling} or the data-splitting method \citep{Dai2020DS}. 
After choosing a proper data-dependent cutoff, we select the features with mirror statistics larger than the cutoff.
The FDR control is achieved by exploiting the symmetric property of the mirror statistic associated with any null feature.
Our FDR control framework enjoys a scale-free property in the sense that any constant rescaling of all the mirror statistics does not change the selection result. 

We consider two asymptotic regimes for GLMs in this paper. The moderate-dimensional setting concerns the regime where the ratio $p/n \to \kappa \in (0, 1)$. We base the construction of the mirror statistics on the maximal likelihood estimator (MLE) of the true coefficient vector and show that both the Gaussian mirror  and the data-splitting methods achieve an asymptotic FDR control under mild conditions. We note that the classical asymptotic result for the MLE breaks down in this regime, in the sense that the asymptotic normality characterization of the MLE involves two additional bias and variance scaling factors \citep{sur2019modern}. In consequence, BHq faces the challenge of estimating the two scaling factors in order to obtain asymptotically valid p-values, which remains an open problem for GLMs except for the logistic/probit regression model. In contrast, our FDR control framework is scale-free and does not require the knowledge of the aforementioned scaling factors, 
thus can be easily and validly applied to all GLMs.

The high-dimensional setting concerns the regime of $p \gg n$. We restrict ourselves to the data-splitting method for the consideration of computational feasibility.  
We construct the mirror statistics using the debiased Lasso \citep{van2014asymptotically}, and theoretically justify our approach by showing the desired FDR control property under proper sparsity and regularity conditions.
\citet{ma2020global} introduced a BHq procedure for the logistic regression model, which relies on the asymptotic normality of the debiased Lasso estimator in order to obtain asymptotically valid p-values. In contrast, 
for the purpose of controlling the FDR, the data-splitting method only requires the asymptotic symmetric property of the debiased Lasso estimator regardless of its scale, thus we expect it to be more robust in finite-sample cases.

The rest of the paper is structured as follows. Section \ref{sec:review-FDR} introduces our  FDR control framework as well as two basic methods for  constructing the mirror statistics:  Gaussian mirror and data splitting.
Section \ref{sec:GLM-moderate-dimension} concerns the GLMs in the moderate-dimensional setting. We specify the construction of the mirror statistics using the MLE, and establish the desired FDR control property for both the Gaussian mirror and the data-splitting methods.
Section \ref{sec:GLM-high-dimension} concerns the GLMs in the high-dimensional setting. By incorporating the debiasing approach, we show that the data-splitting method enjoys an asymptotic FDR control.
Sections \ref{subsec:numeric-moderate-dimension} and \ref{subsec:numeric-high-dimension} demonstrate the competitive performances of our proposed methods through simulation studies on popular GLMs including the logistic regression model and the negative binomial regression model.
Section \ref{subsec:real-data} applies the data-splitting method to a single-cell RNA sequencing data for the purpose of selecting relevant genes with respect to the glucocorticoid response.
Section \ref{sec:conclusion} concludes with a few final remarks. The proofs as well as additional numerical results are given in the Appendix.

\section{FDR Control via Mirror Statistics}
\label{sec:review-FDR}
For a given response variable $y$, we consider a set of $p$ candidate features $\{X_1,\ldots, X_p\}$. Let $X_{n\times p}$ be the design matrix, in which each row, denoted as $x_i^\intercal$ for $i\in[n]$, is an independent realization of these features. Let $y = (y_1, \cdots, y_n)^\intercal$ be the associated response vector. We assume that the response variable $y$ only depends on a subset of features, and the corresponding index set is denoted as $S_1$. Let $p_1 = |S_1|$ and $p_0 = p - p_1$. We refer the feature $X_j$ as  relevant (non-null) if $j\in S_1$; otherwise we call it a null feature. The index set of null features are denoted as $S_0$. The goal is to identify as many relevant features as possible with the FDR under control. Throughout, we denote the power of a selection procedure as the expected proportion of the successfully identified relevant features, i.e., 
\begin{equation}
\text{Power} = \mathbbm{E}\big[\#\{j: j \in S_1,  j \in \widehat{S}\}/p_1\big],
\end{equation}
in which $\widehat{S}$ denotes the index set of selected features.

The FDR control framework we consider here requires constructing a mirror statistic $M_j$ for each feature $X_j$, which satisfies the following two properties.
\begin{enumerate}[({A}1)]
\label{Property:Mirror}
\item A feature with a larger mirror statistic is more likely to be a relevant feature.
\vspace{-0.2cm}
\item The sampling distribution of the mirror statistic associated with any null feature is (asymptotically) symmetric about 0.
\end{enumerate}
Property (A1) enables us to rank the importance of features by their associated mirror statistics. Given the FDR control level $q$, it remains to choose a proper cutoff $\tau_q$, and select the set of features $\{j: M_j > \tau_q\}$. For any cutoff $t > 0$, Property (A2) suggests an approximate upper bound on the number of false positives, 
\begin{equation}
\label{eq:FDP}
\text{FDP}(t) = \frac{ \# \{j: j \notin S_1, M_j > t\}}{ \# \{j: M_j > t\}} \lesssim  \frac{ \# \{j: M_j < -t\}}{ \# \{j: M_j > t\}},
\end{equation}
which leads to the following FDR control framework.
\begin{algorithm*}
\caption{The FDR control framework.}
\begin{enumerate}
\item Calculate the mirror statistic $M_j$ for $j \in [p]$.
\item Given a designated FDR level $q \in (0, 1)$, calculate the cutoff $\tau_q$ as:
\begin{equation}\nonumber
\tau_q = \inf\left\{t > 0: \widehat{\text{FDP}}(t) = \frac{ \# \{j: M_j < -t\}}{ \# \{j: M_j > t\}} \leq q\right\}.
\end{equation}
\vspace{-0.6cm}
\item Set $\widehat{S} = \{j: M_j > \tau_q\}$.
\end{enumerate}
\label{alg:FDR-control-framework}
\end{algorithm*}

A general recipe for constructing the mirror statistics in the regression setting is as follows. For $j \in [p]$, we first obtain two estimates, $\widehat{\beta}^{(1)}_j$ and $\widehat{\beta}^{(2)}_j$, of the true coefficient $\beta_j^\star$. In order for the resulting mirror statistics to have comparable variances, the two estimates generally require further standardization by the corresponding (asymptotic) standard deviations. Denote $T_j^{(1)}$ and $T_j^{(2)}$ as the normalized estimates,
which should satisfy the following conditions.
\begin{condition}
\label{cond:FDR-regression-coefficient-estimate}
\text{}
\begin{itemize}
\item (Independence) The two regression coefficients are (asymptotically) independent. 
\item (Symmetry) For any null feature $j \in S_0$, the sampling distribution of either of the two regression coefficients is (asymptotically) symmetric about 0. 
\end{itemize}
\end{condition}
\noindent The mirror statistic $M_j$ then takes a general form of 
\begin{equation}
\label{eq:mirror-statistic}
M_j = \text{sign}\big(T_j^{(1)} T_j^{(2)}\big)f\big(|T_j^{(1)}|, |T_j^{(2)}|\big),
\end{equation}
in which $f(u,v)$ is a user-specified bivariate function satisfying the following conditions:
\begin{condition}
\label{cond:mirror-statistics-f}
$f(u,v)$ is non-negative, symmetric and monotonically increasing in $u$ and $v$.
\end{condition}

Condition \ref{cond:FDR-regression-coefficient-estimate} and \ref{cond:mirror-statistics-f} together implies Property (A1) and (A2). For a relevant feature $j \in S_1$, the two regression coefficients $T_j^{(1)}$ and $T_j^{(2)}$ tend to be large (in the absolute value) and have the same sign if the estimation procedures are reasonably accurate. Since $f(u, v)$ is monotonically increasing in both $u$ and $v$, the mirror statistic $M_j$ is likely to be positive and relatively large, which implies Property (A1). On the other hand, for a null feature $j\in S_0$, Property (A2) holds given Condition \ref{cond:FDR-regression-coefficient-estimate} because $T_j^{(1)}$ and $T_j^{(2)}$ are (asymptotically) independent, and one of them is (asymptotically) symmetric about 0. 

Three possible choices of $f(u, v)$ are
\begin{equation}\label{contrast_choice}
f(u, v) = 2 \min(u,v) ,\ \ \ f(u,v) = uv,\ \ \ f(u, v) = u + v.
\end{equation}
The first choice is equal to the mirror statistic proposed in \citet{xing2019controlling}, and the third choice is nearly optimal as shown by the following proposition.
\begin{proposition}
\label{prop:optimality-mirror-statistics}
Suppose the set of normalized regression coefficients $\{T_j\}_{j\in [p]}$ are asymptotically independent across the feature index $j$. For each null feature $j\in S_0$, we assume that $T_j$ asymptotically follows $N(0, 1)$, whereas for each relevant feature $j\in S_1$, we assume that  $T_j$ asymptotically follows $N(\omega, 1)$ with $\omega > 0$. In addition, we assume that $p_0/p$ converges to a fixed constant in $(0,1)$. Then $f(u, v) = u + v$ is the nearly optimal choice satisfying Condition \ref{cond:mirror-statistics-f} that yields the highest power, when $\omega$ is sufficiently large.
\end{proposition}

\begin{remark}
The proof of Proposition \ref{prop:optimality-mirror-statistics} might be of interest on its own. We essentially rephrase the FDR control problem under the hypothesis testing framework, and prove the optimality using the Neymann-Pearson Lemma. The form $f(u, v) = u + v$ is derived based on the rejection rule of the corresponding likelihood ratio test. In practice, however, we found that the three choices listed in (\ref{contrast_choice}) have no significant differences in most cases, and the second one $f(u,v)=uv$, which will be used in our simulation studies,  sometimes can even perform 
slightly better. 
\end{remark}

The following subsections review two recently proposed methods, Gaussian mirror \citep{xing2019controlling} and data splitting \citep{Dai2020DS},
for constructing the two regression coefficients $T_j^{(1)}$ and $T_j^{(2)}$ that satisfy Condition \ref{cond:FDR-regression-coefficient-estimate}.

\subsection{Gaussian mirror}
\label{subsec:gaussian-mirror}
For an easy illustration, we restrict ourselves to low-dimensional ($n > p$) linear models. For  high-dimensional settings, we refer the readers to \citet{xing2019controlling}. 
The key idea  of Gaussian mirror is to create a pair of perturbed mirror features, 
\begin{equation}
\label{eq:Gaussian-mirror-perturbed-features}
X_j^+ = X_j + c_jZ_j,\ \ \ X_j^- = X_j - c_jZ_j,
\end{equation}
in which $c_j$ is an adjustable scalar and $Z_j$ follows $N(0,1)$ independently. The linear model can then be equivalently recasted in the following way,
\begin{equation}
\label{eq:linear-Gaussian-mirror-model-reparametrization}
y =  \frac{\beta_j^\star}{2}X_j^+ + \frac{\beta_j^\star}{2}X_j^- + X_{-j}\beta^\star_{-j} + \epsilon.
\end{equation}
As we are in the low-dimensional setting, we obtain $\widehat{\beta}^+$ and $\widehat{\beta}^-$, as well as the normalized estimates $T_j^+$ and $T_j^-$, via the ordinary least squares (OLS). For any null feature $j\in S_0$, both $T_j^+$ and $T_j^-$ follow a $t$ distribution centering at zero, thus the symmetric requirement in Condition \ref{cond:FDR-regression-coefficient-estimate} is fulfilled.
Furthermore, we can properly set $c_j$ as follows so that $T_j^+$ and $T_j^-$ are asymptotically independent,
\begin{equation}
\label{eq:Gaussian-mirror-scaling}
c_j = {||P^\perp_{-j}X_j||}/{||P^\perp_{-j}Z_j||},
\end{equation}
where $P_{-j}^\perp$ is the projection matrix onto the orthogonal complement of the column space spanned by $X_{-j}$.

\subsection{Data splitting}
\label{subsec:data-splitting}
A simple way to obtain two independent regression coefficients is data splitting. More precisely, we randomly split the data into two halves, $(y^{(1)}, X^{(1)})$ and $(y^{(2)}, X^{(2)})$, and estimate $\widehat{\beta}^{(1)}_j$ and $\widehat{\beta}^{(2)}_j$, as well as the normalized versions $T_j^{(1)}$ and $T_j^{(2)}$, on each part of the data. Without loss of generality, we assume that the sample sizes of the two parts of the data are the same.
The independence between the two estimates is naturally implied by data splitting. The symmetric requirement in Condition \ref{cond:FDR-regression-coefficient-estimate} can be satisfied if, for any null feature, either of the estimates is (asymptotically) normal and centered at 0. As we will see, desirable estimates can be constructed for GLMs under proper conditions.

The potential power loss is a major concern of using data splitting. \citet{Dai2020DS} introduced a multiple data-splitting (MDS) method to remedy this issue, which also helps in stabilizing the selection result. 
The idea is to obtain multiple selection results via repeated data splits, and rank the importance of features by their inclusion rates defined as below,
\begin{equation}
\widehat{I}_j = \frac{1}{m}\sum_{k = 1}^m\frac{\mathbbm{1}(j \in \widehat{S}^{(k)})}{|\widehat{S}^{(k)}|\vee 1},
\label{eq:inclustion-rate}
\end{equation}
in which $m$ is the total number of data splits, and $\widehat{S}^{(k)}$ is the index set of the selected features in the $k$-th data split.
We select the features with inclusion rates larger than a properly chosen cutoff so as to maintain the FDR control.
\citet{Dai2020DS} also show that, for the independent Gaussian means problem, MDS can recover the full information in the sense that when $m \to \infty$, the inclusion rates provide the same ranking of features as the p-values calculated based on the full data.
We outline the MDS procedure in Algorithm \ref{alg:multiple-data-splits}, which can be applied on top of the single data-splitting methods designed for GLMs (see Sections \ref{sec:GLM-moderate-dimension} and \ref{sec:GLM-high-dimension}).
\begin{algorithm*}
\caption{Aggregating selection results from multiple data splits.}
\label{alg:multiple-data-splits}
\begin{enumerate}
\item Sort the features with respect to their inclusion rates in the multiple selections. Denote the sorted inclusion rates as $0 \leq \widehat{I}_{(1)} \leq \widehat{I}_{(2)} \leq \cdots \leq \widehat{I}_{(p)}$.
\item Given a designated FDR level $q \in (0, 1)$, find the largest $\ell \in [p]$ such that
$\widehat{I}_{(1)} + \cdots + \widehat{I}_{(\ell)} \leq q$.
\item Set $\widehat{S} = \{j: \widehat{I}_j > \widehat{I}_{(\ell)}\}$.
\end{enumerate}
\end{algorithm*}

\section{Generalized Linear Models  in Moderate Dimensions}
\label{sec:GLM-moderate-dimension}
We consider the following  generalized linear model (GLM) with a canonical link function $\rho$:
\begin{equation}
\label{eq:GLM-moderate-dimension}
p(y|X, \beta^\star) = \prod_{i=1}^n c(y_i)\exp\left(y_ix_i^\intercal\beta^\star - \rho(x_i^\intercal\beta^\star)\right),
\end{equation}
where $\beta^\star$ denotes the true coefficient vector.
In the moderate-dimensional setting, we assume that $p/n \to \kappa \in (0, 1)$.  Note that $\kappa = 0$ corresponds to the classical setting with fixed $p$. We consider a random design setting, in which we assume that the $x_i$'s are i.i.d observations from $N(0,\Sigma)$. We assume that the signal strength also has an asymptotic limit, i.e., $\text{Var}(x_i^\intercal\beta^\star) \to \gamma^2$. 

Let $\rho(X\beta) = (\rho(x_1^\intercal\beta), \cdots, \rho(x_n^\intercal\beta))^\intercal$. 
We base the mirror statistics on the MLE,
\begin{equation}
\label{eq:GLM-MLE-moderate-dimension}
\widehat{\beta}  = \argmin_{\beta\in\mathbbm{R}^p} \left\{\frac{1}{n} 1^\intercal \rho(X\beta) - \frac{1}{n}y^\intercal X\beta\right\}.
\end{equation}
The  MLE can behave very differently in the moderate-dimensional setting compared to the classical setting with fixed $p$. First of all, the MLE may not exist. For instance, for the logistic regression model, the MLE does not exist if the two classes become well separated. More precisely, \citet{candes2020phase} derived the phase transition curve for the existence of MLE, parametrized in terms of $\kappa$ and $\gamma$.
In this section, we restrict ourselves to the regime where the MLE exists, otherwise we can consider the debiased regularization method detailed in Section \ref{sec:GLM-high-dimension}.
Second, when the MLE exists, it is still asymptotically biased and rescaled. 
Based upon the results of  \citet{zhao2020asymptotic} and \citet{salehi2019impact}, we obtain the following characterization of the asymptotic distribution of the MLE. 
\begin{proposition}
\label{prop:GLM-asymptotics-moderate-dimension}
Consider a GLM defined in \eqref{eq:GLM-moderate-dimension} in the moderate-dimensional setting as specified above. Assume that the MLE exists asymptotically. Then, for any $j\in[p]$ with $\sqrt{n}\tau_j\beta^\star_j = O(1)$, we have
\begin{equation}
\label{eq:GLM-asymptotics-moderate-dimension}
\frac{\sqrt{n}(\widehat{\beta}_j - \alpha_\star\beta^\star_j)}{\sigma_\star/\tau_j} \overset{d}{\to} N(0, 1),
\end{equation}
where $\tau_j^2 = 1/\Theta_{jj}$ with $\Theta = \Sigma^{-1}$, and $\alpha_\star, \sigma_\star$ are two universal constants depending on the model $\rho$, the true  coefficient vector $\beta^\star$, the signal strength $\gamma$ and the ratio of dimension to sample size $\kappa$.
\end{proposition}
\begin{remark}
The proof of Proposition \ref{prop:GLM-asymptotics-moderate-dimension} relies on the stochastic representation of the MLE in \citet{zhao2020asymptotic} and the Convex Gaussian Min-max (CGMT) Theorem. 
The constant pair
$(\alpha_\star, \sigma^2_\star)$ is the limit of $(\alpha_n, \sigma_n^2)$, where 
\begin{equation}
\label{eq:definition-alpha-sigma}
\alpha_n = \frac{\langle\widehat\beta, \beta^\star\rangle}{||\widehat\beta||^2},\ \ \ \  \sigma_n^2 = ||P_{\beta^\star}^\perp\widehat\beta||.
\end{equation}
$P_{\beta^\star}^\perp$ is the projection matrix onto the orthogonal complement of $\beta^\star$.
The convergence of $(\alpha_n, \sigma^2_n)$ follows from a routine application of CGMT by transforming the primary optimization problem (PO) into an easy-to-analyze auxiliary optimization problem (AO). 
More details can be found in the proof of Lemma \ref{lemma:glm-moderate-parameter-convergence}.  \end{remark}

\begin{remark}
\label{remark:GLM-conditional-variance-moderate-dimension}
Note that $\tau_j^2$ is simply the conditional variance $\text{Var}(X_j|X_{-j})$, and thus can be estimated either using the inverse of the sample covariance matrix, i.e., $\widehat{\tau}^2_j = 1/(X^\intercal X/n)^{-1}_{jj}$, or via a node-wise regression approach, that is to  regress $X_j$ onto $X_{-j}$ and obtain the residual sum of squares $\text{RSS}_j$. An unbiased estimator of $\tau_j^2$ is then $\widehat{\tau}^2_j = \text{RSS}_j/(n - p + 1)$.
\end{remark}

Proposition \ref{prop:GLM-asymptotics-moderate-dimension} implies that the classical asymptotic normality based on the Fisher information does not apply to the MLE in the moderate-dimensional setting. In addition, in order to obtain asymptotically valid p-values, it requires to estimate the bias and variance scaling factors $\alpha_\star$ and $\sigma_\star$. 
This is in general a challenging task, as it requires to first estimate the signal strength $\gamma$. \citet{sur2019modern} proposed the \textit{ProbFrontier} method to estimate $\gamma$ using the phase transition curve that calibrates the existence of the MLE. However, to the best of our knowledge, 
there is no unified approach to derive the curve for a general GLM and
the existing literature only covers the results for the logistic/probit regression model. Therefore, it remains challenging to apply BHq moving beyond these two models. 

Even for the logistic/probit regression model, we empirically observe that the asymptotic normality characterized in Proposition \ref{prop:GLM-asymptotics-moderate-dimension} might kick in slowly, so that the resulting p-value of the null feature appears non-uniform in finite-sample cases. In contrast, the asymptotic symmetry required by our FDR control framework can kick in much faster compared to the asymptotic normality. Numerical comparisons can be found in Section \ref{subsub:moderate-dimensional-logistic}.

\subsection{FDR control via data splitting}
\label{subsec:GLM-data-splitting-moderate-dimension}
Contrast to BHq, the data-splitting method outlined in Algorithm \ref{alg:GLM-data-splitting-moderate-dimension} is free of estimating the scaling factors $(\alpha_\star, \sigma_\star)$, thus is applicable for all GLMs in the moderate-dimensional setting. According to the asymptotic characterization in Equation \eqref{eq:GLM-asymptotics-moderate-dimension}, we normalize the two independent MLEs $\widehat{\beta}^{(1)}$ and $\widehat{\beta}^{(2)}$ as below,
\begin{equation}
\label{eq:GLM-normalized-estimator-moderate-dimension-data-splitting}
{T}_j^{(1)} = \widehat{\tau}_j^{(1)}\widehat{\beta}^{(1)}_j,\ \ \ {T}_j^{(2)} = \widehat{\tau}_j^{(2)}\widehat{\beta}^{(2)}_j.
\end{equation}
Here $\widehat{\tau}_j^{(1)}$ and $\widehat{\tau}_j^{(2)}$ are two independent estimates of the conditional variance $\text{Var}(X_j|X_{-j})$ obtained via either the node-wise regression or the inverse of the sample covaraince matrix. Although the asymptotic standard deviation of the MLE is $\sigma_\star/\tau_j$, we can safely drop the constant $\sigma_\star$, because our FDR control framework outlined in Algorithm \ref{alg:FDR-control-framework} maintains the same selection result under an arbitrary rescaling of all the mirror statistics. Furthermore, the data-splitting method does not require the knowledge of the bias scaling factor $\alpha_\star$ either. For any null feature $j\in S_0$, since $\alpha_\star\beta_j^\star = 0$, the symmetric requirement in Condition \ref{cond:FDR-regression-coefficient-estimate} is asymptotically fulfilled according to Proposition \ref{prop:GLM-asymptotics-moderate-dimension}. 

\begin{algorithm*}
\caption{The data-splitting method for GLMs in the moderate-dimensional setting.}
\label{alg:GLM-data-splitting-moderate-dimension}
\begin{enumerate}
\item Split the data set into two equal-sized halves $(y^{(1)}, X^{(1)})$ and $(y^{(2)}, X^{(2)})$.  
\item For $j\in[p]$, regress $X^{(1)}_j$ onto $X^{(1)}_{-j}$, and regress $X^{(2)}_j$ onto $X^{(2)}_{-j}$. Let
$${\widehat{\tau}^2_j}{}^{(1)} = \frac{\text{RSS}^{(1)}_j}{n/2 - p + 1},\ \ \ {\widehat{\tau}^2_j}{}^{(2)} = \frac{\text{RSS}^{(2)}_j}{n/2 - p + 1},$$ 
in which $\text{RSS}_j$ is the residual sum of squares.
\item Find the MLEs $\widehat{\beta}^{(1)}$ and $\widehat{\beta}^{(2)}$ on each part of the data. For $j\in[p]$, calculate the mirror statistic $M_j$ based on $T^{(1)}_j$ and $T^{(2)}_j$ defined in Equation \eqref{eq:GLM-normalized-estimator-moderate-dimension-data-splitting}.
\item Select the features using Algorithm \ref{alg:FDR-control-framework}. 
\end{enumerate}
\end{algorithm*}

We require the following assumptions to theoretically justify our approach. Note that we impose  no assumptions on either the sparsity level or the signal magnitude of the true coefficient vector $\beta^\star$.
\begin{assumption}
\label{assump:GLM-data-splitting-moderate-dimension}
\text{}
\begin{enumerate}[(1)]
\item There exists a constant $C > 0$, such that
$1/C\leq \sigma_{\min}(\Sigma)\leq\sigma_{\max}(\Sigma)\leq C$.
\item 
The required number of null features $p_0\to\infty$ as $n, p\to\infty$.
\end{enumerate}
\end{assumption}

\begin{remark}
Assumption \ref{assump:GLM-data-splitting-moderate-dimension} (1) also appears in \citet{zhao2020asymptotic}, in which $\sigma_{\min}(\Sigma)$ and $\sigma_{\max}(\Sigma)$ refer to the minimum and maximum of the eigenvalues of the covariance matrix $\Sigma$.
Assumption \ref{assump:GLM-data-splitting-moderate-dimension} (2) is straightforward because otherwise the asymptotic FDR control problem becomes trivial: we can simply select all  features and the corresponding  FDR converges to 0.
\end{remark}

Under Assumption \ref{assump:GLM-data-splitting-moderate-dimension}, the following proposition shows that the data-splitting method achieves an asymptotic FDR control for a GLM in the moderate-dimensional setting.
\begin{proposition}
\label{prop:GLM-data-splitting-FDR-moderate-dimension}
Consider a GLM defined in \eqref{eq:GLM-moderate-dimension} in the moderate-dimensional setting. For any given FDR control level $q \in (0,1)$, we assume that the pointwise limit $\text{FDP}^\infty(t)$ of $\text{FDP}(t)$ exists for all $t > 0$, and there is a $t_q > 0$ such that $\text{FDP}^\infty(t_q) \leq q.$ Then, under Assumption \ref{assump:GLM-data-splitting-moderate-dimension}, we have
\begin{equation}\nonumber
\limsup_{n, p \to\infty}\mathbbm{E}\left[\frac{\#\{j: j \in S_0, j \in \widehat{S}_{\tau_q}\}}{\# \{j: j \in \widehat{S}_{\tau_q}\}}\right] \leq q.
\end{equation}
using the data-splitting method outlined in Algorithm \ref{alg:GLM-data-splitting-moderate-dimension}.
\end{proposition}

\begin{remark}
The assumption on the existence of a desirable $t_q$ is necessary, as it implies that the asymptotic FDR control is achievable by selecting a proper cutoff for the mirror statistics. 
\end{remark}

\subsection{FDR control via Gaussian mirror}
\label{subsec:GLM-Gaussian-mirror-moderate-dimension}
The data-splitting method is only applicable in the regime $\kappa \in (0, 1/2]$, i.e., $n \geq 2p$; otherwise the MLE does not exist after data splitting.
In fact, for the logistic regression model, even in the regime $\kappa \in (0, 1/2]$, it is still possible, if the signal strength $\gamma$ is sufficiently large, that the MLE exists on the full data but not  on a half of the data (e.g., see Figure 6(a) in \citet{sur2019modern}). To overcome this issue, we consider the Gaussian mirror method, which extends the applicability to $\kappa \in (0,1)$ as long as the MLE exists on the full data.

As discussed in Section \ref{subsec:gaussian-mirror}, we fit a GLM using the response vector $y$ and the augmented set of features $(X_{-j}, X_j^+,  X_j^-)$, to find the MLEs, $\widehat{\beta}^+$ and $\widehat{\beta}^-$, associated with the pair of perturbed mirror features $(X_j^+, X_j^-)$ defined in Equation \eqref{eq:Gaussian-mirror-perturbed-features}. 
Let $\Sigma_{\text{aug}}$ be the covariance matrix of the augmented set of features $(X_{-j}, X_j^+,  X_j^-)$, and let $\Theta_\text{aug} = \Sigma^{-1}_{\text{aug}}$.
We have the following asymptotic characterization.

\begin{proposition}
\label{prop:GLM-asymptotics-mirror-feature-moderate-dimension}
For any $j\in[p]$, consider fitting a GLM using the response vector $y$ and the augmented set of features $(X_{-j}, X_j^+,  X_j^-)$ defined in Equation \eqref{eq:Gaussian-mirror-perturbed-features}. Then, the asymptotic distribution of the MLE $(\widehat{\beta}_j^+, \widehat{\beta}_j^-)$ is:
\begin{equation}
\frac{\sqrt{n}}{\sigma_\star}\left(
\begin{pmatrix}
\widehat{\beta}_j^+ \\
\widehat{\beta}_j^-
\end{pmatrix} 
- \frac{\alpha_\star}{2}
\begin{pmatrix}
\beta_j^\star \\
\beta_j^\star
\end{pmatrix}
\right) 
\overset{d}{\to} N\big(0, \Theta^\ast\big),
\end{equation}
in which $\Theta^\ast$ is the $2\times2$ submatrix at the right bottom of $\Theta_\text{aug}$ corresponding to $(X_j^+, X_j^-)$, and $\alpha_\star, \sigma_\star$ are defined as in Proposition \ref{prop:GLM-asymptotics-moderate-dimension}.
\end{proposition}

According to Proposition \ref{prop:GLM-asymptotics-mirror-feature-moderate-dimension}, we can choose a proper scalar $c_j$ so that $\Theta^\ast_{12} = 0$. This implies that the MLEs $\widehat{\beta}_j^+$ and $\widehat{\beta}_j^-$ are asymptotically independent. 
In practice, we plug in the sample covariance matrix $X^\intercal X/n$ to estimate the population covariance matrix $\Sigma$, leading to the scalar $c_j$ in the form of Equation \eqref{eq:Gaussian-mirror-scaling}.
We note that $\Theta^\ast_{12} = 0$ also implies the asymptotic independence between $X_j^+$ and $X_j^-$ conditioning on $X_{-j}$. Therefore, for our specific choice of $c_j$, the inverse of the asymptotic variances of $\widehat{\beta}_j^+$ and $\widehat{\beta}_j^-$ are 
\begin{equation}
\label{eq:GLM-moderate-dimension-conditional-variance}
\begin{aligned}
& 1/\Theta^\ast_{11} = \text{Var}(X_j^+\mid X_j^-, X_{-j}) \approx \text{Var}(X_j^+\mid  X_{-j}) = \tau_j^2 + c_j^2,\\
& 1/\Theta^\ast_{22} = \text{Var}(X_j^-\mid X_j^+, X_{-j}) \approx \text{Var}(X_j^-\mid  X_{-j}) = \tau_j^2 + c_j^2.
 \end{aligned}
\end{equation}
We thus normalize $\widehat{\beta}_j^+$ and $\widehat{\beta}_j^-$ as 
\begin{equation}
\label{eq:GLM-normalized-estimator-moderate-dimension-gaussian-mirror}
{T}_j^+ = (\widehat{\tau}^2_j + c_j^2)^{1/2}\widehat{\beta}^+_j,\ \ \ {T}_j^- = (\widehat{\tau}^2_j + c_j^2)^{1/2}\widehat{\beta}^-_j,
\end{equation}
with $\widehat{\tau}_j^2$ calculated according to Remark \ref{remark:GLM-conditional-variance-moderate-dimension}. We summarize the Gaussian mirror method in Algorithm \ref{alg:GLM-Gaussian-mirror-moderate-dimension}.

\begin{algorithm*}
\caption{The Gaussian mirror method for GLMs in the moderate-dimensional setting.}
\label{alg:GLM-Gaussian-mirror-moderate-dimension}
\begin{enumerate}
\item For $j \in [p]$, calculate the mirror statistic $M_j$ as follows.
\begin{enumerate}
\item Simulate $Z_j$ from $N(0, I_n)$.
\item Calculate the scaling factor $c_j$ according to Equation \eqref{eq:Gaussian-mirror-scaling}.
\item Fit a GLM using $y$ and $(X_{-j}, X_j^+,  X_j^-)$ to find the MLEs, $\widehat{\beta}_j^+$ and $\widehat{\beta}_j^-$. 
\item Estimate $\widehat{\tau}_j^2$ according to Remark \ref{remark:GLM-conditional-variance-moderate-dimension}.
\item Calculate the mirror statistic $M_j$ based on $T^+_j$ and $T^-_j$ defined in Equation \eqref{eq:GLM-normalized-estimator-moderate-dimension-gaussian-mirror}.
\end{enumerate}
\item Select the features using Algorithm \ref{alg:FDR-control-framework}. 
\end{enumerate}
\end{algorithm*}

\begin{remark}
The Gaussian mirror method is computationally more intensive compared to the data-splitting method. The former requires fitting the GLM $p$ times, whereas the latter only requires fitting the GLM two times.
\end{remark}

Under Assumption \ref{assump:GLM-data-splitting-moderate-dimension}, the following proposition shows that the Gaussian mirror method achieves an asymptotic FDR control for a GLM in the moderate-dimensional setting.
\begin{proposition}
\label{prop:GLM-Gaussian-mirror-FDR-moderate-dimension}
Consider a GLM defined in \eqref{eq:GLM-moderate-dimension} in the moderate-dimensional setting. For any given FDR control level $q \in (0,1)$,
we assume that the pointwise limit $\text{FDP}^\infty(t)$ of $\text{FDP}(t)$ exists for all $t > 0$, and there is a $t_q > 0$ such that $\text{FDP}^\infty(t_q) \leq q.$ Then, under Assumption \ref{assump:GLM-data-splitting-moderate-dimension}, we have
\begin{equation}\nonumber
\limsup_{n, p \to\infty}\mathbbm{E}\left[\frac{\#\{j: j \in S_0, j \in \widehat{S}_{\tau_q}\}}{\# \{j: j \in \widehat{S}_{\tau_q}\}\vee 1}\right] \leq q,
\end{equation}
using the Gaussian mirror method outlined in Algorithm \ref{alg:GLM-Gaussian-mirror-moderate-dimension}.
\end{proposition}

\section{Generalized Linear Models in High Dimensions}
\label{sec:GLM-high-dimension}
In this section, we consider the high-dimensional setting $(p \gg n)$, in which we base the mirror statistics on the regularized estimator instead of the MLE. 
To better illustrate the idea, we first investigate the high-dimensional linear model, upon which we extend the discussion to GLMs. In addition, considering the computational feasibility, we focus on the data-splitting method.

\subsection{Linear models}
\subsubsection{Construction of the mirror statistics}
Assume the true data generating process is a linear model $y = X\beta^\star+\epsilon$ where $\epsilon$ follows $N(0, \sigma^2I_n)$.
We consider the random design setting, where $x_i$'s
are i.i.d. random vectors with population covariance matrix $\Sigma$. Without loss of generality, we assume $\Sigma_{jj} = 1$ for $j\in[p]$. 
In the high-dimensional setting, the mirror statistics are built upon the Lasso estimator \citep{tibshirani1996regression} defined as below,
\begin{equation}
\label{eq:linear-lasso}
\widehat{\beta}(y, X; \lambda) = \argmin_{\beta\in\mathbbm{R}^p}\left\{\frac{1}{2n}||y - X\beta||_2^2 + \lambda||\beta||_1\right\}.
\end{equation}

The Lasso estimator of the null feature is generally biased except for cases with  an orthogonal design matrix. 
In order to asymptotically remove the bias and symmetrize the Lasso estimator, we employ the debiasing approach introduced in \citet{javanmard2014confidence}, \citet{zhang2014confidence}, and \citet{van2014asymptotically}.
The debiased Lasso estimator $\widehat{\beta}^d$ takes the following simple form,
\begin{equation}
\label{eq:linear-debiased-lasso}
\widehat{\beta}^d = \widehat{\beta} + \frac{1}{n}DX^\intercal (y - X\widehat{\beta}),
\end{equation}
in which $D$ is a decorrelating matrix. Plugging in $y = X\beta^\star + \epsilon$, we obtain the following decomposition,
\begin{equation}
\label{eq:linear-decomposition}
\sqrt{n}(\widehat{\beta}^d - \beta^\star) = Z + \Delta,\ \ \ Z|X \sim N(0, \sigma^2D\widehat{\Sigma}D^\intercal),\ \ \ \Delta = \sqrt{n}(D\widehat{\Sigma} - I)(\beta^\star - \widehat{\beta}),
\end{equation}
where $\widehat{\Sigma} = (X^\intercal X)/n$ is the sample covariance matrix. Let $\Lambda = D\widehat{\Sigma}D^\top$. The two terms $\Delta$ and $\sigma^2\Lambda$ calibrate the asymptotic bias and variance of the debiased Lasso estimator, respectively.

Various proposals of the decorrelating matrix $D$ have been documented in the literature. \citet{javanmard2014confidence} proposed an optimization approach in order to simultaneously minimize the bias term $\Delta$ and the variance term $\Lambda$.
In this paper, we follow the approach used in \citet{javanmard2013nearly} and \citet{zhang2014confidence}, and set $D = \widehat{\Theta}$ as an estimator of the precision matrix $\Theta = \Sigma^{-1}$. One natural way to construct $\widehat{\Theta}$ is via regularized node-wise regression as detailed in Algorithm \ref{alg:precision-matrix}, which is based on the fact that $\Theta_{j, -j}$ corresponds to the coefficients of the best linear predictor of $X_j$ using $X_{-j}$. 

\begin{algorithm*}
\caption{Construction of the decorrelating matrix $\widehat{\Theta}$.}
\label{alg:precision-matrix}
\begin{enumerate}
\item Node-wise Lasso regression. For $j\in[p]$, let
\begin{equation}
\begin{aligned}
\text{(Linear model)}\ \ \ &\widehat{\gamma}_j = \argmin_{j\in\mathbbm{R}^{p - 1}}\left\{\frac{1}{2n}||X_j - X_{-j}\gamma||_2^2 + \lambda_j||\gamma||_1\right\};\\
\text{(GLM)}\ \ \ &\widehat{\gamma}_j = \argmin_{j\in\mathbbm{R}^{p - 1}}\left\{\frac{1}{2n}||X_{\widehat{\beta}, j} - X_{\widehat{\beta}, -j}\gamma||_2^2 + \lambda_j||\gamma||_1\right\}.
\end{aligned}
\end{equation}
\item Define $\widehat{C}$ with $\widehat{C}_{j,j} = 1$, and
$\widehat{C}_{j,k} = -\widehat{\gamma}_{j, k}$ for $k\neq j$, where  $\widehat{\gamma}_{j, k}$ is the $k$-th entry of $\widehat{\gamma}_j$. 
\item Let $\widehat{\Theta} = \widehat{G}^{-2}\widehat{C}$, in which $\widehat{G} = \text{diag}(\widehat{\tau}_1^2, \cdots, \widehat{\tau}_p^2)$ with
\begin{equation}
\begin{aligned}
\hspace{-2.7cm}\text{(Linear model)}\ \ \ &\widehat{\tau}_j^2 = (X_j - X_{-j}\widehat{\gamma}_j)^\intercal X_j/n;\\
\hspace{-2.7cm}\text{(GLM)}\ \ \ &\widehat{\tau}_j^2 = (X_{\widehat{\beta}, j} - X_{\widehat{\beta}, -j}\widehat{\gamma}_j)^\intercal X_j/n.
\end{aligned}
\end{equation}
\end{enumerate}
\end{algorithm*}

Under proper conditions, the bias term $\Delta$ vanishes asymptotically. Thus, the symmetry requirement in Condition \ref{cond:FDR-regression-coefficient-estimate} is satisfied since $\sqrt{n}\widehat\beta_j^d$ asymptotically follows $N(0, \sigma^2\Lambda_{jj})$ for $j\in S_0$. After the normalization by the asymptotic variance, we obtain the normalized debiased Lasso estimator:
\begin{equation}
\label{eq:linear-normalized-debiased-Lasso-estimator}
T_j = \widehat{\beta}_j^d/\widehat{\sigma}_j\ \ \ \text{with}\ \ \ \widehat{\sigma}_j^2 = \Lambda_{jj} = \big(\widehat{\Theta}\widehat{\Sigma}\widehat{\Theta}^\top\big)_{jj},\ \ \ \text{for}\ j\in[p].
\end{equation}

\begin{remark}
\label{remark:linear-scaling-high-dimension}
As discussed in Section \ref{subsec:GLM-data-splitting-moderate-dimension}, we can safely drop the constants $\sqrt{n}$ and $\sigma$ in the construction of the mirror statistics. Thus the data-splitting method is free of estimating the noise level $\sigma$. This scale-free property potentially makes our approach more appealing compared to BHq \citep{javanmard2019false}, which generally plugs in a consistent estimator of $\sigma$ (such as the one estimated by the scaled Lasso \citep{sun2012scaled}) in order to obtain asymptotically valid p-values. 
Empirically, we observe that when the features are moderately correlated, the scaled Lasso tends to over estimate the true noise level. In consequence, the asymptotic p-values of the relevant features are right-skewed, leading to a power loss.
Numerical comparisons can be found in Section \ref{subsub:numeric-linear-high-dimension}.
\end{remark}

The data-splitting method then proceeds by first randomly splitting the data into two halves, $(y^{(1)}, X^{(1)})$ and $(y^{(2)}, X^{(2)})$, and then computing the two independent
debiased Lasso estimates, $\widehat{\beta}^{(1, d)}$ and $\widehat{\beta}^{(2, d)}$ following Equation \eqref{eq:linear-debiased-lasso}, where $\widehat{\beta}^{(1)}$ and  $\widehat{\beta}^{(2)}$ are solutions to the optimization problem \eqref{eq:linear-lasso}, and $\widehat{\Theta}^{(1)}$ and $\widehat{\Theta}^{(2)}$ are computed by Algorithm \ref{alg:precision-matrix}. The normalized estimators $T^{(1)}$ and $T^{(2)}$ follow Equation \eqref{eq:linear-normalized-debiased-Lasso-estimator}, and the mirror statistic $M_j$ is constructed based on Equation \eqref{eq:mirror-statistic} using a user-specified function $f(u, v)$ satisfying Condition \ref{cond:mirror-statistics-f}.
A summary of the data-splitting method is given in Algorithm \ref{alg:linear-data-splitting-high-dimension}.

\begin{remark}
\citet{xing2019controlling} and \citet{Dai2020DS} also consider the high-dimensional linear models based on the same FDR control framework described in Section \ref{sec:review-FDR}. \citet{xing2019controlling} proposed to symmetrize the Lasso estimator via the post-selection procedure \citep{lee2016exact}, while \citet{Dai2020DS} introduced a Lasso + OLS  procedure, in which Lasso is first applied to one half of the data, and then OLS is applied to the other half of the data using the subset of features selected  by Lasso. The symmetry requirement in Condition \ref{cond:FDR-regression-coefficient-estimate} is satisfied as long as  all relevant features are selected by Lasso in the first step. However, this may not be justified for GLMs. In contrast, as shown in Section \ref{sec:GLM-high-dimension}, the debiasing approach naturally adapts to high-dimensional GLMs. 
\end{remark}

\begin{algorithm*}
\caption{The data-splitting method for linear models in the high-dimensional setting.}
\label{alg:linear-data-splitting-high-dimension}
\begin{enumerate}
\item Split the data set into two equal-sized halves $(y^{(1)}, X^{(1)})$ and $(y^{(2)}, X^{(2)})$.
\item Construct the normalized debiased Lasso estimator on each part of the data.
\begin{enumerate}
\item Calculate the Lasso estimators $\widehat{\beta}^{(1)}$ and $\widehat{\beta}^{(2)}$ via the optimization problem \eqref{eq:linear-lasso}.
\item Estimate $\widehat{\Theta}^{(1)}$ and $\widehat{\Theta}^{(2)}$ following Algorithm \ref{alg:precision-matrix}. For $j\in [p]$, let 
\begin{equation}
{\widehat{\sigma}^2_j}{}^{(1)}  = \big(\widehat{\Theta}^{(1)}\widehat{\Sigma}^{(1)}\widehat{\Theta}^{(1)} {}^\top\big)_{jj},\ \ \ {\widehat{\sigma}^2_j}{}^{(2)} = \big(\widehat{\Theta}^{(2)}\widehat{\Sigma}^{(2)}{\widehat{\Theta}^{(2)}} {}^\top\big)_{jj},
\end{equation}
in which $\widehat{\Sigma}^{(1)}$ and $\widehat{\Sigma}^{(2)}$ are the sample covariance matrices of $X^{(1)}$ and $X^{(2)}$, respectively. 
\item Calculate the debiased Lasso estimators $\widehat{\beta}^{(1, d)}$ and $\widehat{\beta}^{(2, d)}$ following Equation \eqref{eq:linear-debiased-lasso}.\\ For $j\in[p]$, calculate the mirror statistic $M_j$ based on  
\begin{equation}
T^{(1)}_j = {\widehat{\beta}_j^{(1, d)}}/{\widehat{\sigma}_j}^{(1)},\ \ \ T^{(2)}_j = {\widehat{\beta}_j^{(2, d)}}/{\widehat{\sigma}_j}^{(2)}.
\end{equation}
\end{enumerate}
\item Select the features using Algorithm \ref{alg:FDR-control-framework}. 
\end{enumerate}
\end{algorithm*}

\newpage
\subsubsection{Theoretical justification of the data-splitting method} 
We require the following assumptions. 
\begin{assumption} 
\label{assump:linear-high-dimension}
\text{}
\begin{enumerate}[(1)]
\item The sparsity conditions.
\begin{enumerate}
\item The sparsity condition on $\Theta$: $s = \max_{i\in[p]}|\{j\in[p],\ \Theta_{ij} \neq 0\}| = o(\sqrt{n}/\log p)$.
\item The sparsity condition on the number of signals: $p_1 = |\{j\in[p], \beta_j^\star \neq 0\}| = o(\sqrt{n}/\log p)$.
\end{enumerate}
\item Requirements for the design matrix $X$.
\begin{enumerate}
\item The rows of $X\Theta^{1/2}$ are sub-Gaussian.
\item $1/C\leq \sigma_{\min}(\Sigma)\leq\sigma_{\max}(\Sigma)\leq C$, for some constant $C > 0$. 
\end{enumerate}
\item The sample size requirement: $\sqrt{n}/\log p \to \infty$.
\end{enumerate}
\end{assumption}

\begin{remark}
In contrast to the moderate-dimensional setting, we require proper sparsity conditions on the true coefficient vector $\beta^\star$ as well as the precision matrix $\Theta$. The former is required to ensure that the Lasso estimator enjoys a fast convergence rate \citep{bickel2009simultaneous}. The latter also appears in \citet{javanmard2013nearly} and \citet{van2014asymptotically}, which implies that 
$||\widehat{\Theta} - \Theta||_\infty = o_p(1/\sqrt{\log p})$. The two sparsity conditions, along with proper conditions on the design matrix, ensure that the bias term $\Delta$ vanishes asymptotically in the sense that $||\Delta||_\infty = O_p(p_1\log p/\sqrt{n}) = o_p(1)$. 
\end{remark}

The following proposition shows that the data-splitting method achieves an asymptotic FDR control for high-dimensional linear models.
\begin{proposition}
\label{prop:linear-data-splitting-FDR-high-dimension}
For any given FDR control level $q \in (0,1)$, 
we assume that the pointwise limit $\text{FDP}^\infty(t)$ of $\text{FDP}(t)$ exists for all $t > 0$, and there is a $t_q > 0$ such that $\text{FDP}^\infty(t_q) \leq q.$ Then,
under Assumption \ref{assump:linear-high-dimension}, we have
\begin{equation}\nonumber
\limsup_{n, p \to\infty}\mathbbm{E}\left[\frac{\#\{j: j \in S_0, j \in \widehat{S}_{\tau_q}\}}{\# \{j: j \in \widehat{S}_{\tau_q}\}\vee 1}\right] \leq q,
\end{equation}
using the data-splitting method outlined in Algorithm \ref{alg:linear-data-splitting-high-dimension}.
\end{proposition}

\subsection{Generalized linear models}
\subsubsection{Construction of the mirror statistics}
In this section, we adapt the debiasing approach to the GLM defined in \eqref{eq:GLM-moderate-dimension}. The Lasso estimator of the true coefficient vector $\beta^\star$ in \eqref{eq:GLM-moderate-dimension} is 
\begin{equation}
\label{eq:GLM-lasso}
\widehat{\beta}(y, X; \lambda) = \argmin_{\beta\in\mathbbm{R}^p}\left\{\frac{1}{2n}\sum_{i = 1}^n\ell(y_i, x_i^\intercal\beta) + \lambda||\beta||_1\right\}.
\end{equation}
We refer to $\ell(u, v) = -uv + \rho(v)$ as the loss function associated with the GLM, which is essentially the negative log-likelihood up to an additive constant.

For the ease of presentation, we introduce the following notations. The first and second derivatives of $\ell(u,v)$ with respect to $v$ are denoted as $\dot{\ell}(u, v)$ and $\ddot{\ell}(u, v)$, respectively. 
The gradient and the Hessian of $\ell(y, x^\intercal\beta)$ with respect to $\beta$ are denoted as $\dot{\ell}_{\beta}(y, x)$ and $\ddot{\ell}_\beta(y, x)$, respectively.\footnote{To clarify, both $\dot{\ell}(u, v)$ and $\ddot{\ell}(u, v)$ are scalar, whereas $\dot{\ell}_{\beta}(y, x)$ is a $p\times1$ vector and $\ddot{\ell}_\beta(y, x)$ is a $p\times p$ matrix.}
For a general mapping $g$ defined on an arbitrary data point $(y, x)$, let $P_ng = \sum_{i = 1}^ng(y_i, x_i)/n$ and $Pg = \mathbbm{E}\left[P_ng\right]$, in which the expectation is taken with respect to the randomness in both the response variable $y_i$ and the features $x_i$.
Let $W_\beta$ be a $n\times n$ diagonal matrix with $W^2_{i.i} = \ddot{\rho}( x_i^\intercal\beta)$.
Then the sample version of the Hessian matrix can be written as $P_n\ddot{\ell}_\beta = X_{\beta}^\intercal X_{\beta}/n$, in which $X_{\beta} = W_\beta X$ is the weighted design matrix. 

We consider the following debiased Lasso estimator for high-dimensional GLMs \citep{van2014asymptotically},
\begin{equation}
\label{eq:GLM-debiased-lasso}
\widehat{\beta}^d = \widehat{\beta} - \widehat{\Theta}P_n\dot{\ell}_{\widehat{\beta}}.
\end{equation}
This is a natural generalization of the debiased Lasso estimator for linear models (see Equation \eqref{eq:linear-debiased-lasso}), where $\widehat{\Theta}$ serves as the decorrelating matrix $D$, and $\dot{\ell}_{\widehat\beta}(y, x)$ simplifies to $-x(y - x^\intercal\widehat\beta)$ in the linear model. 
Let $\Sigma = \mathbbm{E}[X_{\beta^\star}^\intercal X_{\beta^\star}]/n$ be the population Hessian matrix evaluated at the true coefficient vector $\beta^\star$. Similar to the linear model, we set $\widehat{\Theta}$ as an estimator of $\Theta = \Sigma^{-1}$, which is again constructed via regularized node-wise regression (see Algorithm \ref{alg:precision-matrix}). 
By analogy with the linear model, we have a similar decomposition as Equation \eqref{eq:linear-decomposition},
\begin{equation}
\label{eq:GLM-decomposition}
\sqrt{n}(\widehat\beta_j^d - \beta_j^\star) = Z_j + \Delta_j,\ \ \ \text{for}\ j\in[p],
\end{equation}
in which $Z_j$ is the asymptotically dominant term defined as below, 
\begin{equation}
Z_j = -{\sqrt{n}\Theta_{j, \cdot}P_n\dot{\ell}_{\beta^\star}} = -{\sqrt{n}}\sum_{i = 1}^n\Theta_{j,\cdot}x_i[-y_i + \dot{\rho}( x_i^\intercal\beta^\star)].
\end{equation}
$\Theta_{j,\cdot}$ denotes the $j$-th row of $\Theta$. One major difference between GLMs and linear models is that, conditioning on the design matrix, $Z_j$ is not exactly normal but only asymptotically normal by the central limit theorem. Fortunately, we can easily quantify the discrepancy between the law of $Z_j$ and the normal distribution using the Berry-Essen theorem. For the bias term $\Delta$, we show that it can be asymptotically ignored under proper conditions.

The decomposition in Equation \eqref{eq:GLM-decomposition} suggests a normalized debiased Lasso estimator as below,
\begin{equation}
T_j = \widehat{\beta}_j^d/\widehat{\sigma}_j\ \ \ \text{with}\ \ \ \widehat{\sigma}_j^2 = (\widehat{\Theta}P_n\dot{\ell}_{\widehat{\beta}}\dot{\ell}_{\widehat{\beta}}^\top\widehat{\Theta}^\top)_{jj},\ \ \ \text{for}\ j\in[p],
\end{equation}
in which ${\widehat{\sigma}^2_j}$ is a consistent estimator of the asymptotic variance 
$$\sigma^2_j = (\Theta \mathbbm{E}[P_n\dot{\ell}_{\beta^\star}\dot{\ell}_{\beta^\star}^\top]\Theta)_{jj} = (\Theta\Sigma\Theta)_{jj} = \Theta_{jj}.$$ 
The symmetric requirement in Condition \ref{cond:FDR-regression-coefficient-estimate} is satisfied since for a null feature $j\in S_0$, $T_j$ is asymptotically centered at 0. A summary of the data-splitting method for high-dimensional GLMs is given in Algorithm \ref{alg:GLM-data-splitting-high-dimension}.

\begin{algorithm*}
\caption{The data-splitting method for GLMs in the high-dimensional setting.}
\label{alg:GLM-data-splitting-high-dimension}
\begin{enumerate}
\item Split the data set into two equal-sized halves $(y^{(1)}, X^{(1)})$ and $(y^{(2)}, X^{(2)})$.
\item Construct the normalized debiased Lasso estimator on each part of the data.
\begin{enumerate}
\item Calculate the Lasso estimators $\widehat{\beta}^{(1)}$ and $\widehat{\beta}^{(2)}$ via the optimization problem \eqref{eq:GLM-lasso}.
\item Estimate $\widehat{\Theta}^{(1)}$ and $\widehat{\Theta}^{(2)}$ following Algorithm \ref{alg:precision-matrix}. For $j\in [p]$, let
\begin{equation}
{\widehat{\sigma}^2_j}{}^{(1)} = \big(\widehat{\Theta}^{(1)}P_n\dot{\ell}_{\widehat{\beta}^{(1)}}\dot{\ell}_{\widehat{\beta}^{(1)}}^\top\widehat{\Theta}^{(1)} {}^\top\big)_{jj},\ \ \ {\widehat{\sigma}^2_j}{}^{(2)} = \big(\widehat{\Theta}^{(2)}P_n\dot{\ell}_{\widehat{\beta}^{(2)}}\dot{\ell}_{\widehat{\beta}^{(2)}}^\top\widehat{\Theta}^{(2)} {}^\top\big)_{jj}.
\end{equation}
\item Calculate the debiased Lasso estimators $\widehat{\beta}^{(1, d)}$ and $\widehat{\beta}^{(2, d)}$ following Equation \eqref{eq:GLM-debiased-lasso}. \\
For $j\in[p]$, calculate the mirror statistic $M_j$ based on
\begin{equation}
T^{(1)}_j = {\widehat{\beta}_j^{(1, d)}}/{\widehat{\sigma}_j}^{(1)},\ \ \ T^{(2)}_j = {\widehat{\beta}_j^{(2, d)}}/{\widehat{\sigma}_j}^{(2)}.
\end{equation}
\end{enumerate}
\item Select the features using Algorithm \ref{alg:FDR-control-framework}. 
\end{enumerate}
\end{algorithm*}

\subsubsection{Theoretical justification of the data-splitting method}
We require the following assumptions.
\begin{assumption} 
\label{assump:GLM-high-dimension}
\text{}
\begin{enumerate}[(1)]
\item The sparsity conditions.
\begin{enumerate}
\item The sparsity condition on $\Theta$. $s = \max_{i\in[p]}|\{j\in[p],\ \Theta_{ij} \neq 0\}| = o(\sqrt{n}/\log p)$.
\item The sparsity condition on the number of signals. $p_1 = |\{j\in[p], \beta_j^\star \neq 0\}| = o(\sqrt{n}/\log p)$.
\end{enumerate}
\newpage
\item Requirements for the design matrix $X$ and the weighted design matrix $X_{\beta^\star}$. 
\begin{enumerate}
\item There exists a constant $C_1 > 0$ such that
\begin{equation}
||X||_{\infty} \leq C_1,\ \ \ ||X\beta^\star||_{\infty} \leq C_1,\ \ \ ||X_{\beta^\star}||_{\infty} \leq C_1,\ \ \ ||X_{\beta^\star, -j}\gamma_j||_{\infty} \leq C_1,\footnote{$\gamma_j$ corresponds to the coefficient vector of the best linear predictor of $X_{\beta^\star, j}$ using $X_{\beta^\star, -j}$. More precisely, we have
$\gamma_j = \argmin_{\gamma\in\mathbbm{R}^{p - 1}}\mathbbm{E}\left[||X_{\beta^\star, j} - X_{\beta^\star, -j}\gamma||_2^2\right]$.}\ \ \ \forall j\in[p].
\end{equation}
\item $1/C_2\leq \sigma_{\min}(\Sigma)\leq\sigma_{\max}(\Sigma)\leq C_2$, for some constant $C_2 > 0$. 
\end{enumerate}
\item The regularity conditions on the link function $\rho$.
\begin{enumerate}
\item $\ddot{\rho}(v)$ is Lipschitz continuous for $|v| \leq C_1$. 
\item $|\dot{\rho}(v)|$ and $|\ddot{\rho}(v)|$ are upper bounded for $|v| \leq C_1$.
\end{enumerate}
\item Requirements for the sample size: $\sqrt{n}/\log p \to \infty$.
\end{enumerate}
\end{assumption}

\begin{remark}
For mathematical convenience, we consider bounded (weighted) design matrices in this paper, although the theoretical results can be possibly generalized to more general cases such as sub-Gaussian (weighted) design  matrices.
Similar regularity conditions on the link function $\rho$ also appears in \citet{van2014asymptotically}, and hold for popular GLMs including the logistic regression model, the Poisson regression model, and the negative binomial regression
model.
\end{remark}

The following propositions show that the bias term $\Delta$ vanishes asymptotically, and the data-splitting method achieves an asymptotic FDR control for high-dimensional GLMs.
\begin{proposition}
\label{prop:GLM-bias-high-dimension}
Under Assumption \ref{assump:GLM-high-dimension}, we have $||\Delta||_\infty = o_p(1)$.
\end{proposition}

\begin{proposition}
\label{prop:GLM-data-splitting-FDR-high-dimension}
For any given FDR control level $q \in (0,1)$, 
we assume that the pointwise limit $\text{FDP}^\infty(t)$ of $\text{FDP}(t)$ exists for all $t > 0$, and there is a $t_q > 0$ such that $\text{FDP}^\infty(t_q) \leq q.$ Then,
under Assumption \ref{assump:GLM-high-dimension}, we have
\begin{equation}\nonumber
\limsup_{n,p\to\infty}\mathbbm{E}\left[\frac{\#\{j: j \in S_0, j \in \widehat{S}_{\tau_q}\}}{\# \{j: j \in \widehat{S}_{\tau_q}\}\vee 1}\right] \leq q,
\end{equation}
using the data-splitting method outlined in Algorithm \ref{alg:GLM-data-splitting-high-dimension}.
\end{proposition}

\subsection{Comparison with existing methods}
\label{subsec:GLM-comparison-high-dimension}
The knockoff filter is a class of recently developed methods, which achieves the FDR control by creating knockoff features in a similar spirit as  spike-ins in biological experiments. The knockoff filter does not require calculating individual p-values, and can be applied to fairly general settings without having to know the underlying true relationship between the response variable and the associated features. In particular, the model-X knockoff filter \citep{candes2018panning}, as well as some further developments including the DeepPINK filter \citep{lu2018deeppink} and the conditional knockoff filter \citep{huang2019relaxing}, can be applied to select features for an arbitrary GLM. 
Compared to the data-splitting method, one major limitation of the knockoff filter is that it requires the knowledge of the joint distribution of the features. In addition, we empirically observe that the power of the knockoff filter can deteriorate rapidly as correlations among features increase.

BHq is also potentially applicable for the FDR control in GLMs once we obtain the asymptotic p-value for each feature based on the asymptotic normality of the debiased Lasso estimator. Developments along this line include \citet{javanmard2019false} and \citet{ma2020global}. The former focuses on high-dimensional linear models, while the latter focuses on high-dimensional logistic regression models. Although both the data-splitting method and BHq rely on the asymptotic property of the debiased Lasso estimator, the symmetry requirement is less stringent and more likely to be satisfied in  finite-sample cases compared to the normality requirement. In particular, we empirically observe that the scale of the normalized debiased Lasso estimator, i.e., the scale of the Z-score, can be quite off compared to the scale of the standard normal distribution. We note that if the scaling factor is 
under-estimated, BHq is at the risk of losing the FDR control, since the resulting p-values of the null features will skew to the left. On the other hand, if the scaling factor is over-estimated, BHq can be over conservative, 
leading to a possible power loss. In contrast, the data-splitting method is scale free, i.e., the scaling factor does not materially change the selection result, thus is potentially more robust  compared to BHq.

\begin{figure*}
\begin{center}
\includegraphics[width=0.33\columnwidth]{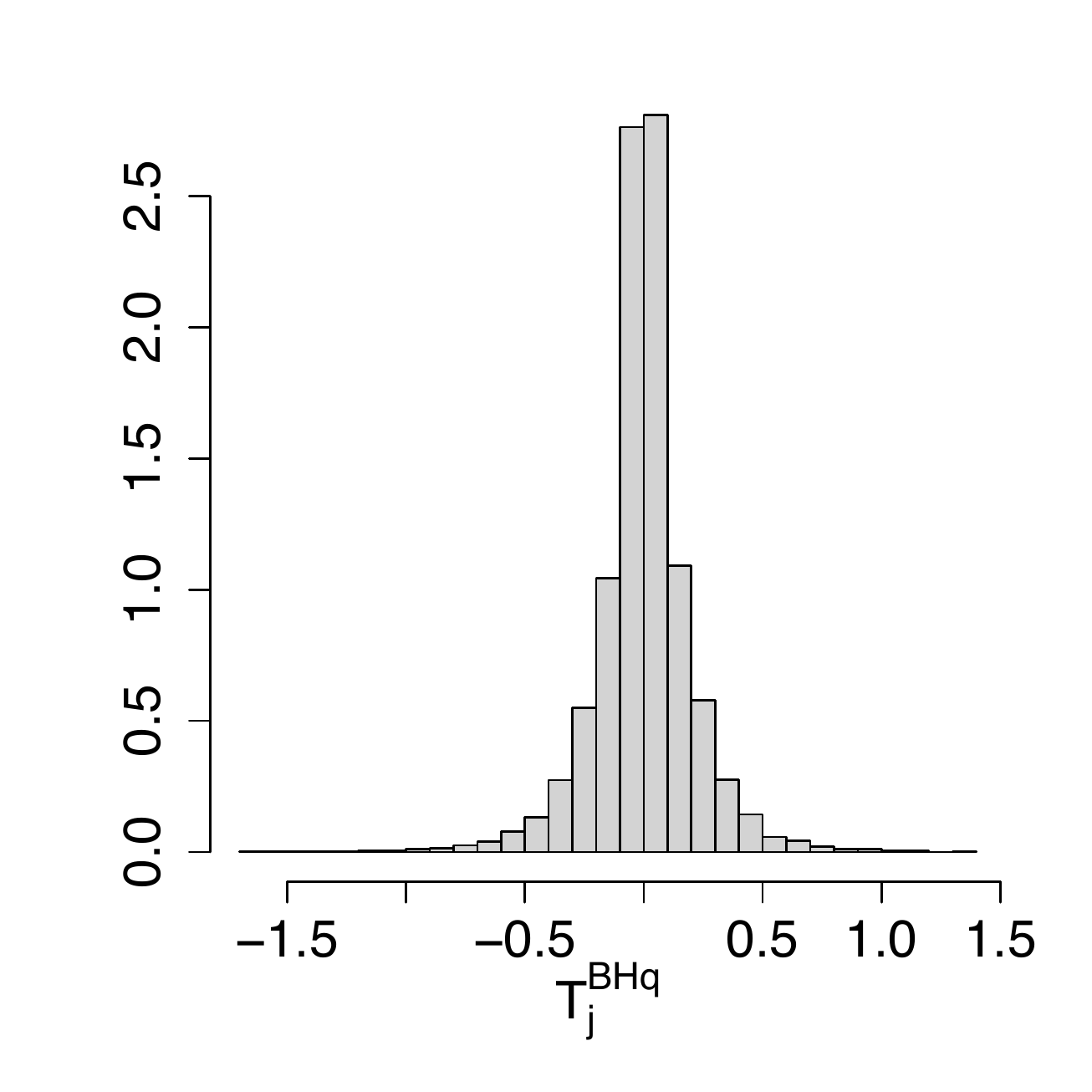}
\includegraphics[width=0.33\columnwidth]{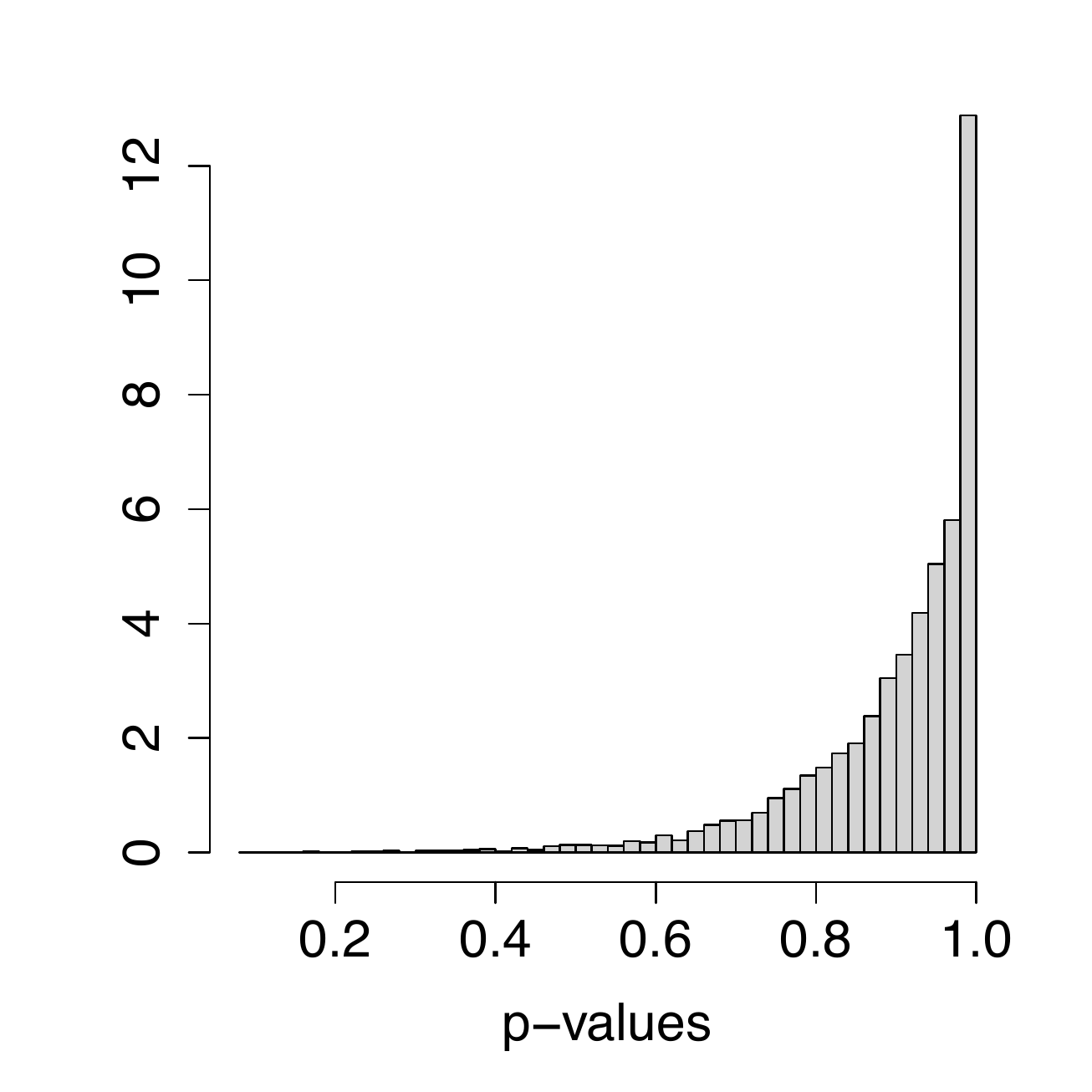}
\includegraphics[width=0.33\columnwidth]{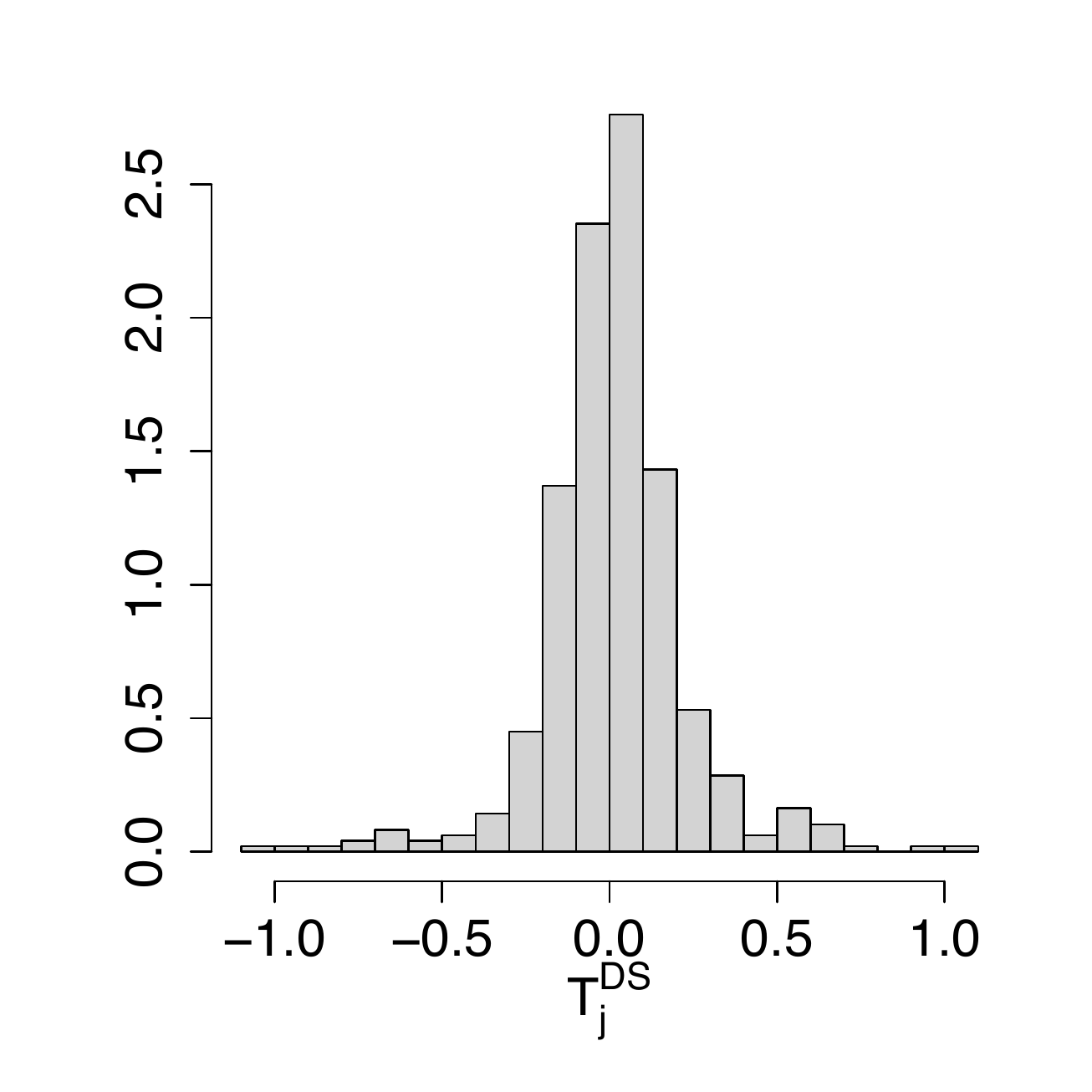}
\includegraphics[width=0.33\columnwidth]{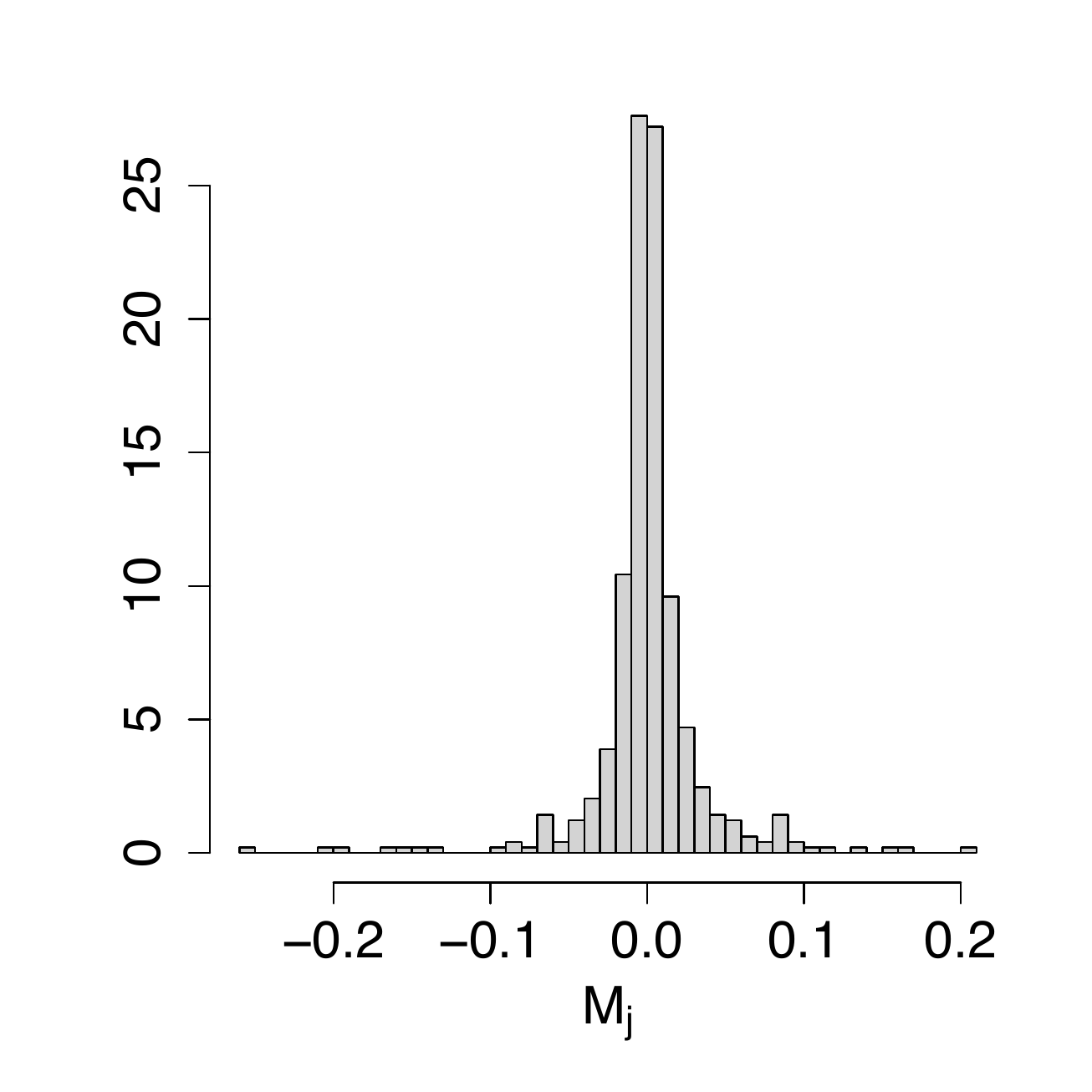}
\end{center}
\caption{A logistic regression model with $n = 250$, $p = 500$ and $p_1 = 10$. Each row of the design matrix is generated from $N(0, I_p)$. The true coefficient for $j\in S_1$ is set to be $\beta_j^\star = \pm 4$ with equal probability. Top left: The normalized debiased Lasso estimators $T_j^{\text{BHq}}$ of the null features. Top right: The asymptotic p-values of the null features. Both histograms are generated based on 20 independent runs of the algorithm in \citet{ma2020global}. Bottom left: The normalized debiased Lasso estimators $T_j^{\text{DS}}$ of the null features. Bottom right: The mirror statistics of the null features. Both histograms are generated based on a single run of Algorithm \ref{alg:GLM-data-splitting-high-dimension}.}
\label{fig:logistic-pvalues-high-dimension}
\end{figure*}

Figure \ref{fig:logistic-pvalues-high-dimension} illustrates the above discussion in a  logistic regression model with $n = 250$, $p = 500$ and $p_1 = 10$. Detailed algorithmic settings can be found in the Figure caption. We obtain the normalized debiased Lasso estimator $T_j^{\text{BHq}}$ and $T_j^{\text{DS}}$ via the method proposed in \citet{ma2020global} and Algorithm \ref{alg:GLM-data-splitting-high-dimension}, respectively. 
We see that, for BHq, the scale of $T_j^{\text{BHq}}$ is much smaller compared to the scale of the standard normal distribution, thus the resulting p-values of the null features are significantly right-skewed. In contrast, for the data-splitting method, the symmetry requirement in Condition \ref{cond:FDR-regression-coefficient-estimate} is well satisfied.

\section{Numerical Illustrations}
\label{sec:numerical-illustration}
To remind the readers, we use the following abbreviations DS, MDS, BHq, GM, and Knockoff to denote the single data-splitting method, the multiple data-splitting method, the Benjamini-Hochberg procedure, the Gaussian mirror method, and the model-X knockoff filter,\footnote{We use the R package \textit{knockoff} to implement the model-X knockoff filter. See \url{https://cran.r-project.org/web/packages/knockoff} for the documentation.} respectively. We will clarify the exact implementations of these methods for different numerical examples. 
For MDS, we repeat the selection for 50 times and aggregate the results using Algorithm \ref{alg:multiple-data-splits}.
In addition, for all the synthetic examples except for the linear model in Section \ref{subsub:numeric-linear-high-dimension}, we keep $|\beta_j^\star|$ the same across relevant features $j\in S_1$, and randomly generate their signs with equal probability. The elements of $S_1$ are randomly drawn from $\{1,\ldots,p\}$.
Furthermore, with a bit abuse of terminology, we refer to $|\beta_j^\star|$ for $j\in S_1$ as the signal strength throughout these synthetic examples. 
For all the examples in Section \ref{sec:numerical-illustration}, we construct the mirror statistic following Equation \eqref{eq:mirror-statistic} with $f(u, v) = uv$. The designated FDR control level is set to be $q = 0.1$ henceforth. 

\subsection{The moderate-dimensional setting}
\label{subsec:numeric-moderate-dimension}
\subsubsection{Logistic regression}
\label{subsub:moderate-dimensional-logistic}
We consider two moderate-dimensional settings for the logistic regression model. The first setting is the classical small-$n$-and-$p$ setting, in which  sample size $n = 500$ and dimension $p = 60$, so that the dimension-to-sample ratio of $\kappa = p/n = 0.12$. The second setting concerns with the large-$n$-and-$p$ setting, in which $n = 3000$,  $p = 500$, and  $\kappa  = 1/6$.
The number of relevant features $p_1=p-p_0$ is 30 in the small-$n$-and-$p$ setting, and 50 in the large-$n$-and-$p$ setting. 
In both settings, we consider six competing methods based on the MLE, including DS, MDS, GM, BHq along with its adjusted version ABHq, and Knockoff. The implementation details of DS and GM are given in Algorithm \ref{alg:GLM-data-splitting-moderate-dimension} and \ref{alg:GLM-Gaussian-mirror-moderate-dimension}, respectively. 
BHq utilizes the classical asymptotic p-values calculated via the Fisher information, whereas ABHq is based on the adjusted asymptotic p-values derived recently by \cite{sur2019modern}. 

Each row of the design matrix is independently drawn from the multivariate normal distribution $N(0, \Sigma)$ with a Toeplitz correlation structure, i.e., $\Sigma_{ij} = r^{|i - j|}$.
The variance of each feature is then standardized to be $1/n$.
In Section \ref{subsub:logistic-moderate-dimension-additional-result} of the Appendix, we report additional results for different types of covariance matrix $\Sigma$, including the case where features have constant pairwise correlation. 

In this example, we vary (a) the correlation $r$; (b) the signal strength $|\beta_j^\star|$ for $j\in S_1$. 
In the small-$n$-and-$p$ setting, for scenario (a), we fix the signal strength at $|\beta_j^\star| = 6.5$ for $j\in S_1$, and vary the correlation $r$ from 0.0 to 0.4, whereas for scenario (b), we fix the correlation at $r = 0.2$, and vary the signal strength from 4.5 to 6.5.
In the large-$n$-and-$p$ setting, for scenario (a), we fix the signal strength at $|\beta_j^\star| = 11$ for $j\in S_1$, and vary the correlation $r$ from 0.0 to 0.4, whereas for scenario (b), we fix the correlation at $r = 0.2$, and vary the signal strength from 8 to 12.

\begin{figure*}
\begin{center}
\includegraphics[width=0.45\columnwidth]{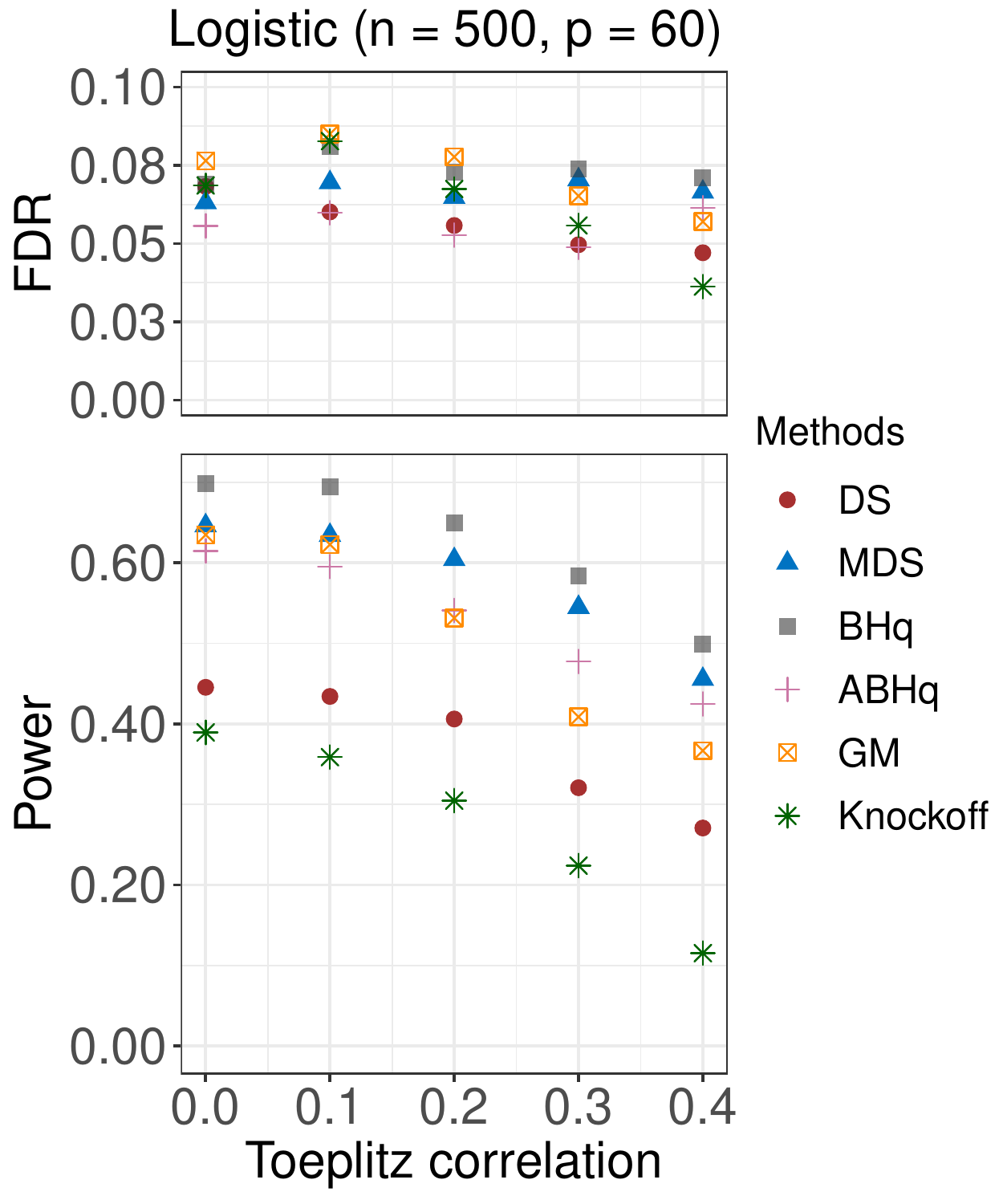}
\includegraphics[width=0.45\columnwidth]{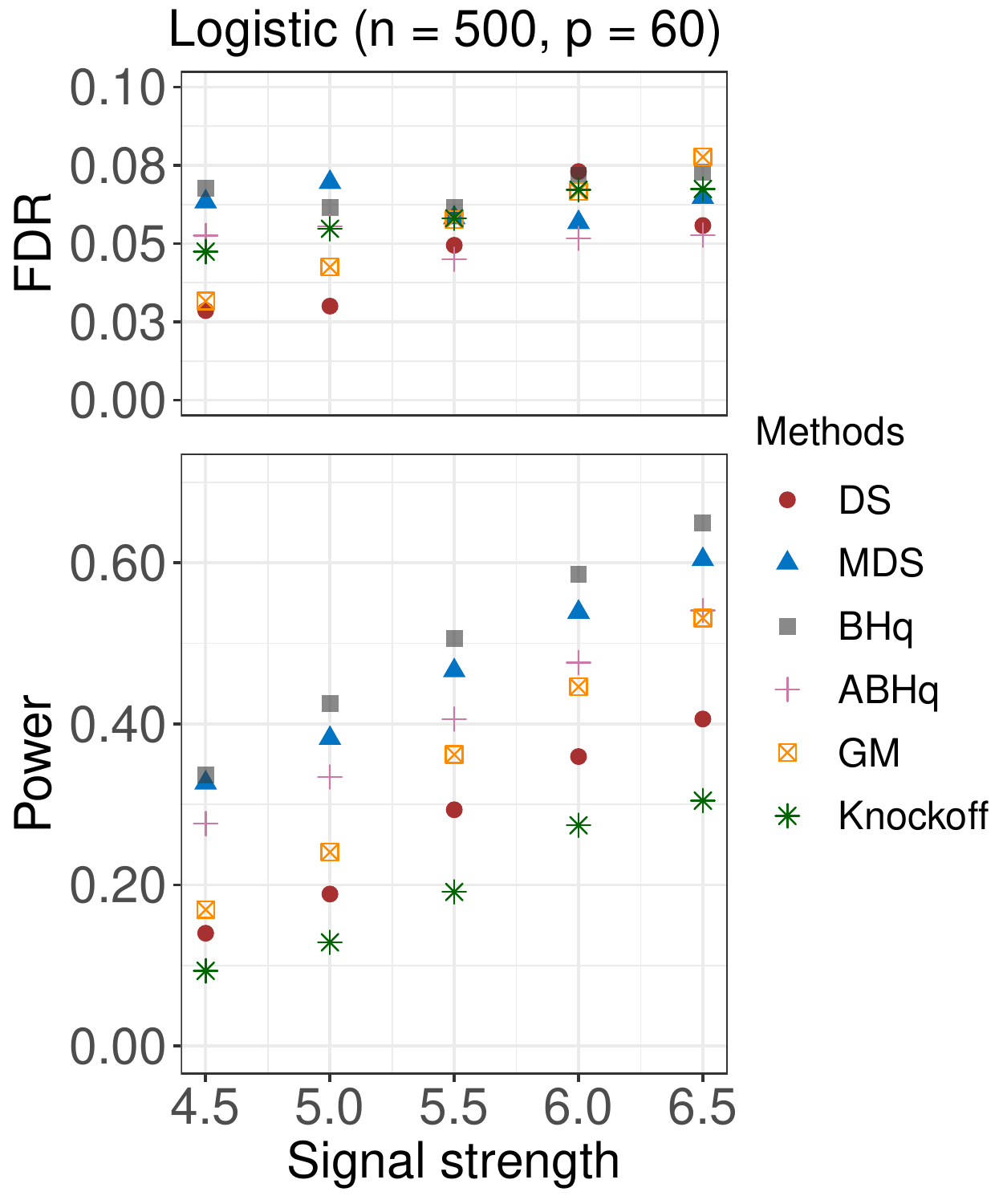}
\end{center}
\caption{Empirical FDRs and powers for the logistic regression model in the small-$n$-and-$p$ setting. 
Each row of the design matrix is first independently drawn from $N(0, \Sigma)$ with a Toeplitz correlation structure, i.e., $\Sigma_{ij} = r^{|i - j|}$.
The variance of each feature is then standardized to be $1/n$.
The power is defined as the ratio of  the identified versus all relevant features.
In the left panel, we fix the signal strength at $|\beta_j^\star| = 6.5$ for $j\in S_1$ and vary the correlation $r$.
In the right panel, we fix the correlation at $r = 0.2$ and vary the signal strength.
The number of relevant features is 30 across all settings, and the designated FDR control level is $q = 0.1$. 
Each dot represents the average from  50 independent runs.}
\label{fig:moderate-dimensional-logistic-small-n-and-p-toeplitz-correlation}
\end{figure*}

The empirical FDRs and powers of different methods in the small-$n$-and-$p$ setting are summarized in Figure \ref{fig:moderate-dimensional-logistic-small-n-and-p-toeplitz-correlation}. The FDRs of the six competing methods are under control across all settings. 
In terms of the power, BHq is the leading method, and performs the best in all cases.
MDS is the second best method, having a slightly lower power than BHq.
We observe that ABHq is less powerful than BHq, indicating  that the asymptotics for the p-value adjustment is not ready to kick in when the sample size $n$ and the dimension $p$ are relatively small. 

\begin{figure*}
\begin{center}
\includegraphics[width=0.45\columnwidth]{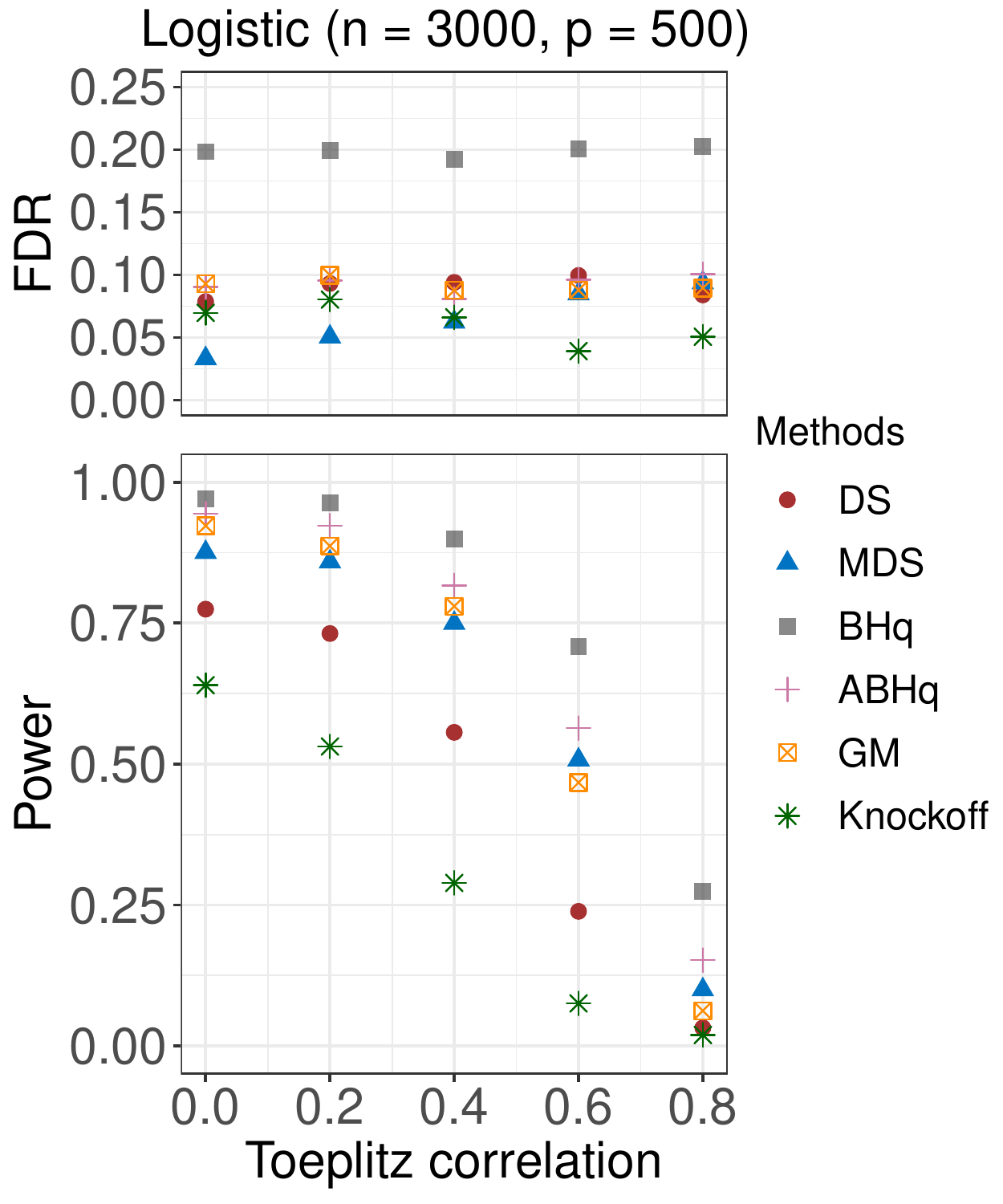}
\includegraphics[width=0.45\columnwidth]{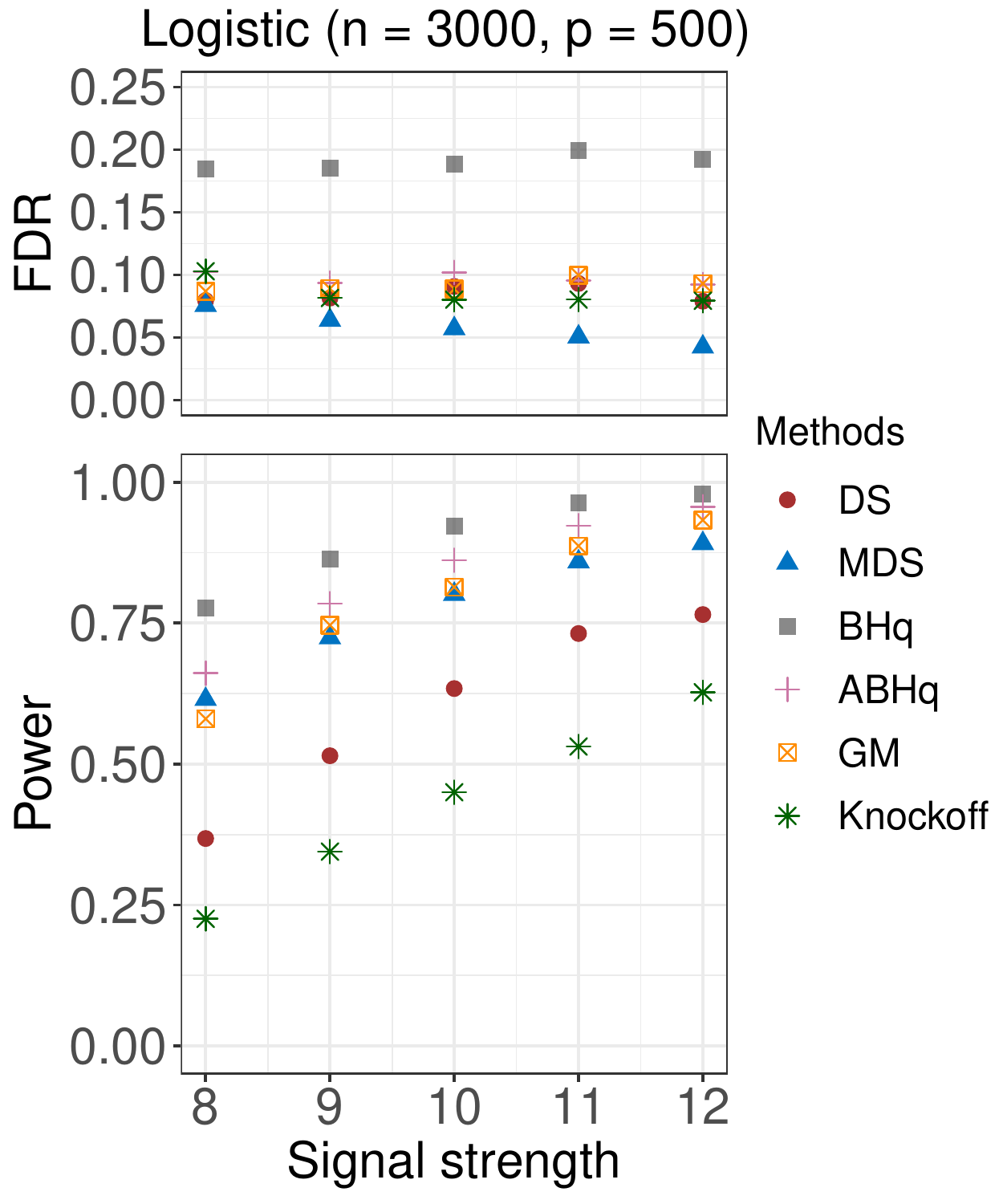}
\end{center}
\caption{Empirical FDRs and powers for the logistic regression model in the large-$n$-and-$p$ setting. 
The design matrix is simulated similarly as per Figure \ref{fig:moderate-dimensional-logistic-small-n-and-p-toeplitz-correlation}.
In the left panel, we fix the signal strength at $|\beta_j^\star| = 11$ for $j\in S_1$, and vary the correlation $r$.
In the right panel, we fix the correlation at $r = 0.2$, and vary the signal strength.
The number of relevant features is 50 across all settings, and
the designated FDR control level is $q = 0.1$. 
Each dot represents the average from  50 independent runs.}
\label{fig:moderate-dimensional-logistic-large-n-and-p-toeplitz-correlation}
\end{figure*}

After adding $p$ knockoff variables created by the second-order method with a James-Stein-type shrinkage applied to the estimated covariance matrix \citep{barber2015controlling}, the sample covariance matrix of the augmented set of features appears almost singular; thus the resulting MLEs are unstable. To overcome this issue, we multiply the vector $s$ output from the SDP program (see Equation (2.4)
in \citet{barber2015controlling}) by a factor 0.9. The power of the resulting knockoff filter, as shown in Figures \ref{fig:moderate-dimensional-logistic-small-n-and-p-toeplitz-correlation} and \ref{fig:moderate-dimensional-logistic-large-n-and-p-toeplitz-correlation},  is still  unsatisfactory. We also tested out the recently proposed conditional knockoff filter \citep{huang2019relaxing}, which is free of the covariance matrix estimation, but observed no improvements in power. 

Except for MDS, all the perturbation-based methods,  such as GM, DS, and Knockoff are not as powerful as the p-value-based methods. 
A possible reason is that when $p$ and $n$ are small, the types of perturbations in these methods may have diluted the signal too much. More interestingly, however, MDS gains back almost all the lost power due to perturbations without sacrificing FDR controls. 

The empirical FDRs and powers of different methods in the large-$n$-and-$p$ setting are summarized in Figure \ref{fig:moderate-dimensional-logistic-large-n-and-p-toeplitz-correlation}. We see that BHq loses the FDR control because the classical asymptotic p-value of the null feature, calculated based on the Fisher information, is non-uniform and skew to the left. In contrast, ABHq still controls  the FDR well and enjoys the best power, which verifies the adjusted asymptotic distribution of the MLE derived in \citet{sur2019modern}. MDS and GM also perform competitively, and have similar but slightly lower power compared to ABHq.
GM shows much improved performances compared with the small $n$-and-$p$ setting.
Knockoff performs poorly for the same reason as mentioned above. 

\subsubsection{Negative binomial regression}
\label{subsub:moderate-dimensional-negative-binomial}
We consider a negative binomial regression model, in which the dispersion parameter is 2, i.e., the target number of successful trials is $2$. We set sample size $n = 3000$, and dimension $p = 500$, resulting in a dimension-to-sample ratio of $\kappa  = 1/6$. 
We simulate the design matrix the same way as in Section \ref{subsub:moderate-dimensional-logistic}, and
 vary (a) the correlation $r$; (b) the signal strength $|\beta_j^\star|$ for $j\in S_1$.
In scenario (a), we fix the signal strength at $|\beta_j^\star| = 6$ for $j\in S_1$, and vary the correlation $r$ from 0.1 to 0.5. 
In scenario (b), we fix the correlation at $r = 0.2$, and vary the signal strength from 4 to 8. 
The number of relevant features is 50 across all settings, i.e., $p_1 = 50$ and $p_0 = 450$. In Section \ref{subsub:negative-binomial-moderate-dimension-additional-result} of the Appendix, we report additional results for the case where features have constant pairwise correlation.

We consider five competing methods based on the MLE, including DS, MDS, BHq, GM and Knockoff. The implementation details of DS and GM are given in Algorithms \ref{alg:GLM-data-splitting-moderate-dimension} and \ref{alg:GLM-Gaussian-mirror-moderate-dimension}, respectively. BHq is based upon the classical asymptotic p-values calculated via the Fisher information. Although we expect such p-values to be non-uniform for the null features, the exact asymptotic distribution of the MLE under this moderate-dimensional setting has not been derived in the literature, thus no proper adjustment of the p-values exists to the best of our knowledge. Knockoff is implemented in the same way as in Section \ref{subsub:moderate-dimensional-logistic} to overcome the degeneracy issue. 

\begin{figure*}
\begin{center}
\includegraphics[width=0.45\columnwidth]{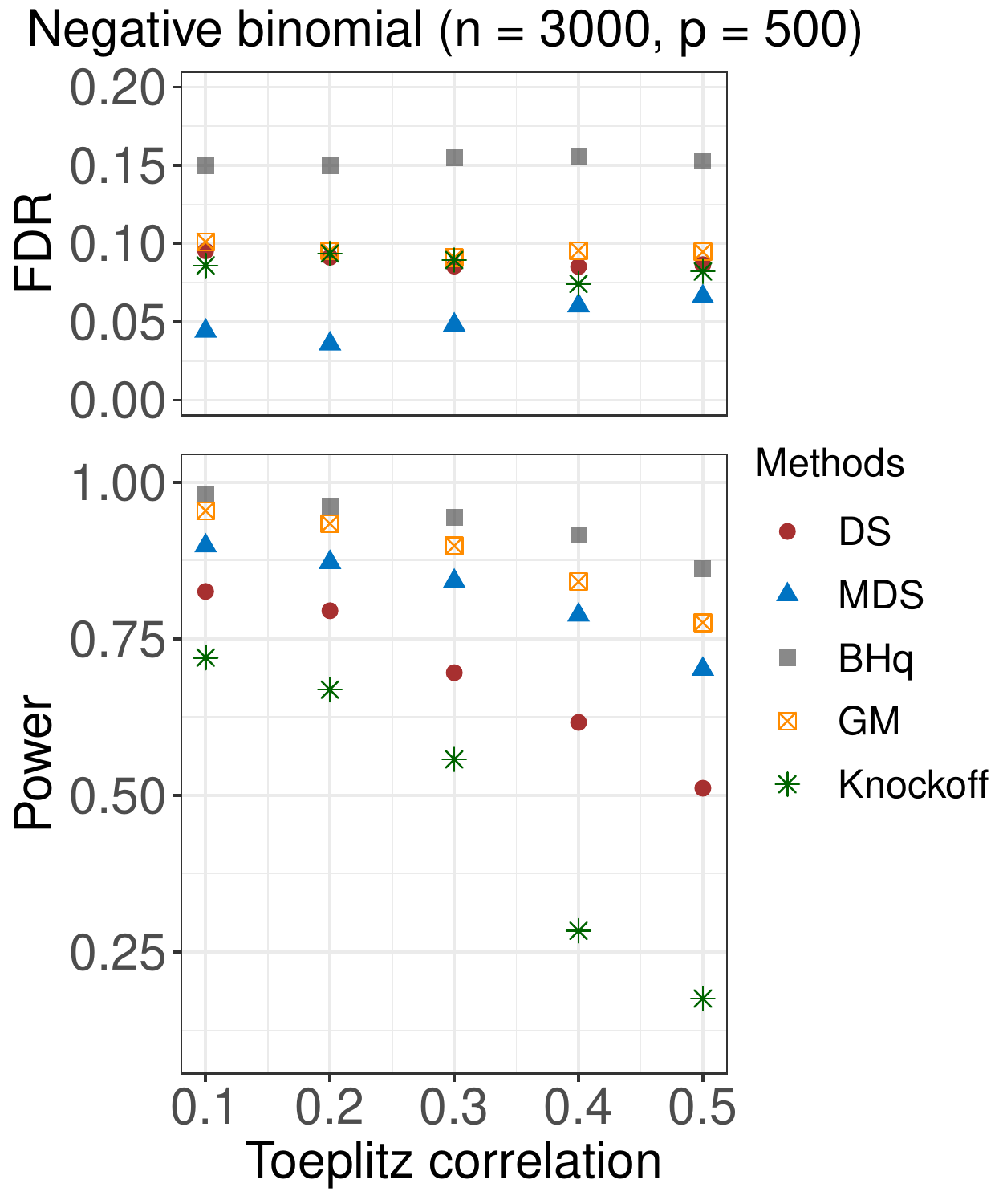}
\includegraphics[width=0.45\columnwidth]{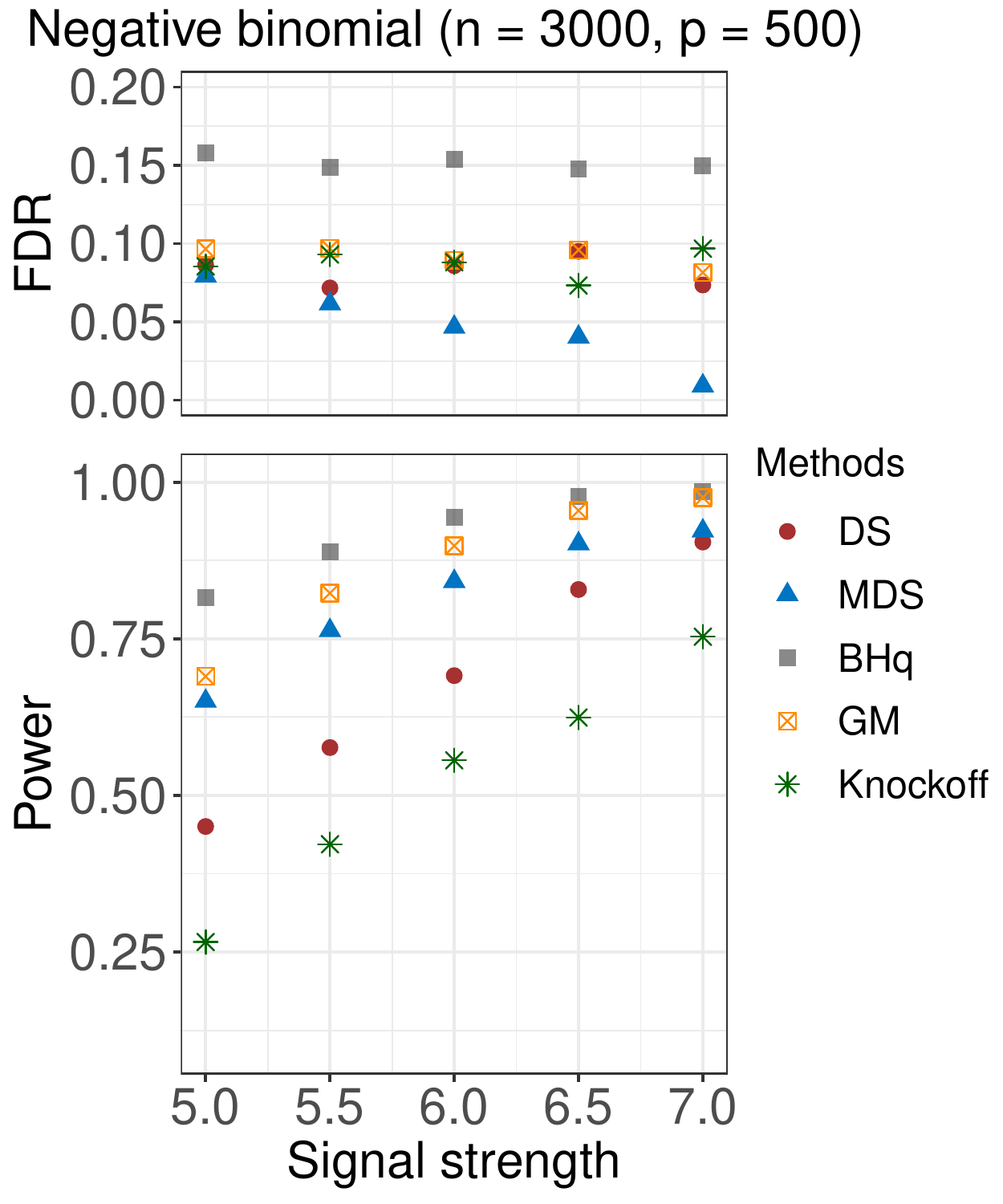}
\end{center}
\caption{Empirical FDRs and powers for the negative binomial regression model. 
The design matrix is simulated similarly as per Figure \ref{fig:moderate-dimensional-logistic-small-n-and-p-toeplitz-correlation}.
In the left panel, we fix the signal strength at $|\beta_j^\star| = 6$ for $j\in S_1$, and vary the correlation $r$.
In the right panel, we fix the correlation at $r = 0.3$, and vary the signal strength.
The number of relevant features is 50 across all setting, and
the designated FDR control level is $q = 0.1$,
Each dot represents the average from 50 independent runs.}
\label{fig:moderate-dimensional-negative-binomial-toeplitz-correlation}
\end{figure*}

The empirical FDRs and powers of different methods are summarized in Figure \ref{fig:moderate-dimensional-negative-binomial-toeplitz-correlation}. We see that BHq is the only method losing the FDR control, because of the non-uniformity (skew to the left) of the p-values for the null features. Among the methods with FDR control, GM and MDS consistently perform the best over different levels of correlation and signal strength. MDS has a silghtly lower power but also a lower FDR compared with GM, and is significantly better than DS in the sense that it simultaneously reduces the FDR and boosts the power. 
Knockoff has the lowest power among all competing methods for the same reason as discussed in Section \ref{subsub:moderate-dimensional-logistic}.

\subsection{The high-dimensional setting}
\label{subsec:numeric-high-dimension}
\subsubsection{Linear regression}
\label{subsub:numeric-linear-high-dimension}
We consider the Gaussian linear model $y = X\beta^\star + \epsilon, \ \epsilon\sim N(0,I_n),$ with sample size $n = 800$ and  dimension $p = 2000$. Each row of the design matrix  is drawn independently from the multivariate Gaussian $N(0, \Sigma)$ with a Toeplitz correlation structure. More precisely, $\Sigma$ is a blockwise diagonal matrix, consisting of 10 identical unit diagonal Toeplitz matrices. The detailed formula involves a correlation factor $r \in (0, 1)$, and is given in Section \ref{subsub:linear-high-dimension-additional-result} of the Appendix.
Features are more correlated with a larger $r$.
We independently sample $\beta_j^\star$ for $j\in S_1$ from a centered normal distribution, of which the standard deviation is referred to as the signal strength. 

In this example, we vary (a) the correlation factor $r$; (b) the signal strength.
In scenario (a), we fix the signal strength at $6\sqrt{\log p/n}$, and vary the correlation factor $r$ from $0.0$ to $0.8$, whereas in scenario (b), we fix the correlation factor at $r = 0.6$, and vary the signal strength from $2$ to $10$ up to a multiplicative constant $\sqrt{\log p/n}$. 
The number of relevant features is 70 across all settings, i.e., $p_1 = 70$. 
In Section \ref{subsub:linear-high-dimension-additional-result} of the Appendix, we report additional results for the case where features have constant pairwise correlation. 

We consider four competing methods, including DS, MDS, the BHq procedure outlined in \citet{javanmard2019false}, and Knockoff. DS and MDS are based on the debiased Lasso estimator, with implementation details given in Algorithm \ref{alg:linear-data-splitting-high-dimension}. 
BHq utilizes the same debiasing approach, and estimates the noise level using scaled Lasso. 
For Knockoff, we use the second-order method to create multivariate normal knockoffs.

\begin{figure*}
\begin{center}
\includegraphics[width=0.45\columnwidth]{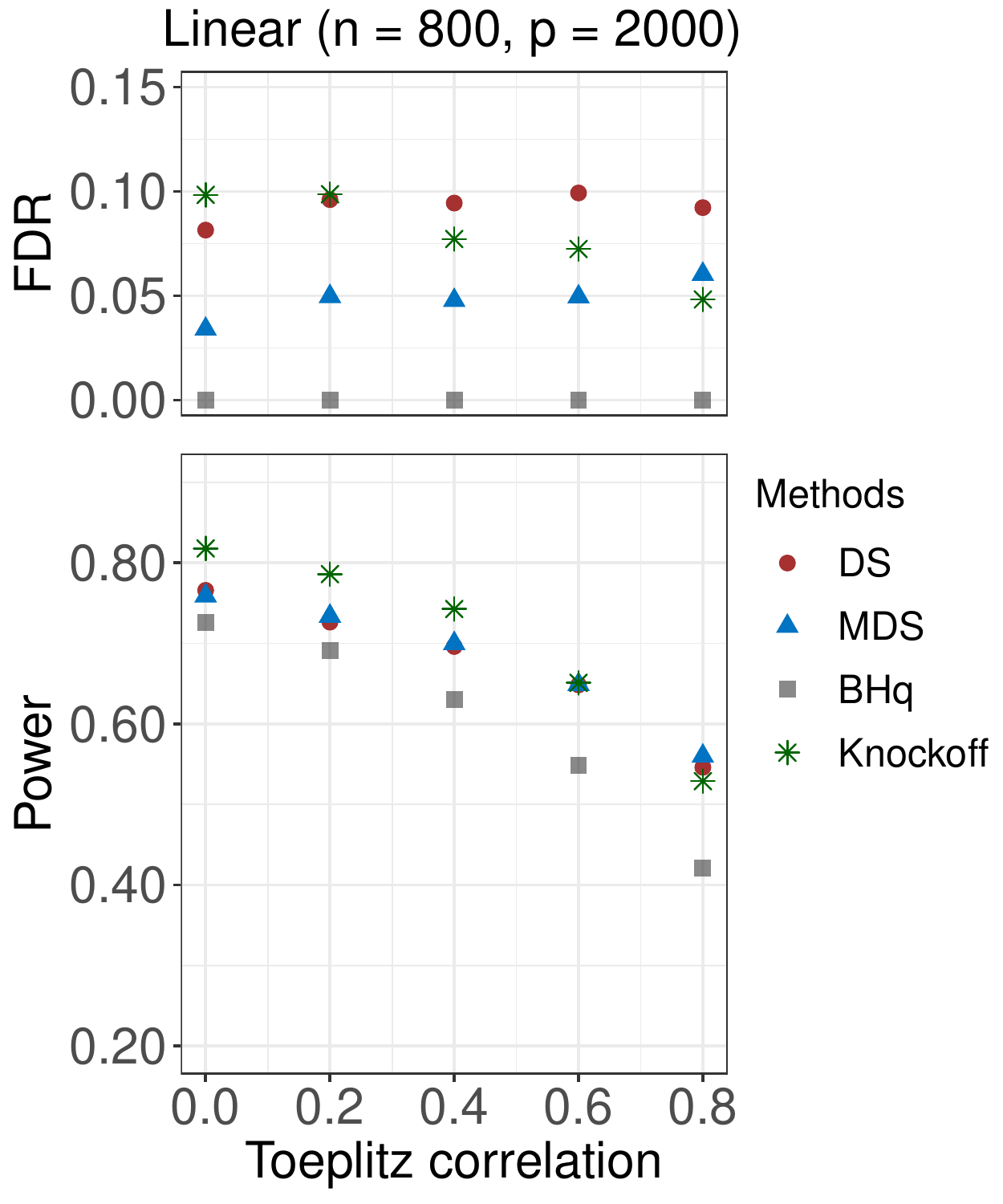}
\includegraphics[width=0.45\columnwidth]{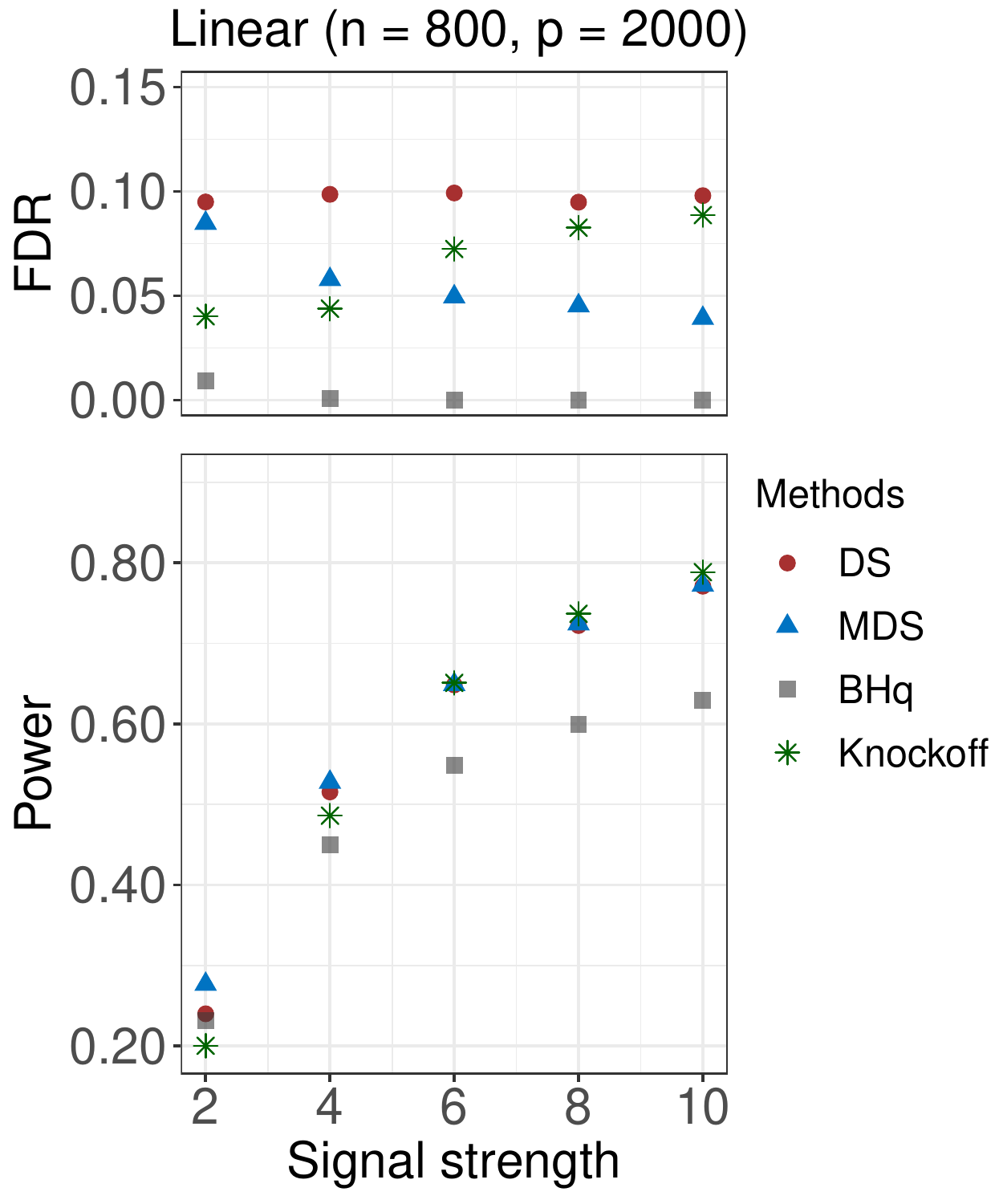}
\end{center}
\caption{Empirical FDRs and powers for the linear model. Each row of the design matrix is independently drawn from $N(0, \Sigma)$ with $\Sigma$ being a blockwise diagonal matrix that consists of 10 identical unit diagonal Toeplitz matrices (see Section \ref{subsub:linear-high-dimension-additional-result} of the Appendix).
$\beta_j^\star$ for $j\in S_1$ are i.i.d. samples from $N(0,s^2)$, where
$s$ 
is referred to as the signal strength. 
The signal strength along the x-axis in the right panel shows multiples of 
$\sqrt{\log p/n}$.
In the left panel, we fix the signal strength at $6\sqrt{\log p/n}$ and vary the correlation factor $r$. 
In the right panel, we fix the correlation factor at $r = 0.6$ and vary the signal strength.
The number of relevant features is 70 across all settings, and
the designated FDR control level is $q = 0.1$.
Each dot in the figure represents the average from 50 independent runs.}
\label{fig:linear-high-dimension-toeplitz-correlation}
\end{figure*}

The empirical FDRs and powers of different methods are summarized in Figure \ref{fig:linear-high-dimension-toeplitz-correlation}. The FDRs of the four competing methods are under control across all settings. In particular, MDS has much lower FDRs than DS, also significantly lower than the nominal level.
In terms of the power, Knockoff achieves the highest power in most cases except when the signal strength is low. 
We also observe that when the correlation among relevant features increases, such as when the design matrix has constant pairwise correlation (Appendix, Section~\ref{subsub:linear-high-dimension-additional-result}),
the power of Knockoff decreases rapidly and can be much lower compared with the other three  methods.
DS and MDS also perform competitively and consistently enjoy a higher power than BHq over different levels of correlation and signal strength.
We see that the empirical FDRs of BHq are nearly zero across all settings. One possible reason is that the scaled Lasso over estimated the noise level, thus the p-values largely skew to the right and make the procedure too conservative. The numerical results suggest that bypassing the task of estimating the noise level enables the data-splitting methods to gain a substantial advantage over BHq.

\subsubsection{Logistic regression}
\label{subsub:high-dimensional-logistic}
We consider a case 
with sample size $n = 800$ and  dimension $p = 2000$. Each row of the design matrix is drawn independently from $N(\bm{0}, \Sigma)$. We consider a similar setup as in \citet{ma2020global}, in which $\Sigma = 0.1\times\Sigma_B$, where $\Sigma_B$ is the blockwise diagonal matrix introduced in Section \ref{subsub:numeric-linear-high-dimension} with $r = 0.1$.
In this example, we vary (a) the sparsity level $p_1$, i.e., the number of relevant features; (b) the signal strength $|\beta^\star_j|$ for $j\in S_1$. In scenario (a), we fix the signal strength at $|\beta_j^\star| = 4$ for $j\in S_1$, and vary the sparsity level $p_1$ from 40 to 80. In scenario (b), we fix the sparsity level at $p_1 = 60$, and vary the signal strength from 2 to 6.

We consider five competing methods, including DS, MDS, two BHq procedures (BHq-I and BHq-II), and Knockoff. DS and MDS are based on the debiased Lasso estimator, with implementation details given in Algorithm \ref{alg:GLM-data-splitting-high-dimension}. BHq-II uses the same debiasing approach, whereas BHq-I, which corresponds to the method proposed in \citet{ma2020global}, utilizes a different debiasing approach. 
For Knockoff, we use the second-order method to create multivariate normal knockoffs.

\begin{figure*}
\begin{center}
\includegraphics[width=0.45\columnwidth]{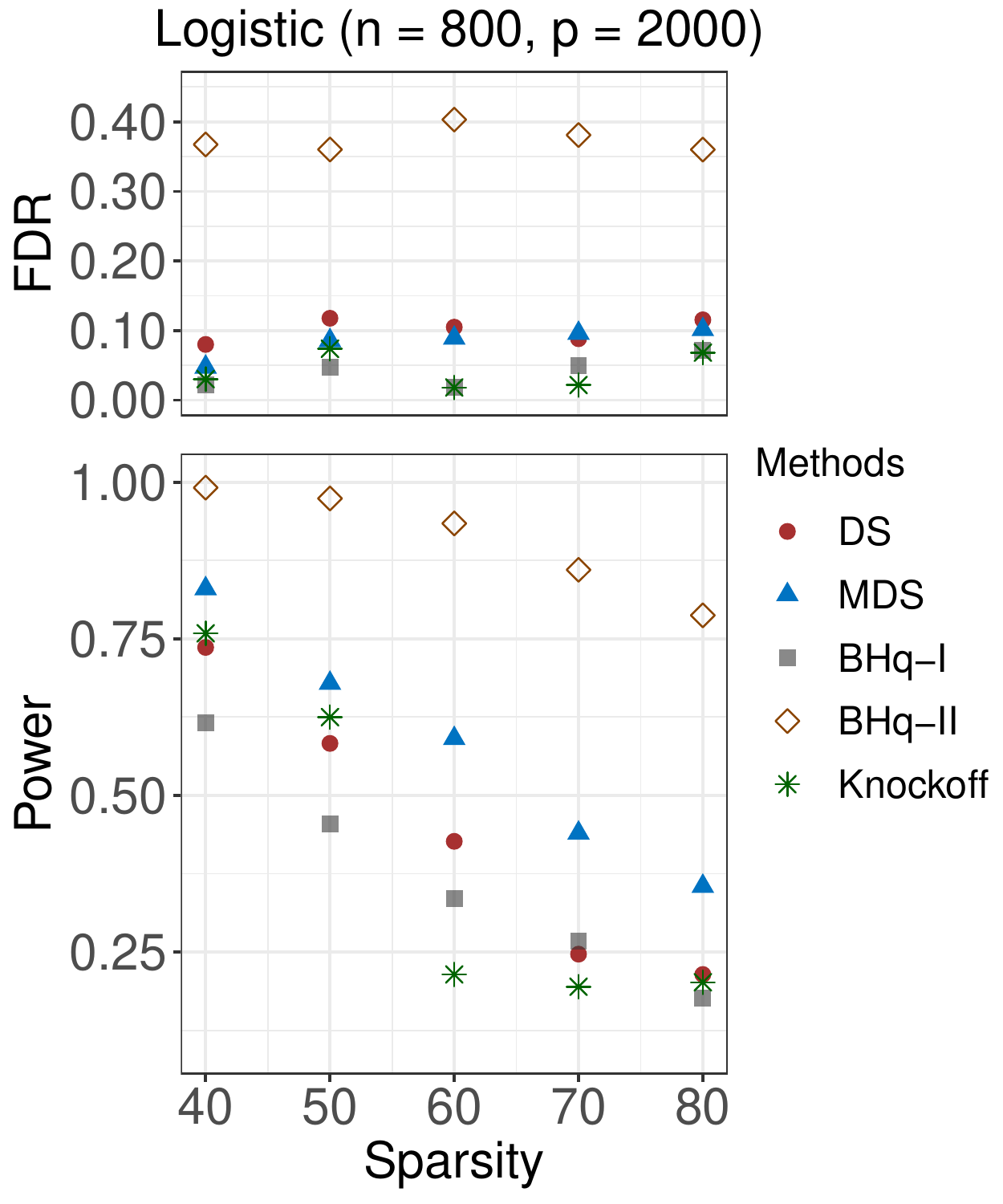}
\includegraphics[width=0.45\columnwidth]{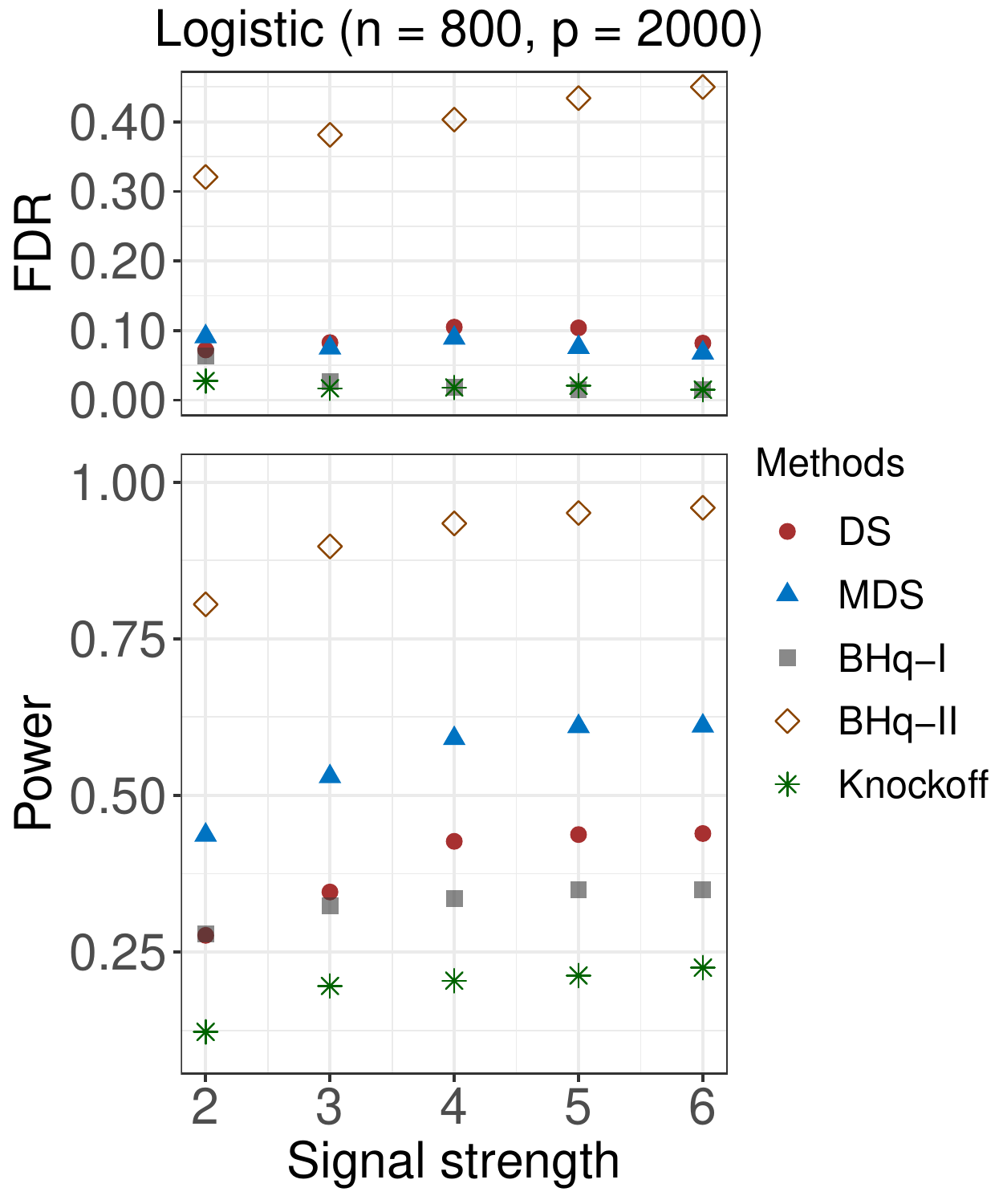}
\end{center}
\caption{Empirical FDRs and powers for the logistic regression model.  
The rows of the design matrix are i.i.d. samples from
$N(0_p, 0.1 \times \Sigma_B)$, where $\Sigma_B$ is a blockwise diagonal matrix that consists of 10 identical unit diagonal Toeplitz matrices (see Section \ref{subsub:linear-high-dimension-additional-result} of the Appendix with $r = 0.1$).
In the left panel, we fix the signal strength at $|\beta_j^\star| = 4$ for $j\in S_1$, and vary the sparsity level $p_1$. 
In the right panel, we fix the sparsity level at $p_1 = 60$, and vary the signal strength.
The designated FDR level is $q = 0.1$. Each dot represents the average 
of 50 independent runs.}
\label{fig:high-dimensional-logistic}
\end{figure*}

The empirical FDRs and powers of different methods are summarized in Figure \ref{fig:high-dimensional-logistic}. We see that  all methods except BHq-II control the FDR successfully. In terms of power, MDS is the leading method across different levels of sparsity and  signal strength, and 
enjoys a significant improvement over DS.
Even DS appears to be  more powerful than BHq-I, suggesting that the p-values constructed following \citet{ma2020global} can be highly non-informative (skew to the right) in finite-sample cases. Knockoff performs competitively when the number of relevant features is small (e.g., $p_1 \leq 50$), but can potentially suffer when the relevant features become denser, i.e., $p_1$ gets larger.

\subsection{Real data application}
\label{subsec:real-data}
Compared with traditional bulk RNA sequencing technologies, single-cell RNA sequencing (scRNAseq) allows researchers to examine the sequence information of each individual cell, which promises to lead to new biological discoveries ranging from cancer genomics to metagenomics. In this section, we consider the task of selecting relevant genes with respect to the glucocorticoid response in a human breast cancer cell line, using the scRNAseq  data in \citet{hoffman2020single}. A total of 400 T47D A1–2 human breast cancer cells were treated with 100 nM synthetic glucocorticoid dexamethasone (Dex) at 1, 2, 4, 8, and 18h time points. 
An scRNASeq experiment  was performed at each time point, which results in a total of 2,000 samples of gene expression for the treatment group. For the control group, there are 400 vehicle-treated control cells. An scRNAseq experiment was performed at the 18h timepoint to obtain the corresponding profile of gene expression. After proper normalization, the final scRNAseq dataset\footnote{The dataset is available at \url{https://www.ncbi.nlm.nih.gov/geo/query/acc.cgi?acc=GSE141834}.} contains 2,400 samples, each with 32,049 gene expressions. To further reduce the dimensionality, we first screen out the genes detected in fewer than 10\% of cells, and then pick up the top 500 most variable genes following \citet{hoffman2020single}.

We consider a logistic regression model with $n = 2,400$ and $p = 500$. Since the MLE does not exist for this dataset,  we cannot use the method in \citet{sur2019modern} to obtain relevant p-values.
Instead, we use the debiased Lasso estimator and apply DS outlined in Algorithm \ref{alg:GLM-data-splitting-high-dimension} to this dataset. As the sample size $n$ is  larger than the dimension $p$, we directly estimate the precision matrix $\Theta$ (see Equation \eqref{eq:GLM-debiased-lasso} and the discussion therein) by inverting the sample Hessian matrix $\widehat{\Sigma}$, instead of using the regularized node-wise regression outlined in Algorithm \ref{alg:precision-matrix}.
Two other methods are tested out as well, including the BHq procedure outlined in \citet{ma2020global} and Knockoff.

\begin{figure*}
\begin{center}
\includegraphics[width=0.4\columnwidth]{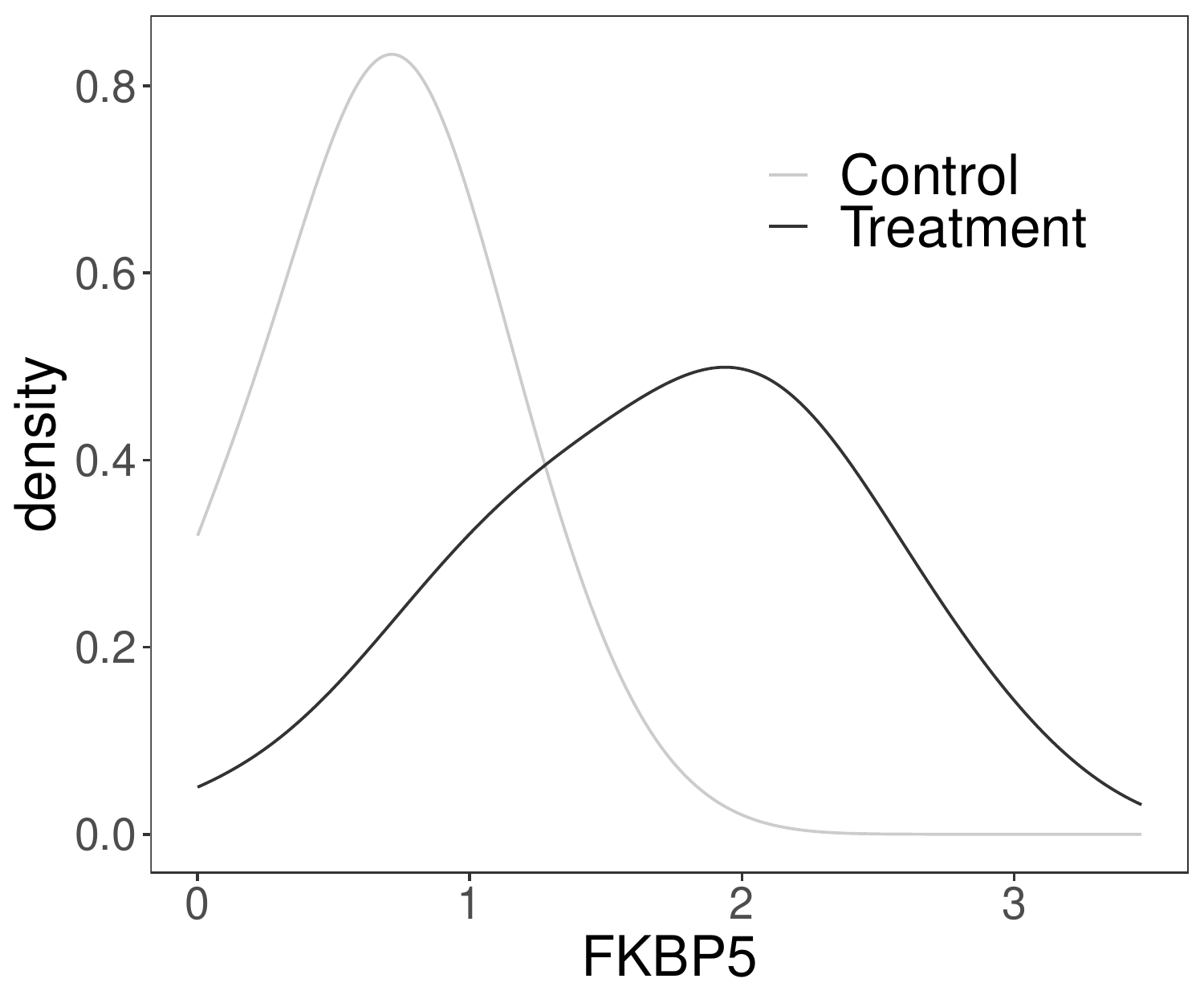}
\includegraphics[width=0.4\columnwidth]{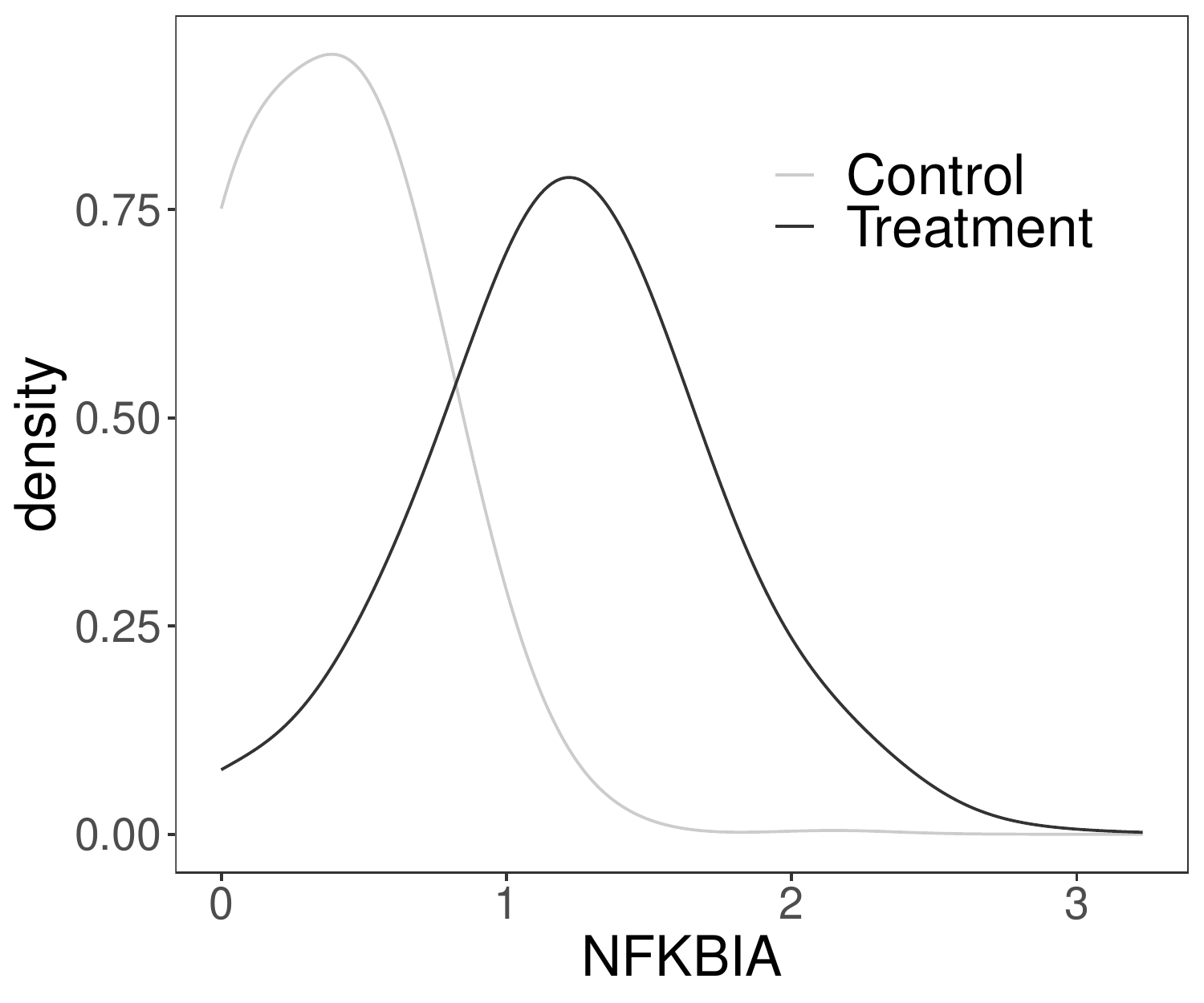}
\includegraphics[width=0.4\columnwidth]{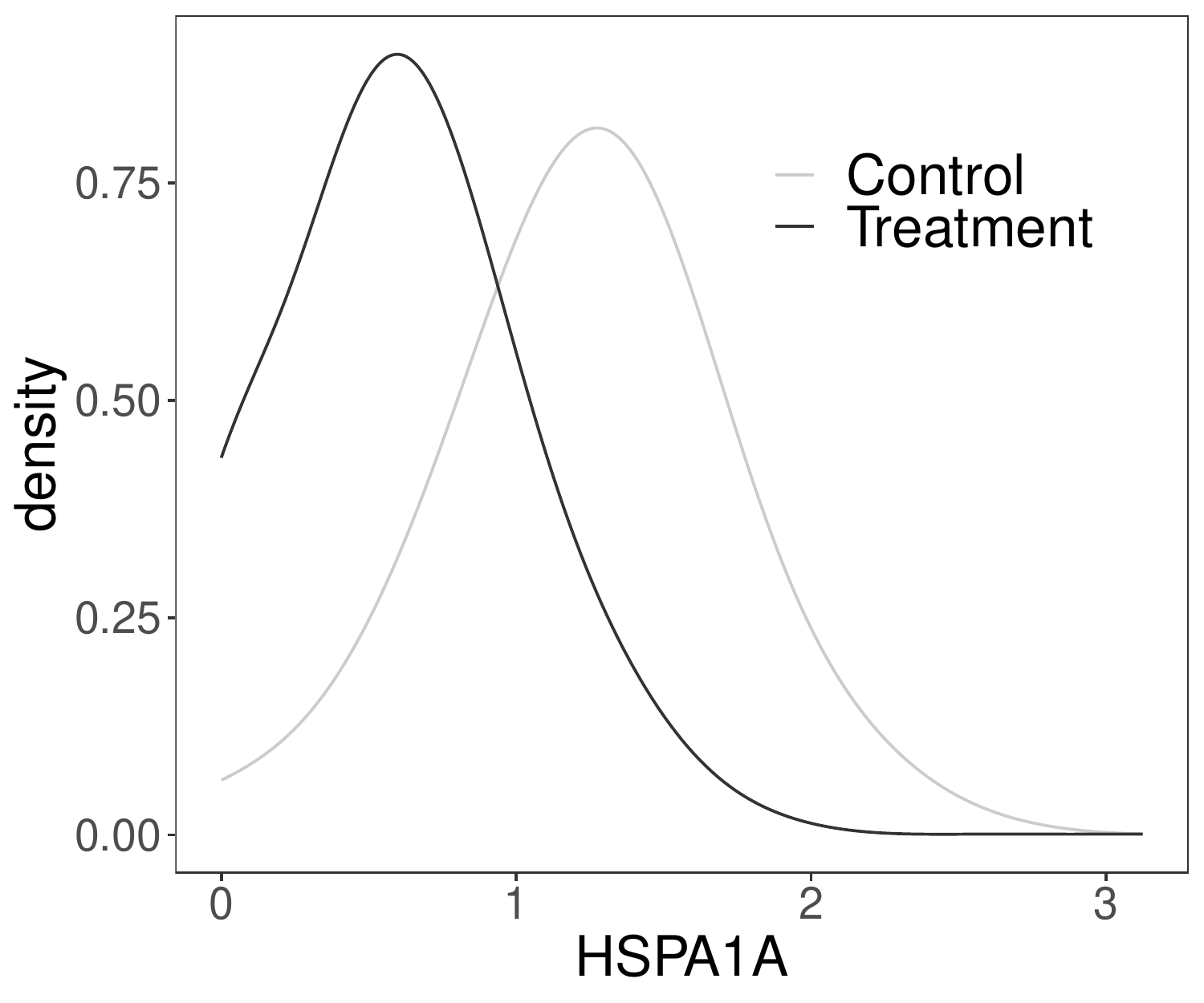}
\includegraphics[width=0.4\columnwidth]{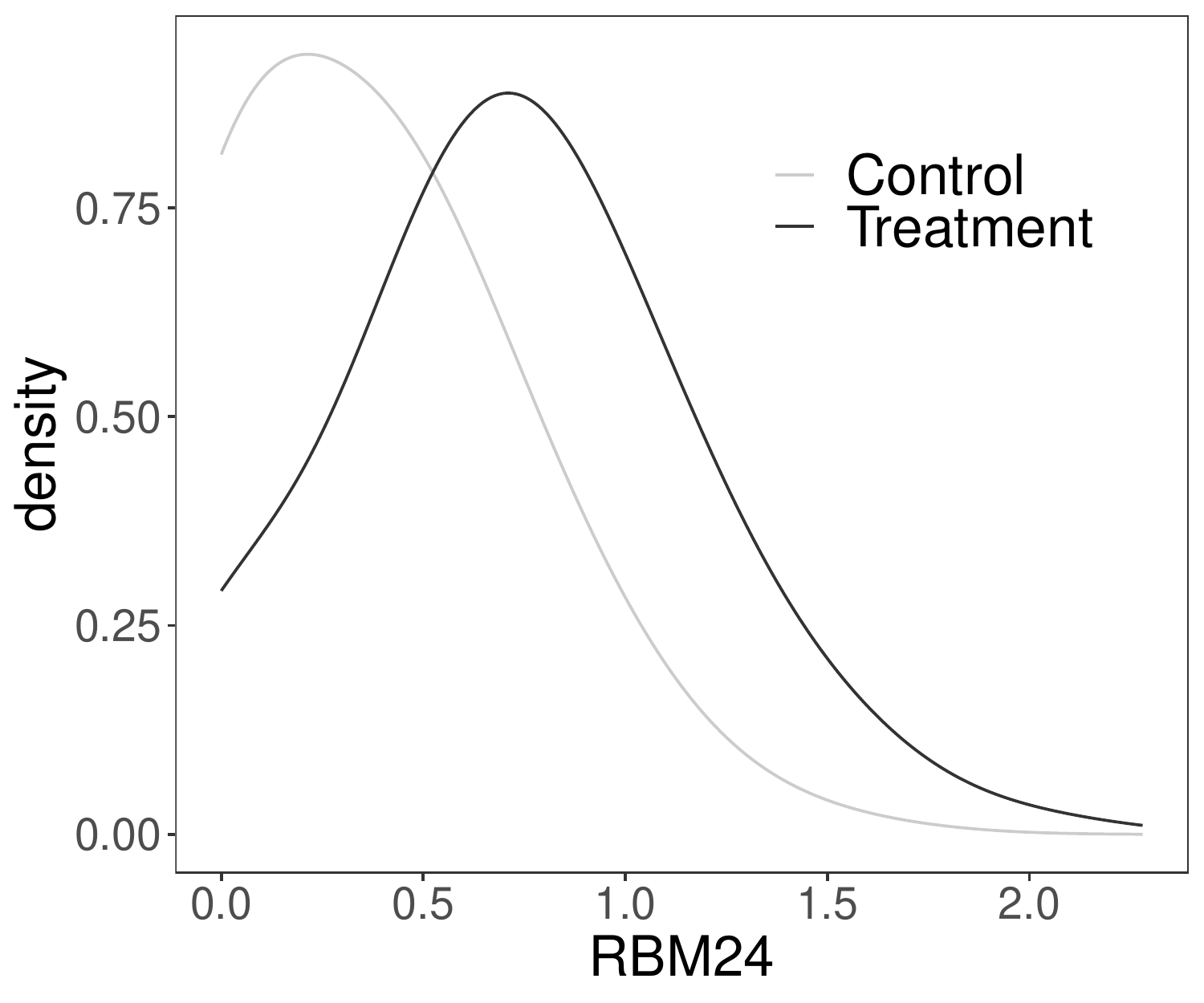}
\end{center}
\caption{Comparison of the gene expression distributions between the synthetic glucocorticoid dexamethasone treatment group and the vehicle-treated control group for 4 genes. FKBP5 and NFKBIA are selected by both MDS and the model-X knockoff filter, whereas HSPA1A and RBM24 are only selected by MDS.}
\label{fig:gene-expression-histogram}
\end{figure*}

With the nominal FDR control level set at $q = 0.1$, MDS performs significantly more powerful than the other two competing methods. More precisely, MDS stably selects approximately 30 genes (see Table \ref{tab:gene-reference}), in the sense that the selection results obtained via two independent runs of MDS only differ by one or two genes. In contrast, Knockoff approximately selects 13 genes, forming a subset of the genes selected by MDS (see Table \ref{tab:gene-reference}). BHq only selects 1 gene, RPL10, which is also consistently selected by MDS and Knockoff. Figure \ref{fig:gene-expression-histogram} demonstrates the sharp difference in the gene expression distributions between the treatment group and the control group for 4 genes: FKBP5, NFKBIA,  HSPA1A, and RBM24 selected by MDS, of which the first two are also selected by Knockoff.

The existing literature confirms the interactions between the glucocorticoid receptor (GR) and many  genes selected by both MDS and Knokoff, as well as the majority selected only by MDS,  thus backing up our selection results to some extent. We highlight the following supporting evidences, and provide a summary of the references associated with the selected genes in Table \ref{tab:gene-reference}.

\begin{itemize}
\item Genes selected by both MDS and Knockoff.
\begin{enumerate}[(i)]
\item The SERPINA6 gene is a coding gene for the protein corticosteroid-binding globulin (CBG), which is a major transport protein for glucocorticoids and progestins in the blood of almost all vertebrate species \citep{zhou2008s}. Among its related pathways are Glucocorticoid Pathway (Peripheral Tissue), Pharmacodynamics.
\item The FKBP5 gene encodes the FK506 binding protein 51, a co-chaperone in the heat shock protein 90 (Hsp90) and steroid complex, which regulates GR sensitivity \citep{nair1997molecular}.
\item The NFKBIA gene is a coding gene for the protein NF-kappa-B inhibitor alpha, and the glucocorticoids are potent inhibitors of nuclear factor kappa B (NF-kappa B) activation \citep{auphan1995immunosuppression, deroo2001glucocorticoid}.
\item \citet{williamson2020plexinb1} showed that activation of PlexinB1 by SEMA3C and SEMA4D promotes nuclear translocation of the GR.
\item The HSPB1 gene encodes the heat shock protein 27 (Hsp27), and the synthesis of Hsp27 is regulated by glucocorticoids. The HSPA1A gene also interacts with Hsp70. {\it In vitro} it acts as an ATP-independent chaperone by inhibiting protein aggregation and by stabilizing partially denatured proteins, which ensures refolding by the Hsp70-complex \citep{barr1999glucocorticoids}.
\end{enumerate}
\item Genes selected  by MDS only.
\begin{enumerate}[(i)]
\item The HSPA8 gene is a coding gene for the protein Hsc70. The cooperation of Hsc70 with Hsp90 regulates the GR activation and signaling pathway \citep{furay2006region}.
\item The HSPA1A gene encodes a 70kDa heat shock protein, which is a member of the heat shock protein 70 (Hsp70) family. Hsp70 promotes GR ligand release and inactivation by inducing partial unfolding \citep{kirschke2014glucocorticoid}.
\item \citet{goodman2016steroid} reported that glucocorticoids, along with mineralocorticoids, effectively expand the cytokeratin-5-positive cell population induced by 3-ketosteroids, which requires induction of the transcriptional repressor BCL6 based on suppression of BCL6.
\item The ATF4 gene is a coding gene for the activating transcription factor 4, which is repressed by glucocorticoids and induced by insulin \citep{adams2007role}.
\item The IGFBP4 gene encodes the insulin-like growth factor-binding protein 4, regulated by glucocorticoids (e.g. dexamethasone inhibits IGFBP4 expression) \citep{cheung1994glucocorticoid, okazaki1994glucocorticoid, conover1995effect}.
\item The YWHAQ gene encodes the 14--3-3 protein theta, and 14--3-3 protein interacts with the ligand-activated GR (e.g., overexpression of the 14–3-3 protein enhances the transcriptional activity of glucocorticoid receptor in transient transfection experiments) \citep{wakui1997interaction, zilliacus2001regulation}.
\item The DDIT4 gene encodes the DNA damage-inducible transcript 4 protein, and among its inducers are glucocorticoids \citep{wang2003dexamethasone, boldizsar2006low, wolff2014redd1}.
\item The RBM24 gene encodes the RNA-binding protein 24, and is antagonistically regulated by glucocorticoid dexamethasone \citep{whirledge2013global}.
\end{enumerate}
\end{itemize}

\begin{table*}
\begin{center}
\caption{The references 
in support of the genes selected by  MDS.}
\begin{tabular}{|l|c|l|} 
\hline
Gene & Knockoff & Reference\\
\hline
SERPINA6 & $\cmark$ & \href{https://www.genecards.org/Search/Keyword?queryString=SERPINA6}{GeneCards-SERPINA6}, \citet{zhou2008s}\\
\hline
FKBP5 & $\cmark$ &
\href{http://atlasgeneticsoncology.org/Genes/GC_FKBP5.html}{AGCOH-FKBP5}, \citet{nair1997molecular}\\
\hline
NFKBIA & $\cmark$ &
\citet{auphan1995immunosuppression, deroo2001glucocorticoid}\\
\hline
RPL10 & $\cmark$ & \citet{zorzatto2015nik1}\\
\hline
SEMA3C & $\cmark$ &\citet{williamson2020plexinb1}\\
\hline
HSPB1 & $\cmark$ & \citet{barr1999glucocorticoids, tuckermann1999dna}\\
\hline
RBBP7 & $\cmark$ & \citet{jangani2014methyltransferase}\\
\hline
EIF4EBP1 & $\cmark$ & \citet{watson2012cell}\\
\hline
S100A11 & $\cmark$ & \href{https://string-db.org/network/9606.ENSP00000271638}{String-S100A11}, \citet{erica2009etal}\\
\hline
NUPR1 & $\cmark$ & \href{https://www.wikigenes.org/e/ref/e/8175759.html}{Wikigenes-NUPR1}, \citet{Mukaida1994}\\
\hline
MSX2 & $\cmark$ & \citet{Jaskoll1998}\\
\hline
LY6E & $\cmark$ & --\\
\hline
BLOC1S1 & $\cmark$ & --\\
\hline
HSPA8 & $\xmark$ & 
\citet{furay2006region}\\
\hline
HSPA1A & $\xmark$ &
\citet{kirschke2014glucocorticoid}\\
\hline
EEF1A1 & $\xmark$ & \href{https://www.ncbi.nlm.nih.gov/gene/1915}{NCBI-EEF1A1}\\
\hline
BCL6 & $\xmark$ & \citet{goodman2016steroid}\\
\hline
ATF4 & $\xmark$ &\citet{adams2007role}\\
\hline
IGFBP4 & $\xmark$ & \citet{cheung1994glucocorticoid, okazaki1994glucocorticoid, conover1995effect}\\
\hline
YWHAQ & $\xmark$ & \citet{wakui1997interaction, zilliacus2001regulation}\\
\hline
DDIT4 & $\xmark$ & \citet{wang2003dexamethasone, boldizsar2006low, wolff2014redd1}\\
\hline
IRX2 & $\xmark$ &\citet{lambert2013brain} \\
\hline
GATA3-AS1 & $\xmark$ & \citet{liberman2009glucocorticoids}\\
\hline
RBM24 & $\xmark$ & \citet{whirledge2013global}\\
\hline
TACSTD2 & $\xmark$ & \citet{mcdougall2017}\\
\hline
DSCAM-AS1 & $\xmark$ & \citet{zhao2016g, chen2020dscam}\\
\hline
C1QBP & $\xmark$ & \citet{sheppard1994calcium, cote2008pkc,zhang2013interactome}\\
\hline
SNHG19 & $\xmark$ & --\\
\hline
PRR15L & $\xmark$ & --\\
\hline
RPLP0P6 & $\xmark$ & --\\
\hline
UHMK1 & $\xmark$ & --\\
\hline
\end{tabular}
\justify{
The second column indicates whether the gene is also selected by the model-X knockoff filter. There are 6 genes that we do not find direct supporting evidence in the existing literature for their interaction with the glucocorticoid receptor (GR), which might be of interest for further investigations. 
The red hyperlinks point to the documented information of the corresponding genes in some widely referred databases including GeneCards, Strings, Wikigenes, the National Center for Biotechnology Information (NCBI), and Atlas of Genetics and Cytogenetics in Oncology and Harmatology (AGCOH).}
\label{tab:gene-reference}
\end{center}
\end{table*}

\section{Conclusion}
\label{sec:conclusion}
We have described a general framework for the task of feature selection in GLMs with FDR asymptotically under control. 
In particular, we detail the construction of the mirror statistics under two asymptotic regimes, including the moderate-dimensional setting ($p/n\to\kappa\in(0,1)$) and the high-dimensional setting ($p\gg n$). Compared to BHq, the proposed methodology enjoys a wider applicability and improved robustness thanks to its scale-free property.
Compared to the knockoff filter, the proposed methodology does not require the knowledge of the joint distribution of the features, which is the norm for many practical problems, and is less affected by the correlations among the features.

We conclude by pointing out several directions for future work. First, it is of immediate interest to generalize the proposed methods in order to cover the case where the set of explanatory features exhibit a group structure.
Second, we would like to investigate the potential applicability of our FDR control framework to dependent observations (e.g., stationary time series data). 
These two types of data structures appear a lot in practice including genetic studies and financial engineering.
Third, moving beyond the parametric models, we would like to consider the FDR control problem in semiparametric single-index models, in which the link function becomes unknown.

\newpage
\bibliographystyle{apalike}
\bibliography{GLM-FDR.bib}

\vspace{.5in}

\noindent{\Large\bf Appendix}

\appendix

\section{Proof}
Let $Q(t) = 1 - \Phi(t)$, where $\Phi$ is the CDF of the standard normal distribution. Let $\phi(t)$ be the probability density function of the standard normal distribution. 
We consider a general form of the mirror statistic defined in Equation \eqref{eq:mirror-statistic} with $f(u,v)$ satisfying Condition \ref{cond:mirror-statistics-f}. For $t\in\mathbbm{R}$, let $H(t) = \mathbbm{P}(\text{sign}(Z_1Z_2)f(Z_1, Z_2) > t)$, where $Z_1$ and $Z_2$ are independent, following the standard normal distribution. For any $t > 0$ and $v\geq 0$, let 
\begin{equation}
\label{eq:inverse-f}
\mathcal{I}_t(v) = \inf\{u\geq 0: f(u, v) > t\}
\end{equation}
with the convention $\inf\varnothing = +\infty$.

\subsection{Proof of Proposition \ref{prop:optimality-mirror-statistics}}
Let $r = \lim p_1/p_0$. Without loss of generality, we assume the designated FDR control level $q\in(0,1)$ satisfying $rq/(1 - q) < 1$, otherwise selecting all the features would achieve an asymptotic FDR control.

Denote $f^\text{opt}(u, v)$ as the optimal choice, and denote $\widehat{S}^{\text{opt}}$ as the optimal selection result that enjoys an asymptotic FDR control. By the law of large number, we have
\begin{equation}
\label{eq:optimality-LLN}
\lim_{p\to\infty}\frac{\#\{j: j \in S_0, j \in \widehat{S}^{\text{opt}}\}}{\# \{j: j \in \widehat{S}^{\text{opt}}\}} = \frac{\mathbbm{P}(j \in \widehat{S}^{\text{opt}}\mid  j\in S_0)}{\mathbbm{P}(j \in \widehat{S}^{\text{opt}} \mid  j\in S_0) + r\mathbbm{P}(j \in \widehat{S}^{\text{opt}} \mid j\in S_1)} \leq q,
\end{equation}
in which the numerator is essentially the type-I error. More precisely,
\begin{equation}
\begin{aligned}
& \mathbbm{P}(j \in \widehat{S}^{\text{opt}}\mid  j\in S_0) =  \mathbbm{P}(\text{sign}(Z_1Z_2)f^{\text{opt}}(|Z_1|, |Z_2|) > t^{\text{opt}}),\\
& \mathbbm{P}(j \in \widehat{S}^{\text{opt}}\mid  j\in S_1) =  \mathbbm{P}(\text{sign}(Z_3Z_4)f^{\text{opt}}(|Z_3|, |Z_4|) > t^{\text{opt}}),
\end{aligned}
\end{equation}
where $Z_1, Z_2$ follow $N(0,1)$, $Z_3, Z_4$ follow $N(\omega, 1)$, and all of them are independent. 
$t^{\text{opt}} > 0$ is the cutoff that maximizes the power $\mathbbm{P}(j \in \widehat{S}^{\text{opt}}\mid  j\in S_1)$, under the constraint that Equation \eqref{eq:optimality-LLN} is satisfied.

We now consider testing whether $X_j$ is a null feature, with the significance level $\alpha$ specified as below,
\begin{equation}
\alpha = \frac{rq}{1 - q}\mathbbm{P}(j \in \widehat{S}^{\text{opt}} \mid j\in S_1) < 1.
\end{equation}
We note that we essentially have two observations $T_j^{(1)}$ and $T_j^{(2)}$, which independently follow $N(0,1)$ or $N(\omega, 1)$ if $X_j$ is a null feature or a relevant feature, respectively. By Equation \eqref{eq:optimality-LLN}, the test which rejects the null hypothesis (i.e., $j\in \widehat{S}^{\text{opt}} $) if
\begin{equation}
\text{sign}(T_j^{(1)}T_j^{(2)})f^{\text{opt}}(|T_j^{(1)}|, |T_j^{(2)}|) > t^{\text{opt}}
\end{equation}
achieves the significance level $\alpha$. 

By the Neymann-Pearson Lemma, the likelihood ratio test is the most powerful test, which rejects the null hypothesis if the following likelihood ratio (LR) is large enough,
\begin{equation}
\text{LR} = \frac{\phi(T_j^{(1)} - \omega)\phi(T_j^{(2)} - \omega)}{\phi(T_j^{(1)})\phi(T_j^{(1)})} \approx \frac{\phi(|T_j^{(1)}| - \omega)\phi(|T_j^{(2)}| - \omega)}{\phi(|T_j^{(1)}|)\phi(|T_j^{(1)}|)} 
\end{equation}
The approximation shall be reasonably accurate when $\omega > 0$ is sufficiently large. 
A simple calculation yields the following rejection rule
\begin{equation}
\begin{aligned}
|T_j^{(1)}| +  |T_j^{(2)}| > t^{\text{lik}},
\end{aligned}
\end{equation}
in which $t^{\text{lik}}$ is a properly chosen cutoff so that the likelihood ratio test achieves the significance level $\alpha$.

We then consider the following rejection rule
\begin{equation}
\begin{aligned}
\text{sign}(T_j^{(1)}T_j^{(2)})f^{\text{lik}}(|T_j^{(1)}|, |T_j^{(2)}|) > t^{\text{lik}},
\end{aligned}
\end{equation}
in which $f^{\text{lik}}(u, v) = u + v$. Denote $\{j \in \widehat{S}^{\text{lik}}\}$ as the event of rejecting the null hypothesis. We note that the type-I error is upper bounded by $\alpha$ since
\begin{equation}
\label{eq:optimality-error-bound}
\begin{aligned}
\mathbbm{P}(\{j \in \widehat{S}^{\text{lik}}\} \mid j\in S_0) \leq  \mathbbm{P}(|T_j^{(1)}| +  |T_j^{(2)}| > t^{\text{lik}}\mid j\in S_0) \leq \alpha.
\end{aligned}
\end{equation}
In addition, since the likelihood ratio test is the optimal test, we have
\begin{equation}
\label{eq:optimality-power-bound}
\begin{aligned}
\mathbbm{P}(j \in \widehat{S}^{\text{lik}}\mid  j\in S_1) & \approx \mathbbm{P}(|T_j^{(1)}| +  |T_j^{(2)}| > t^{\text{lik}}\mid  j\in S_1)\\
& \geq \mathbbm{P}(j \in \widehat{S}^{\text{opt}}\mid  j\in S_1).
\end{aligned}
\end{equation}
when $\omega > 0$ is sufficiently large. Combining Equations \eqref{eq:optimality-error-bound} and \eqref{eq:optimality-power-bound}, we show that the selection set $\widehat{S}^{\text{lik}}$ enjoys an asymptotic FDR control since
\begin{equation}
\label{eq:optimality-LR-FDR}
\begin{aligned}
\lim_{p\to\infty}\frac{\#\{j: j \in S_0, j \in \widehat{S}^{\text{lik}}\}}{\# \{j: j \in \widehat{S}^{\text{lik}}\}} & = \frac{\mathbbm{P}(j \in \widehat{S}^{\text{lik}}\mid  j\in S_0)}{\mathbbm{P}(j \in \widehat{S}^{\text{lik}} \mid  j\in S_0) + r\mathbbm{P}(j \in \widehat{S}^{\text{lik}} \mid j\in S_1)} \leq q.
\end{aligned}
\end{equation}
Since $f^{\text{opt}}$ is optimal, by Equations \eqref{eq:optimality-LR-FDR} and \eqref{eq:optimality-power-bound}, $f^{\text{lik}}$ is also optimal. This concludes the proof of Proposition \ref{prop:optimality-mirror-statistics}.

\subsection{Proof of Proposition \ref{prop:GLM-asymptotics-moderate-dimension}}
\subsubsection{Technical Lemmas}
\begin{lemma}
\label{lemma:glm-moderate-parameter-convergence}
Consider the case $\Sigma = I_{p}$. As $n,p\to\infty$, we have 
\begin{equation}
\sigma_n \overset{p}{\to} \sqrt{\kappa}\sigma_\star,\ \ \ \alpha_{n} \overset{p}{\to} \alpha_{\star},
\end{equation}
in which
$(\sigma_n,\ \alpha_n)$ is defined in Equation \eqref{eq:definition-alpha-sigma},
and $(\sigma_{\star}, \alpha_{\star})$ is the unique optimizer of the following optimization problem,
\begin{equation}
\label{eq:glm-optimization}
\begin{aligned}
\min_{\sigma, \delta > 0, \alpha \in \mathbbm{R}}\max_{r > 0} \bigg\{r{\kappa}\sigma + \frac{r}{2\delta} - &  \frac{1}{2r\delta}\mathbbm{E}(y_i^2) - \alpha\mathbbm{E}(y_ix_i^\intercal\beta^\star) \\
& + \mathbbm{E}\Big[M_\rho\Big(\alpha \gamma Z_1 + \sigma\sqrt{\kappa}Z_2 + \frac{y_i}{r\delta}, \frac{1}{r\delta}\Big)\Big]\bigg\}.
\end{aligned}
\end{equation}
$Z_1$, $Z_2$ are independent (also independent to everything else) random variables following the standard normal distribution, and
\begin{equation}
M_\rho(v, t) = \min_{x\in\mathbbm{R}}\left\{\rho(x) + \frac{1}{2t}(x - v)^2\right\}.
\end{equation}
\end{lemma}

\textit{Proof of Lemma \ref{lemma:glm-moderate-parameter-convergence}}. The proof heavily relies on \citet{Chirstos2018} and \citet{salehi2019impact}, in which the key technical tool is the Convex Gaussian Min-max Theorem(CGMT).
We introduce a new variable $u = X\beta$ and rewrite the optimization problem \eqref{eq:GLM-MLE-moderate-dimension} as 
\begin{equation}
\min_{\beta\in\mathbbm{R}^{p}, u\in\mathbbm{R}^{n}}\frac{1}{n} 1^\intercal \rho(u) - \frac{1}{n}y^\intercal u \ \ \ \ s.t.\  \ \ \ u  = X\beta.
\end{equation}
Based on the method of Lagrange multipliers, the above optimization problem is equivalent to a min-max problem specified as below,
\begin{equation}\label{PO}
(\text{PO})\ \ \ \min_{\beta\in\mathbbm{R}^{p}, u\in\mathbbm{R}^{n}}\max_{v\in\mathbbm{R}^{n}}\frac{1}{n} 1^\intercal \rho(u) - \frac{1}{n}y^\intercal u+\frac{1}{n}v^{\intercal}(u-X\beta).
\end{equation}
We refer to this min-max optimization problem as the primary optimization (PO) problem henceforth. 

We proceed to associate the PO with an auxiliary optimization (AO) problem using CGMT. Similar as in \citet{salehi2019impact}, 
We decompose $\beta = P\beta + P^{\perp}\beta$, in which $P$ is the projection operator projecting onto the column space spanned by $\beta^\star$, and rewrite the PO as below,
\begin{equation}\label{POmodify}
\min_{\beta\in\mathbbm{R}^{p}, u\in\mathbbm{R}^{n}}\max_{v\in\mathbbm{R}^{n}}\frac{1}{n} 1^\intercal \rho(u) - \frac{1}{n}y^\intercal u+\frac{1}{n}v^{\intercal}u-\frac{1}{n}v^{\intercal}XP\beta-\frac{1}{n}v^{\intercal}XP^{\perp}\beta.
\end{equation}
Before we apply CGMT, we remark that we can simply assume the feasible sets of $v, u$ and $\beta$ are convex and compact, following Lemma A.1 and Lemma A.2 in \citet{Chirstos2018}. For simplicity, we omit the details here and write the AO as:
\begin{equation}\label{AO}
(\text{AO})\ \ \ \min_{\beta\in\mathbbm{R}^{p}, u\in\mathbbm{R}^{n}}\max_{v\in\mathbbm{R}^{n}}\frac{1}{n} 1^\intercal \rho(u) - \frac{1}{n}y^\intercal u+\frac{1}{n}v^{\intercal}(u-XP\beta)-\frac{1}{n}(v^{\intercal}h||P^{\perp}\beta||+||v||g^{\intercal}P^{\perp}\beta),
\end{equation}
in which $h$ and $g$ are independent, following $N(0, I_{n})$ and $N(0, I_{p})$, respectively.

We proceed to simplify the AO. Let $r = ||v||/\sqrt{n}$. Maximizing with respect to the direction of $v$, the AO can be written as
\begin{equation}\label{AOMaxV}
\min_{\beta\in\mathbbm{R}^{p}, u\in\mathbbm{R}^{n}}\max_{r\geq 0}\frac{1}{n} 1^\intercal \rho(u) - \frac{1}{n}y^\intercal u -\frac{r}{\sqrt{n}}g^{\intercal}P^{\perp}\beta+r\left|\left|\frac{u}{\sqrt{n}}-\frac{XP\beta}{\sqrt{n}}-\frac{||P^{\perp}\beta||h}{\sqrt{n}}\right|\right|.
\end{equation}
By Lemma A.3 in \citet{Chirstos2018}, we swap the order of min-max in the optimization problem \eqref{AOMaxV}. Let $\sigma = ||P^{\perp}\beta||/\sqrt{\kappa}$, $\alpha = \left<\beta, \beta^{\star}\right>/||\beta^{\star}||^{2}$, then $P\beta = \alpha\beta^{\star}$. Optimizing with respect to the direction of $P^{\perp}\beta$, we can further simplify the optimization problem \eqref{AOMaxV} as
\begin{equation}\label{AOMaxPperpBeta}
\max_{r\geq0}\min_{u\in\mathbbm{R}^{n}, \sigma\geq0, \alpha\in\mathbbm{R}}\frac{1}{n} 1^\intercal \rho(u) - \frac{1}{n}y^\intercal u+\frac{r\sqrt{p}}{n}||g||\sigma+r\left|\left|\frac{u}{\sqrt{n}}-\frac{\alpha X\beta^{\star}}{\sqrt{n}}-\frac{\sqrt{\kappa}\sigma h}{\sqrt{n}}\right|\right|.
\end{equation}
Using the square-root trick, i.e., $\sqrt{x} = \inf_{\delta>0} \left\{\frac{\delta}{2}+\frac{x}{2\delta}\right\}$, we obtain
\begin{equation}\label{AOSRTrick}
\max_{r\geq0} \min_{\substack{u\in\mathbbm{R}^{n}, \sigma\geq0\\ \alpha\in\mathbbm{R},\ \delta > 0}}
\frac{1}{n} 1^\intercal \rho(u) - \frac{1}{n}y^\intercal u+\frac{r\sqrt{p}}{n}||g||\sigma+\frac{\delta r}{2}\left|\left|\frac{u}{\sqrt{n}}-\frac{\alpha X\beta^{\star}}{\sqrt{n}}-\frac{\sqrt{\kappa}\sigma h}{\sqrt{n}}\right|\right|^{2}+\frac{r}{2\delta}.
\end{equation}

We proceed to optimize with respect to $u$. Based on completion of square, we have
\begin{equation}\label{OverU}
\begin{aligned}
-\frac{1}{n}y^{\intercal}u + \frac{\delta r}{2}&\left|\left| \frac{u}{\sqrt{n}}-\frac{\alpha X\beta^{\star}}{\sqrt{n}}-\frac{\sqrt{\kappa}\sigma h}{\sqrt{n}}\right|\right|^{2}\\
& = 
\frac{\delta r}{2n}\left|\left|u-\alpha X\beta^{\star}-\sqrt{\kappa}\sigma h-\frac{y}{\delta r}\right|\right|^{2}-\frac{||y||^{2}}{2n\delta r}-\frac{\alpha}{n}y^{\intercal}X\beta^{\star}-\frac{\sqrt{\kappa}}{n}\sigma y^{\intercal}h.
\end{aligned}
\end{equation}
Optimizing with respect to the direction of $u$, the AO can be simplified as below,
\begin{equation}\label{AOMinU}
\begin{aligned}
\max_{r\geq0}\min_{\sigma\geq0, \alpha\in\mathbbm{R}, \delta>0} \mathcal{R}_{n}(\sigma, \alpha, \delta, r) & : =  \frac{r\sqrt{p}}{n}||g||\sigma  +\frac{r}{2\delta} -\frac{||y||^{2}}{2n\delta r}-\frac{\alpha}{n}y^{\intercal}X\beta^{\star}-\frac{\sqrt{\kappa}}{n}\sigma y^{\intercal}h\\
& \ \ \ \ +\frac{1}{n}M_{\rho}\Big(\alpha X\beta^{\star}+\sigma\sqrt{\kappa}h+\frac{y}{\delta r}, \frac{1}{\delta r}\Big),
\end{aligned}
\end{equation}
in which the function $M_\rho$ is applied in an element-wise fashion. We now reduce the original vector optimization problem into a scalar optimization problem. Since the objective function is convex in $\sigma, \alpha, \delta$ and concave in $r$, we can swap the order of min and max.

We proceed to further simplify the AO by invoking asymptotics. By the law of large numbers, we have the following simple results,
\begin{equation}
\begin{aligned}
\text{(a)}\ \ \ & \frac{1}{n}||y||^{2} \overset{p}{\to} \mathbbm{E}(y^{2});\\
\text{(b)}\ \ \ & \frac{1}{n}y^{\intercal}X\beta^{\star} \overset{p}{\to} \mathbbm{E}(y_{i}X_{i}^{\intercal}\beta^{\star});\\
\text{(c)}\ \ \ & \frac{1}{n}M_{\rho}\Big(\alpha X\beta^{\star}+\sigma\sqrt{\kappa}h+\frac{y}{\delta r}, \frac{1}{\delta r}\Big) \overset{p}{\to} \mathbbm{E}\Big[M_{\rho}\Big(\alpha\gamma Z_{1}+\sigma\sqrt{\kappa}Z_{2}+\frac{y}{\delta r}, \frac{1}{\delta r}\Big)\Big],\\
\text{(d)}\ \ \ & \frac{1}{n}y^{\intercal}h = \frac{1}{\sqrt{n}}\frac{1}{\sqrt{n}}y^{\intercal}h = O_{p}({1}/{\sqrt{n}}),\\
\text{(e)}\ \ \ & \frac{\sqrt{p}}{n} ||g||  = \frac{p}{n}\frac{||g||}{\sqrt{p}}\overset{p}{\to} \frac{p}{n} = \kappa.
\end{aligned}
\end{equation}
Based upon these facts, $\mathcal{R}_{n}$ converges in probability to
\begin{equation}
D(\sigma, \alpha, \delta, r) := r\kappa\sigma+\frac{r}{2\delta} - \frac{1}{2\delta r}\mathbbm{E}(y^{2}) - \alpha\mathbbm{E}(yX^{\intercal}\beta^{\star})+\mathbbm{E}\Big[M_{\rho}\Big(\alpha\gamma Z_{1}+\sigma\sqrt{\kappa}Z_{2}+\frac{y}{\delta r}, \frac{1}{\delta r}\Big)\Big].
\end{equation}
We note that the objective function $D(\sigma, \alpha, \delta, r)$ is convex in $\sigma, \alpha, \delta$ and concave in $r$, since the convexity is preserved through point-wise limit. The covergence of the optimizers can be established based on the same arguments as in Lemma B.1 and Lemma A.5 in \citet{Chirstos2018}. This completes the proof of Lemma \ref{lemma:glm-moderate-parameter-convergence}.

\subsubsection{Proof of Proposition \ref{prop:GLM-asymptotics-moderate-dimension}}
The proof of Proposition \ref{prop:GLM-asymptotics-moderate-dimension} is essentially the same as the proof of Theorem 3.1 in \citet{zhao2020asymptotic}. 
Denote $\Sigma = LL^\intercal$ as the Cholesky decomposition of the covariance matrix $\Sigma$. Let 
\begin{equation}
\label{eq:glm-moderate-transformed-coef}
\theta^\star = L^\intercal\beta^\star,\ \ \ \widehat{\theta} = L^\intercal\widehat{\beta},
\end{equation}
in which $\widehat{\beta}$ is the MLE of the true regression coefficient $\beta^\star$.
By Proposition 2.1 in \citet{zhao2020asymptotic}, $\widehat{\theta}$ has the same distribution as the MLE of the underlying GLM with true regression coefficient $\theta^\star$ and features drawn i.i.d from $N(0, I_p)$.
We note that it is sufficient to consider the case $\Sigma = I_{p}$, as the general result follows from the following relationship,
\begin{equation}
\tau_{j}\frac{\widehat\beta_{j}-\alpha_{\star}\beta_{j}^{\star}}{\sigma_{\star}} = \frac{\widehat\theta_{j}-\alpha_{\star}\theta_{j}^{\star}}{\sigma_{\star}}.
\end{equation}

For $j\in[p]$, we have the following decomposition,
\begin{equation}
\frac{\sqrt{n}(\widehat\theta_{j}-\alpha_{\star}\theta_{j}^{\star})}{\sigma_{\star}} = \frac{\sqrt{n}(\widehat\theta_{j}-\alpha_{n}\theta_{j}^{\star})}{\sigma_{n}}\frac{\sigma_{n}}{\sigma_{\star}}+\frac{\sqrt{n}(\alpha_{n}-\alpha_{\star})\theta_{j}^{\star}}{\sigma_{\star}}.
\end{equation}
Using the same arguments as in the proof of Theorem 3.1 in \citet{zhao2020asymptotic}, we can show that the first term asymptotically behaves as the standard normal distribution, whereas the second term asymptotically vanishes. This completes the proof of Proposition \ref{prop:GLM-asymptotics-moderate-dimension}.

\subsection{Proof of Proposition \ref{prop:GLM-data-splitting-FDR-moderate-dimension}}
In addition to the notations introduced in Equation \eqref{eq:glm-moderate-transformed-coef}, we require the following notations for the ease of presentation. 
Let $\langle u, v\rangle_\Sigma = u^\intercal\Sigma v$ for any $u, v \in\mathbbm{R}^p$, and $||u||^2_\Sigma = u^\intercal\Sigma u$. 
We introduce the following random vector, 
\begin{equation}
\label{eq:glm-moderate-definition-xi}
\begin{aligned}
\xi = W - \frac{1}{\gamma_n^2}\langle W, \beta^\star\rangle_\Sigma \beta^\star,
\end{aligned}
\end{equation}
in which $W$ follows $N(0, \Theta)$, independent to everything else. Recall that $\gamma_n^2 = \text{Var}(x_i^\intercal\beta^\star)$ calibrates the signal strength, and we assume $\gamma_n \to \gamma$ as $n, p \to \infty$. 
We define
\begin{equation}
\label{eq:glm-moderate-dimension-V-definition}
\widetilde{V}_j =  \frac{\sqrt{p}V_j}{||\xi||_\Sigma} \ \ \ \text{with}\ \ \ V_j = \tau_j W_j,
\end{equation}
for $j\in[p]$ where $\tau_j^2 = 1/\Theta_{jj}$. 
The random vector $V$ follows $N(0, R)$ with $R_{ij} = \tau_i\tau_j\Theta_{ij}$. With a bit abuse of notation, we denote $\kappa = 2p/n \in (0, 1)$ as the ratio of the dimension over the sample size after data splitting. 

We define the normalized MLE $T_j$ as well as its approximation $\widetilde{T}_j$ as below,
\begin{equation}
\label{eq:glm-moderate-dimension-T-definition}
T_j = \frac{\sqrt{n}\widehat{\tau}_j\widehat{\beta}_j}{\sigma_\star},\ \ \ 
\widetilde{T}_j = \frac{\sqrt{n\kappa}\tau_j\widehat{\beta}_j}{\sigma_n}
\end{equation}
for $j\in[p]$, in which $\sigma_\star$ is the unique optimizer of the optimization problem defined in \eqref{eq:glm-optimization}.
Without changing the selection result obtained via Algorithm \ref{alg:FDR-control-framework}, we multiply the normalized MLE defined in Equation \eqref{eq:GLM-normalized-estimator-moderate-dimension-data-splitting} by a constant factor so that it follows the standard normal distribution asymptotically.
Let $\widetilde{M}_j$ be the corresponding approximated mirror statistic constructed based upon $\widetilde{T}^{(1)}_j$ and $\widetilde{T}^{(2)}_j$.

\subsubsection{Technical lemmas}
\begin{lemma}
\label{lemma:glm-moderate-stochastic-representation}
$\widetilde{T}_{S_0} \overset{d}{=} \widetilde{V}_{S_0}$.
\end{lemma}

\textit{Proof of Lemma \ref{lemma:glm-moderate-stochastic-representation}}. 
We note that a simple calculation yields
\begin{equation}
L^{-\intercal} P_{\theta^\star}^\perp Z \overset{d}{=} \xi,\ \ \ ||P_{\theta^\star}^\perp Z||_2 = ||\xi||_\Sigma,
\end{equation}
in which $Z$ following $N(0, I_p)$ is independent to everything else. 
Let $\Lambda_{S_0}$ be a $p_0\times p_0$ diagonal matrix, of which the diagonal elements are $\tau_j$ for $j\in S_0$. By the stochastic representation (Lemma 2.1, Proposition A.1) in \citet{zhao2020asymptotic}, we have
\begin{equation}
\begin{aligned}
\widetilde{T}_{S_0} &  = \frac{\sqrt{p}\Lambda_{S_0}\widehat{\beta}_{S_0}}{\sigma_n} = \frac{\sqrt{p}\Lambda_{S_0}(L^{-\intercal}\widehat{\theta})_{S_0}}{\sigma_n}\\ &\overset{d}{=} \frac{\sqrt{p}\Lambda_{S_0}(L^{-\intercal} P_{\theta^\star}^\perp Z)_{S_0}}{||P_{\theta^\star}^\perp Z||_2}  \overset{d}{=} \frac{\sqrt{p}\Lambda_{S_0}\xi_{S_0}}{||\xi||_\Sigma} \overset{d}{=} \widetilde{V}_{S_0}.
\end{aligned}
\end{equation}
This completes the proof of Lemma \ref{lemma:glm-moderate-stochastic-representation}. 

\begin{lemma}
\label{lemma:glm-moderate-tau}
Under Assumption \ref{assump:GLM-data-splitting-moderate-dimension}, there exists a constant $c > 0$ such that for $\epsilon \in (0, 1)$, we have
\begin{equation}
\mathbbm{P}\left(\max_{j\in[p]}\left|{\widehat{\tau}_j^2}/{\tau_j^2} - 1\right| > \epsilon\right) \leq p\exp(-c(n/2 - p + 1)\epsilon^2).
\end{equation}
\end{lemma}

\textit{Proof of Lemma \ref{lemma:glm-moderate-tau}}. By Assumption \ref{assump:GLM-data-splitting-moderate-dimension}, we have
\begin{equation}
\label{eq:bound-tau_j}
\begin{aligned}
& \min_{j\in[p]}{\tau_j^2} = 1/\max_{j\in[p]}\Theta_{jj} \geq 1/\sigma_{\max}(\Theta) = \sigma_{\min}(\Sigma) \geq 1/C > 0,\\
& \max_{j\in[p]}{\tau_j^2} = 1/\min_{j\in[p]}\Theta_{jj} \leq 1/\sigma_{\min}(\Theta) = \sigma_{\max}(\Sigma) \leq C < \infty.
\end{aligned}
\end{equation}
Thus, it is sufficient for us to consider $\max_{j\in[p]}\left|\widehat{\tau}_j^2 - \tau_j^2\right|$. Since $\text{RSS}_j\sim\tau_j^2\chi^2_{n/2 - p + 1}$, by the union bound and a  Bernstein-type inequality, for any $\epsilon \in (0, 1)$, we have
\begin{equation}
\begin{aligned}
\mathbbm{P}\left(\max_{j\in[p]}\left|{\widehat{\tau}_j^2} - \tau_j^2\right| > \epsilon\right) & \leq \sum_{j = 1}^p\mathbbm{P}\left(\left|{\widehat{\tau}_j^2} - \tau_j^2\right| > \epsilon\right)\\
& \leq p\exp(-c(n/2 - p + 1)\epsilon^2)
\end{aligned}
\end{equation}
for some constant $c > 0$.

\begin{lemma}
\label{lemma:glm-moderate-approx-mirror}
Under Assumption \ref{assump:GLM-data-splitting-moderate-dimension}, as $n, p\to\infty$, we have
\begin{equation}
\label{eq:T_j-normal-approximation}
    \sup_{t\in\mathbbm{R}, j\in S_0}|\mathbbm{P}(T_j > t)-Q(t)|\longrightarrow 0 
\end{equation}
\end{lemma}

\textit{Proof of Lemma \ref{lemma:glm-moderate-approx-mirror}}. Without loss of generality, we assume $t>0$. For $\epsilon \in (0, 1)$, we condition on the event $E_1\cap E_2$, in which
\begin{equation}
\begin{aligned}
& E_1 = \{|\sigma_n/\sqrt{\kappa} - \sigma_\star| < \epsilon\},\\
& E_2 = \big\{\max_{j\in[p]}\left|{\widehat{\tau}_j}/{\tau_j} - 1\right| < \epsilon\big\}. 
\end{aligned}
\end{equation}
For large enough $n$ and $p$, $E_1$ and $E_2$ hold with high probability according to Lemma \ref{lemma:glm-moderate-parameter-convergence} and Lemma \ref{lemma:glm-moderate-tau}, respectively. 
Conditioning on the events $E_1$ and $E_2$, we have
\begin{equation}
\begin{aligned}
\max_{j\in[p]}|T_j/\widetilde{T}_j-1| & \leq \max_{j\in[p]}|\widehat{\tau}_j/\tau_j||\sigma_n/\sqrt{\kappa} - \sigma_\star|/\sigma_\star + \max_{j\in[p]}\left|{\widehat{\tau}_j}/{\tau_j} - 1\right|\\
& \leq (\epsilon + 1)\epsilon/\sigma_\star + \epsilon.    
\end{aligned}
\end{equation}
Consequently, we have
\begin{equation}
\label{eq:T_j-overtildeT_j}
\max_{j\in[p]}|T_j/\widetilde{T}_j-1|\overset{p}{\to} 0.
\end{equation}

We proceed to show that 
\begin{equation}
\label{eq:tildeZ_joverZ_j}
\max_{j\in[p]}\left|\widetilde V_j/V_j - 1\right| = \left|\frac{\sqrt{p}}{||\xi||_\Sigma}-1\right|\overset{p}{\to} 0,
\end{equation}
in which by definition, $V_j$ follows the standard normal distribution. Indeed, by the definition of $\xi$ in Equation \eqref{eq:glm-moderate-definition-xi}, we have
\begin{equation}
\begin{aligned}
\frac{||\xi||^2_\Sigma}{p} & = \frac{||W||^2_\Sigma}{p} - \frac{1}{p}\langle W, \frac{\beta^\star}{\gamma_n}\rangle_\Sigma^2.
\end{aligned}
\end{equation}
The second term converges to 0 in probability, since $\langle W, {\beta^\star}/{\gamma_n}\rangle_\Sigma$ follows the standard normal distribution. For the first term, we have
\begin{equation}
\frac{||W||^2_\Sigma}{p} = \frac{1}{p}{W}^\intercal\Sigma W = \frac{Z^\intercal Z}{p}\overset{p}{\longrightarrow} 1.
\end{equation}

It follows that
\begin{equation}
\begin{aligned}
\sup_{t > 0,\ j\in S_0}\left|\mathbbm{P}(T_j>t)-Q(t)\right|&\leq \sup_{t > 0,\ j\in S_0}\left|\mathbbm{P}(T_j>t)-\mathbbm{P}(\widetilde T_j>t)\right| \\
& \hspace{0.05cm} + \sup_{t > 0,\ j\in S_0}\left|\mathbbm{P}(\widetilde T_j>t)-Q(t)\right| \\
& = \sup_{t > 0,\ j\in S_0}\left|\mathbbm{P}(T_j>t)-\mathbbm{P}(\widetilde T_j>t)\right| \\
& \hspace{0.05cm} + \sup_{t > 0,\ j\in S_0}\left|\mathbbm{P}(\widetilde V_j>t)-\mathbbm{P}( V_j>t)\right| \\
& \overset{p}{\longrightarrow} 0.
\end{aligned}
\end{equation}
The equality follows from Lemma \ref{lemma:glm-moderate-stochastic-representation}. The convergence in the last line follows from Equations \eqref{eq:T_j-overtildeT_j} and \eqref{eq:tildeZ_joverZ_j} (detailed arguments can be found in Equations \eqref{eq:inequality-T-and-Z-lower-bound} and \eqref{eq:inequality-T-and-Z-upper-bound}).
This completes the proof of Lemma \ref{lemma:glm-moderate-approx-mirror}.

\begin{lemma}
\label{lemma:mirror-statistic-bias-bound}
If Equation \eqref{eq:T_j-normal-approximation} holds, as $n,p\to\infty$, we have
\begin{equation}
\begin{aligned}
& \sup_{t\in\mathbbm{R},\ j\in S_0}\left|\mathbbm{P}(M_j > t) - H(t)\right| \longrightarrow 0.
\end{aligned}
\end{equation}
\end{lemma}

\textit{Proof of Lemma} \ref{lemma:mirror-statistic-bias-bound}. Without loss of generality, we assume $t > 0$. 
Let $Z^{(1)}$ and $Z^{(2)}$ be two independent random variables following the standard normal distribution, which are also independent to $T_j^{(1)}$ and $T_j^{(2)}$. 
For any $\epsilon>0$, for large enough $n$ and $p$, we have
\begin{equation}
\begin{aligned}
\mathbbm{P}(M_j > t) & = \mathbbm{P}\Big(T_j^{(2)} > I_t\big(T_j^{(1)}\big), \ T_j^{(1)} > 0\Big) + \mathbbm{P}\Big(T_j^{(2)} < -I_t\big(T_j^{(1)}\big),\ T_j^{(1)} < 0\Big)\\
& \leq \mathbbm{P}\Big(Z^{(2)} > I_t\big(T_j^{(1)}\big),\ T_j^{(1)} > 0\Big) + \mathbbm{P}\Big(Z^{(2)} < -I_t\big(T_j^{(1)}\big),\ T_j^{(1)} < 0\Big) + \epsilon\\
& = \mathbbm{P}\Big(\text{sign}\big(T_j^{(1)}Z^{(2)}\big)f\big(T_j^{(1)}, Z^{(2)}\big) > t\Big) + \epsilon\\
& = \mathbbm{P}\Big(T_j^{(1)} > I_t\big(Z^{(2)}\big), \ Z^{(2)} > 0\Big) + \mathbbm{P}\Big(T_j^{(1)} < -I_t\big(Z^{(2)}\big),\ Z^{(2)} < 0\Big) + \epsilon\\
& \leq \mathbbm{P}\Big(Z^{(1)} > I_t\big(Z^{(2)}\big),\ Z^{(2)} > 0\Big) + \mathbbm{P}\Big(Z^{(1)} < -I_t\big(Z^{(2)}\big),\ Z^{(2)} < 0\Big) + 2\epsilon\\
& = H(t) + 2\epsilon,
\end{aligned}
\end{equation}
where all the equalities follow from the fact that $f(u, v)$ is monotonically increasing with respect to $|u|$ and $|v|$, and the two inequalities follow from Equation \eqref{eq:T_j-normal-approximation} as well as the independence between the random variables.
Similarly, we can show that $\mathbbm{P}(M_j > t) \geq H(t) - 2\epsilon$ for large enough $n$ and $p$. This completes the proof of Lemma \ref{lemma:mirror-statistic-bias-bound}.

\begin{lemma}
\label{lemma:glm-moderate-approx-mirror-two}
Under Assumption \ref{assump:GLM-data-splitting-moderate-dimension}, as $n, p\to\infty$, we have 
\begin{equation}
\sup_{i,j\in S_0,\ t_1, t_2\in \mathbbm{R}}\left|\mathbbm{P}(T_i > t_1, T_j > t_2)-\mathbbm{P}(V_i > t_1, V_j > t_2)\right|\longrightarrow 0.
\end{equation}
\end{lemma}

\textit{Proof of Lemma} \ref{lemma:glm-moderate-approx-mirror-two}. 
Without loss of generality, we assume $t_1 >0$ and $t_2 > 0$. For any given $\epsilon > 0$, denote 
\begin{equation}
\begin{aligned}
& E_1 = \Big\{\max_{k\in[p]}|T_k/\widetilde{T}_k-1|\leq \epsilon\Big\},\ \ \ E_2 = \Big\{\max_{k\in[p]}|\widetilde{V}_k/V_k-1| \leq \epsilon\Big\}.
\end{aligned}
\end{equation}
By Equations \eqref{eq:T_j-overtildeT_j} and \eqref{eq:tildeZ_joverZ_j}, for large enough $n$ and $p$, we have $\mathbbm{P}(E_1)\geq 1-\epsilon$ and $\mathbbm{P}(E_2)\geq 1-\epsilon$. For $i, j \in S_0$, we have
\begin{equation}
\label{eq:inequality-T-and-Z-lower-bound}
\begin{aligned}
\mathbbm{P}(T_i > t_1, T_j > t_2) & \geq \mathbbm{P}(T_i > t_1, T_j > t_2 \mid  E_1)\mathbbm{P}(E_1)\\
& \geq \mathbbm{P}\Big(\widetilde{T}_i > \frac{t_1}{1-\epsilon}, \widetilde{T}_j > \frac{t_2}{1-\epsilon}\ \big|\ E_1\Big)\mathbbm{P}(E_1)\\
& \geq \mathbbm{P}\Big(\widetilde{T}_i > \frac{t_1}{1-\epsilon}, \widetilde{T}_j > \frac{t_2}{1-\epsilon}\Big) - \epsilon\\
& = \mathbbm{P}\Big(\widetilde{V}_i > \frac{t_1}{1-\epsilon}, \widetilde{V}_j > \frac{t_2}{1-\epsilon}\Big) - \epsilon\\
& \geq \mathbbm{P}\Big(\widetilde{V}_i > \frac{t_1}{1-\epsilon}, \widetilde{V}_j > \frac{t_2}{1-\epsilon}\ \big|\ E_2\Big)\mathbbm{P}(E_2) - \epsilon\\
& \geq \mathbbm{P}\Big(V_i > \frac{t_1}{(1-\epsilon)^2}, V_j > \frac{t_2}{(1-\epsilon)^2}\ \big|\ E_2\Big)\mathbbm{P}(E_2) - \epsilon\\
& \geq \mathbbm{P}\Big(V_i > \frac{t_1}{(1-\epsilon)^2}, V_j > \frac{t_2}{(1-\epsilon)^2}\Big) - 2\epsilon,
\end{aligned}
\end{equation}
in which the equality in the fourth line follows from Lemma \ref{lemma:glm-moderate-stochastic-representation}. Similarly, we have
\begin{equation}
\label{eq:inequality-T-and-Z-upper-bound}
\mathbbm{P}(T_i > t_1, T_j > t_2) \leq \mathbbm{P}(V_i > (1-\epsilon)^2t_1, V_j > (1-\epsilon)^2t_2) + 2\epsilon.
\end{equation}
Combining Equations \eqref{eq:inequality-T-and-Z-lower-bound} and \eqref{eq:inequality-T-and-Z-upper-bound} leads to the claim in Lemma \ref{lemma:glm-moderate-approx-mirror-two}.

\begin{corollary}
\label{cor:borel-set-convergence}
Under Assumption \ref{assump:GLM-data-splitting-moderate-dimension}, for any Borel set $B$ in $\mathbbm{R}^2$, as $n, p\to\infty$, we have 
\begin{equation}
\sup_{i,j\in S_0\in \mathbbm{R}}\left|\mathbbm{P}((T_i, T_j)\in B)-\mathbbm{P}((V_i, V_j)\in B)\right|\longrightarrow 0.
\end{equation}
\end{corollary}

\textit{Proof of Corollary \ref{cor:borel-set-convergence}.} The proof follows immediately from Lemma \ref{lemma:glm-moderate-approx-mirror-two} and Example 2.3 in \citet{billingsley2013convergence}.

\begin{lemma}
\label{lemma:weakly-correlated-normal}
Under Assumption \ref{assump:GLM-data-splitting-moderate-dimension}, as $n, p\to\infty$, we have 
\begin{equation}
\sup_{t\in\mathbbm{R}}\textnormal{Var}\bigg(\frac{1}{p_0}\sum_{j\in S_0}\mathbbm{1}(M_j > t)\bigg)\leq \frac{1}{4p_0}+ O(||R_{S_0}||_1) + o(1),
\end{equation}
in which $||R_{S_0}||_1 = \sum_{i,j\in S_0}R_{ij}/p_0^2$.
\end{lemma}

\textit{Proof of Lemma \ref{lemma:weakly-correlated-normal}}.
Without loss of generality, we assume $t > 0$. We have
\begin{equation}
\begin{aligned}
\sup_{t\in\mathbbm{R}}\textnormal{Var}\bigg(\frac{1}{p_0}\sum_{j\in S_0}\mathbbm{1}(M_j > t)\bigg) & \leq \frac{1}{p_0^2}\sum_{j\in S_0}\sup_{t\in\mathbbm{R}}\text{Var}(\mathbbm{1}(M_j > t))\\
& + \frac{1}{p_0^2}\sum_{i\neq j\in S_0}\sup_{t\in\mathbbm{R}}\text{Cov}(\mathbbm{1}(M_i > t), \mathbbm{1}(M_j > t)).
\end{aligned}
\end{equation}
Using the Cauchy-Schwartz inequality, the first term is upper bounded by $1/(4p_0)$. 
For the second term, since
\begin{equation}
\begin{aligned}
\text{Cov}(\mathbbm{1}(M_i > t), \mathbbm{1}(M_j > t)) & \leq \left|\mathbbm{P}(M_i > t, M_j > t) - H^2(t)\right| \\
& \hspace{0.05cm} + \left|\mathbbm{P}(M_i > t)\mathbbm{P}(M_j > t) - H^2(t)\right|,
\end{aligned}
\end{equation}
by Lemma \ref{lemma:mirror-statistic-bias-bound}, it is sufficient for us to show that for $i,j\in S_0$ and $i \neq j$,
\begin{equation}
\sup_{t\in\mathbbm{R}}\left|\mathbbm{P}(M_i > t, M_j > t) - H^2(t)\right| \leq O(|R_{ij}|) + o(1).
\end{equation}

By Condition \ref{cond:mirror-statistics-f}, we have the following decomposition based on the signs of $T^{(1)}_i$ and $T^{(1)}_j$,
\begin{equation}
\label{eq:mirror-statistic-variance-bound-eq1}
\begin{aligned}
\mathbbm{P}(M_i > t, M_j > t) & = \mathbbm{P}\left(T_i^{(2)} > I_t\big(T_i^{(1)} \big),\ \ T_j^{(2)} > I_t\big(T_j^{(1)}\big),\ \ T_i^{(1)} > 0, T_j^{(1)} > 0\right)\\
& +\hspace{0.05cm} \mathbbm{P}\left(T_i^{(2)} > I_t\big(T_i^{(1)} \big),\ \ T_j^{(2)} <- I_t\big(T_j^{(1)}\big), T_i^{(1)} > 0, T_j^{(1)} < 0\right)\\
& +\hspace{0.05cm} \mathbbm{P}\left(T_i^{(2)} < -I_t\big(T_i^{(1)} \big), T_j^{(2)} > I_t\big(T_j^{(1)}\big),\ \ T_i^{(1)} < 0, T_j^{(1)} > 0\right)\\
& +\hspace{0.05cm} \mathbbm{P}\left(T_i^{(2)} < -I_t\big(T_i^{(1)} \big), T_j^{(2)} <- I_t\big(T_j^{(1)}\big), T_i^{(1)} < 0, T_j^{(1)} < 0\right)\\
&:= I_1 + I_2 + I_3 + I_4.
\end{aligned}
\end{equation}
Denote $\Phi_r$ and $\phi_r$ as the CDF and pdf of the bivariate normal distribution with variances 1 and correlation $r$, respectively. Let  $\phi^{(n)}$ be the $n$-th derivative of $\phi$. 
Recall the Mehler's identity \citep{kotz2000bivariate}, that is, for any $t_1, t_2 \in\mathbbm{R}$,
\begin{equation}
\label{eq:mehler}
\Phi_r(t_1, t_2) = \Phi(t_1)\Phi(t_2) + \sum_{n = 1}^\infty \frac{r^n}{n!}\phi^{(n - 1)}(t_1)\phi^{(n - 1)}(t_2).
\end{equation}
For $I_1$, we have the following upper bound,
\begin{equation}
\label{eq:glm-moderate-mirror-joint-bound-eq1}
\begin{aligned}
I_{1} & = \mathbbm{E}\left[\mathbbm{P}\big(T_i^{(2)} > I_t(x),\ T_j^{(2)} > I_t(y)\big) \ \big|\ T_i^{(1)} = x > 0,\ T_j^{(1)} = y > 0\right]\\
& = \mathbbm{E}\left[\mathbbm{P}\big(V_i^{(2)} > I_t(x),\ V_j^{(2)} > I_t(y)\big) \ \big|\ T_i^{(1)} = x > 0,\ T_j^{(1)} = y > 0\right] + o(1)\\
& \leq \mathbbm{E}\left[Q(I_t(x))Q(I_t(y)) \ \big|\ T_i^{(1)} = x > 0,\ T_j^{(1)} = y > 0\right] + O(|R_{ij}|) + o(1).\\
\end{aligned}
\end{equation}
The second line follows from Lemma \ref{lemma:glm-moderate-approx-mirror-two}.
The third line follows from the Mehler's identity and Lemma 1 in \citet{azriel2015empirical}, i.e.,
\begin{equation}
\sum_{n = 1}^\infty \frac{\left[\sup_{t\in\mathbbm{R}}\phi^{(n - 1)}(t)\right]^2}{n!} < \infty.
\end{equation}

Similarly, we can upper bound $I_2$, $I_3$ and $I_4$. Combining the four upper bounds together, we obtain an upper bound on $\mathbbm{P}(M_i > t, M_j > t)$ specified as below,
\begin{equation}
\label{eq:glm-moderate-mirror-joint-bound-eq2}
\mathbbm{P}\left(\text{sign}(Z_i^{(2)}T_i^{(1)})f(Z_i^{(2)}, T_i^{(1)}) > t, \text{sign}(Z_j^{(2)}T_j^{(1)})f(Z_j^{(2)}, T_j^{(1)}) > t\right) + O(|R_{ij}|) + o(1),
\end{equation}
in which $Z_i^{(2)}$ and $Z_j^{(2)}$ are two independent random variables (also independent to everything else) following the standard normal distribution. We can further decompose the first term in Equation \eqref{eq:glm-moderate-mirror-joint-bound-eq2} into four terms as Equation \eqref{eq:mirror-statistic-variance-bound-eq1} by conditioning on the signs of $Z_i^{(2)}$ and $Z_j^{(2)}$, and repeat the upper bound in Equation \eqref{eq:glm-moderate-mirror-joint-bound-eq1}. This leads to
\begin{equation}
\mathbbm{P}(M_i > t, M_j > t) \leq H^2(t) + O(|R_{ij}|) + o(1).
\end{equation}
Similarly, we can establish the corresponding lower bound.
This completes the proof of Lemma \ref{lemma:weakly-correlated-normal}.

\begin{corollary}
\label{cor:convergence-of-mirror}
Under Assumption \ref{assump:GLM-data-splitting-moderate-dimension}, as $n,p\to\infty$, for any $t\in\mathbbm{R}$, we have
\begin{equation}
\bigg|\frac{1}p_0\sum_{j\in S_0}\mathbbm{1}(M_j > t)-H(t)\bigg|\overset{p}{\to}0.
\end{equation}
\end{corollary}

\textit{Proof of Corollary \ref{cor:convergence-of-mirror}}. By Lemma \ref{lemma:weakly-correlated-normal} and Lemma \ref{lemma:mirror-statistic-bias-bound}, we have
\begin{equation}
\mathbbm{E}\bigg[\bigg(\frac{1}{p_0}\sum_{j\in S_0}\mathbbm{1}(M_j > t)-H(t)\bigg)^2\bigg]\leq \frac{1}{4p_0}+O(||R_{S_0}||_1) + o(1).
\end{equation}
Under Assumption \ref{assump:GLM-data-splitting-moderate-dimension}, we have 
$||R_{S_0}||_1\to 0$ following the arguments in the proof of
Lemma A.2 in \citet{zhao2020asymptotic}. Thus we complete the proof of Corollary \ref{cor:convergence-of-mirror} using the Markov inequality.

\subsubsection{Proof of Proposition \ref{prop:GLM-data-splitting-FDR-moderate-dimension}}
\label{subsubsec:proof-glm-moderate-FDR}
For the ease of presentation, we introduce the following notations. For $t \in \mathbbm{R}$, denote
\begin{equation}
\begin{aligned}
&\widehat{G}^0_{p}(t) = \frac{1}{p_0}\sum_{j \in S_0}\mathbbm{1}(M_j > t),\ \ \ \widehat{G}^1_{p}(t) = \frac{1}{p_1}\sum_{j \in S^\star}\mathbbm{1}(M_j > t),\ \ \
\widehat{V}^0_{p}(t) = \frac{1}{p_0}\sum_{j \in S_0}\mathbbm{1}(M_j < - t).
\end{aligned}
\end{equation}
Let $r_{p} = p_1/p_0$. Denote
\begin{equation}
\begin{aligned}
&\text{FDP}_p(t)  = \frac{\widehat{G}_{p}^0(t)}{\widehat{G}_{p}^0(t) + r_{p}\widehat{G}_{p}^1(t)},\ \ \ 
\text{FDP}^\dagger_p(t)  = \frac{\widehat{V}_{p}^0(t)}{\widehat{G}_{p}^0(t) + r_{p}\widehat{G}_{p}^1(t)},\ \ \ 
\overline{\text{FDP}}_p(t)  = \frac{H(t)}{H(t) + r_{p}\widehat{G}_{p}^1(t)}.
\end{aligned}
\end{equation}

\begin{lemma}
\label{lemma:glm-moderate-supreme-convergence}
We have as $n,p\to\infty$, 
\begin{equation}
\begin{aligned}
&\sup_{t\in\mathbbm{R}} \left|\widehat{G}^0_{p}(t) - H(t)\right| \overset{p}{\longrightarrow} 0,\\
&\sup_{t\in\mathbbm{R}} \left|\widehat{V}^0_{p}(t) - H(t)\right| \overset{p}{\longrightarrow} 0.
\end{aligned}
\end{equation}
\end{lemma}

\textit{Proof of Lemma \ref{lemma:glm-moderate-supreme-convergence}}. 
We prove the first claim based on an $\epsilon$-net argument. The second claim follows similarly. 
For any $\epsilon \in (0, 1)$, denote $-\infty = \alpha^{p}_0 < \alpha^{p}_1 < \cdots < \alpha^{p}_{N_\epsilon} = \infty$ in which $N_\epsilon = \lceil2/\epsilon\rceil$, such that $H(\alpha^{p}_{k - 1}) - H(\alpha^{p}_k) \leq \epsilon /2$ for $k\in[N_\epsilon]$. Such a sequence $\{\alpha_k^p\}$ exists because $H(t)$ is continuous and in the range of $[0, 1]$.
We have
\begin{equation}
\label{eq:lemma-2-1}
\begin{aligned}
\mathbbm{P}\left(\sup_{t\in\mathbbm{R}}\widehat{G}^0_{p}(t) - H(t) > \epsilon\right) & \leq \mathbbm{P}\left(\bigcup_{k = 1}^{N_\epsilon}\sup_{t \in\left[\alpha^{p}_{k - 1}, \alpha^{p}_k\right)}\widehat{G}^0_{p}(t) - H(t) > \epsilon\right)\\
& \leq \sum_{k = 1}^{N_{\epsilon}}\mathbbm{P}\left(\sup_{t \in\left[\alpha^{p}_{k - 1}, \alpha^{p}_k\right)}\widehat{G}^0_{p}(t) - H(t) > \epsilon\right).
\end{aligned}
\end{equation}
We note that both $\widehat{G}^0_{p}(t)$ and $H(t)$ are monotonic decreasing function. Therefore, for any $k \in [N_\epsilon]$, we have
\begin{equation}
\begin{aligned}
\sup_{t \in\left[\alpha^{p}_{k - 1}, \alpha^{p}_k\right)}\widehat{G}^0_{p}(t) - H(t) & \leq \widehat{G}^0_{p}(\alpha^{p}_{k - 1}) - H(\alpha^{p}_{k})\\
& \leq \widehat{G}^0_{p}(\alpha^{p}_{k - 1}) - H(\alpha^{p}_{k - 1}) + \epsilon/2.
\end{aligned}
\end{equation}
By Corollary \ref{cor:convergence-of-mirror}, we have
\begin{equation}
\label{eq:supreme-bound}
\begin{aligned}
\mathbbm{P}\left(\sup_{t\in\mathbbm{R}}\widehat{G}^0_{p}(t) - H(t) > \epsilon\right) & \leq \sum_{k = 1}^{N_{\epsilon}}\mathbbm{P}\left(\widehat{G}^0_{p}(\alpha^{p}_{k - 1}) - H(\alpha^{p}_{k - 1}) > \frac{\epsilon}{2}\right)\\
& \leq N_\epsilon \max_{k\in[N_\epsilon]}\mathbbm{P}\left(\widehat{G}^0_{p}(\alpha^{p}_{k - 1}) - H(\alpha^{p}_{k - 1}) > \frac{\epsilon}{2}\right)\\
& \longrightarrow 0,
\end{aligned}
\end{equation}
as $n,p\to\infty$.
Similarly, we can show that
\begin{equation}
\begin{aligned}
\mathbbm{P}\left(\inf_{t\in\mathbbm{R}}\widehat{G}^0_{p}(t) - H(t) < -\epsilon\right) &\leq \sum_{k = 1}^{N_{\epsilon}}\mathbbm{P}\left(\widehat{G}^0_{p}(\alpha^{p}_k) - H(\alpha^{p}_k) < -\frac{\epsilon}{2}\right)\\
&  \leq  N_\epsilon \max_{k\in[N_\epsilon]}\mathbbm{P}\left(\widehat{G}^0_{p}(\alpha^{p}_k) - H(\alpha^{p}_k) < -\frac{\epsilon}{2}\right)\\
&\longrightarrow 0.
\end{aligned}
\end{equation}
This concludes the proof of the first claim in Lemma \ref{lemma:glm-moderate-supreme-convergence}.

\vspace{0.5cm}
\noindent\textit{Proof of Proposition \ref{prop:GLM-data-splitting-FDR-moderate-dimension}}.
We first show that for any $\epsilon \in (0, q)$, we have
\begin{equation}
\label{lemma:t-boundness}
\mathbbm{P}(\tau_q\leq t_{q-\epsilon}) \geq 1 - \epsilon,
\end{equation}
in which $t_{q-\epsilon} > 0$ satisfying $\text{FDP}^\infty(t_{q-\epsilon}) \leq q - \epsilon$. 
By Lemma \ref{lemma:glm-moderate-supreme-convergence}, for any fixed $t \in\mathbbm{R}$, we have 
\begin{equation}
|\text{FDP}^\dagger_p(t)-\text{FDP}_p(t)| \overset{p}{\to} 0.
\end{equation}
It follows that for any fixed $t\in\mathbbm{R}$, we have
\begin{equation}
|\text{FDP}^\dagger_p(t)-\text{FDP}^\infty(t)|\leq |\text{FDP}^\dagger_p(t)-\text{FDP}_p(t)|+|\text{FDP}_p(t)-\text{FDP}^\infty(t)| \overset{p}{\to} 0.
\end{equation}
By the definition of $\tau_q$, i.e., $\tau_q = \inf\{t > 0: \text{FDP}^\dagger_p(t)\leq q\}$, we have
\begin{equation}
\begin{aligned}
\mathbbm{P}(\tau_q\leq t_{q-\epsilon}) & \geq \mathbbm{P}(\text{FDP}^\dagger_p(t_{q-\epsilon}) \leq q)\\
& \geq \mathbbm{P}(|\text{FDP}^\dagger_p(t_{q-\epsilon}) -\text{FDP}^\infty(t_{q-\epsilon})| \leq \epsilon)\\
& \geq 1 - \epsilon
\end{aligned}
\end{equation}
for $p$ large enough. Conditioning on the event $\tau_q\leq t_{q-\epsilon}$, we have
\begin{equation}
\begin{aligned}
\limsup_{p\to\infty}
\mathbbm{E}\left[\text{FDP}_p\left(\tau_q\right)\right] 
& \leq \limsup_{p\to\infty}
\mathbbm{E}\left[\text{FDP}_p\left(\tau_q\right)\mid \tau_q\leq t_{q-\epsilon}\right] \mathbbm{P}(\tau_q\leq t_{q-\epsilon}) + \epsilon\\
& \leq \limsup_{p\to\infty} \mathbbm{E}\Big[\big|\text{FDP}_p\left(\tau_q\right) - \overline{\text{FDP}}_p\left(\tau_q\right)\big|\ \big\vert\ \tau_q\leq t_{q-\epsilon}\Big]\mathbbm{P}(\tau_q\leq t_{q-\epsilon}) \\
& \hspace{0.05cm} + \limsup_{p\to\infty} \mathbbm{E}\left[\big|\text{FDP}^\dagger_p\left(\tau_q\right) - \overline{\text{FDP}}_p\left(\tau_q\right)\big|\ \big\vert\ \tau_q\leq t_{q-\epsilon}\right]\mathbbm{P}(\tau_q\leq t_{q-\epsilon})\\
& \hspace{0.05cm} + \limsup_{p\to\infty} \mathbbm{E}\left[\text{FDP}^\dagger_p\left(\tau_q\right) \ \big\vert\ \tau_q\leq t_{q-\epsilon}\right]\mathbbm{P}(\tau_q\leq t_{q-\epsilon}) + \epsilon\\
& \leq \limsup_{p\to\infty} \mathbbm{E}\Big[\sup_{0 < t \leq t_{q-\epsilon}}\left|\text{FDP}_p(t) - \overline{\text{FDP}}_p(t)\right|\Big]\\
& \hspace{0.05cm}  + \limsup_{p\to\infty} \mathbbm{E}\Big[\sup_{0<t\leq t_{q-\epsilon}}\left|\text{FDP}^\dagger_p(t) - \overline{\text{FDP}}_p(t)\right|\Big] \\
& \hspace{0.05cm} + \limsup_{p\to\infty} \mathbbm{E}\left[\text{FDP}^\dagger_p\left(\tau_q\right)\right] + \epsilon.
\end{aligned}
\end{equation}
The first two terms are 0 based on Lemma \ref{lemma:glm-moderate-supreme-convergence} and the dominated convergence theorem. For the third term, we have
$\text{FDP}^\dagger_p\left(\tau_q\right)\leq q$ almost surely
based on the definition of $\tau_q$. This concludes the proof of Proposition \ref{prop:GLM-data-splitting-FDR-moderate-dimension}.

\subsection{Proof of Proposition \ref{prop:GLM-asymptotics-mirror-feature-moderate-dimension}}
Similar to Equation \eqref{eq:linear-Gaussian-mirror-model-reparametrization}, we can reparametrize the GLM with respect to the features $(X_{-j}, X_j^+, X_j^-)$ so that the corresponding ture regression coefficients are $\beta^\star_{-j},\ \beta^\star_j/2$ and $\beta^\star_j/2$, respectively. In addition, the asymptotic signal strength $\gamma$ and the sampling ratio $\kappa$ remain the same, whereas the condition variance becomes
\begin{equation}
\text{Var}(X_j^ + \mid X_{-j}, X_j^-) = \text{Var}(X_j^ - \mid X_{-j}, X_j^+) = 1/\Theta^\ast_{11},
\end{equation}
in which $\Theta^\ast$ is defined in Proposition \ref{prop:GLM-asymptotics-mirror-feature-moderate-dimension}. The proof of Proposition \ref{prop:GLM-asymptotics-mirror-feature-moderate-dimension} thus follows from Lemma \ref{lemma:glm-moderate-parameter-convergence} and the proof of Theorem 3.1 in \citet{zhao2020asymptotic}.

\subsection{Proof of Proposition \ref{prop:GLM-Gaussian-mirror-FDR-moderate-dimension}}
The proof of Proposition \ref{prop:GLM-Gaussian-mirror-FDR-moderate-dimension} is essentially the same as the proof of Proposition \ref{prop:GLM-data-splitting-FDR-moderate-dimension}, once we have the following Lemmas (in particular, Lemma \ref{lemma:glm-moderate-mirror-covariance}).

\begin{lemma}
\label{lemma:glm-mirror-close}
For the response vector $y$, we consider fitting two GLMs with respect to the set of features $X = (Z, X_1,\ldots, X_p)\in \mathbbm{R}^{n\times(p+1)}$ (full model), and $\widetilde X = (X_1, \ldots, X_p)\in \mathbbm{R}^{n\times p}$ (reduced model), respectively, in which each row of $\widetilde X$ are i.i.d. samples from $N(0, I_p)$, and $Z$ is a null feature following $N(0, I_n)$.
Denote the MLEs for the full model and the reduced model as $\widehat\beta \in \mathbbm{R}^{p + 1}$ and $\widetilde\beta \in \mathbbm{R}^p$, respectively.
Denote the difference between the two MLEs as $\Delta = \widehat\beta_{2:(p+1)}-\widetilde\beta$, then we have 
$$||\Delta||_\infty = O_p\big(n^{-1+o(1)}\big).$$
\end{lemma}

\textit{Proof of Lemma \ref{lemma:glm-mirror-close}.} The proof relies heavily on the theoretical results derived in \citet{Pragya2019LikelihoodRatio} (see Section 7 therein), and the main technical tool is the leave-one-out analysis. In the following, we sketch the proof of Lemma \ref{lemma:glm-mirror-close}, and refer the readers to the results in \citet{Pragya2019LikelihoodRatio} for complete details. 

We first construct a surrogate of $\widehat\beta$ as below following Equation (86) in \citet{Pragya2019LikelihoodRatio},
\begin{equation}
\widetilde b = \left[
\begin{matrix}
   0 \\
   \widetilde\beta
\end{matrix}
\right]+\widetilde b_1\left[
\begin{matrix}
   1\\
   -\widetilde G^{-1}w
\end{matrix}\right],
\end{equation}
in which
\begin{equation}
\widetilde G = \frac{1}{n}\widetilde X^\intercal D_{\widetilde\beta} \widetilde X,\ \ \ \ \ \ w = \frac{1}{n}\widetilde X^\intercal D_{\widetilde\beta}Z,
\end{equation}
where $D_{\widetilde\beta}$ is a $n\times n$ diagonal matrix with entries $\rho^{\prime\prime}(\widetilde x_i^\intercal\widetilde\beta)$ for $i\in[n]$, and $\widetilde b_1$ is defined as below following Equation (91) in \citet{Pragya2019LikelihoodRatio},
\begin{equation}
\widetilde b_1 = \frac{Z^\intercal \widetilde r}{Z^\intercal D_{\widetilde\beta}^{1/2}HD_{\widetilde\beta}^{1/2}Z},    
\end{equation}
where 
\begin{equation}
H = I - \frac{1}{n}   D_{\widetilde\beta}^{1/2}\widetilde X\widetilde G^{-1} \widetilde X^\intercal D_{\widetilde\beta}^{1/2}\ \ \ \text{and}\ \ \ \widetilde r_i = y_i - \rho^\prime(\widetilde x_i^\intercal\widetilde\beta),\ \ \ i\in[n].
\end{equation}
By Theorem 8 and similar arguments in Section 7.4 in \citet{Pragya2019LikelihoodRatio}, we have
\begin{equation}
\begin{aligned}
||\Delta||_\infty & \leq ||\widehat\beta-\widetilde b||_\infty +||\widetilde b_1\widetilde G^{-1}w||_\infty \\
& \lesssim n^{-1+o(1)}+n^{-1/2+o(1)}||\widetilde G^{-1}w||_\infty.
\end{aligned}
\end{equation}
Thus it remains to bound $||\widetilde G^{-1}w||_\infty$. 

We note that both $\widetilde X$ and $D_{\widetilde\beta}$ are independent to $Z$. Therefore, conditioning on $\widetilde X$, $\widetilde G^{-1}w$ follows a multivariate normal distribution
with mean 0, and covariance matrix $\Sigma_Z$ specified as below,
\begin{equation}
\begin{aligned}
\Sigma_Z = \frac{1}{n^2}\widetilde G^{-1}\widetilde X^\intercal D_{\widetilde\beta}^2\widetilde X\widetilde G^{-1} &\prec \sup_{i\in[n]}|\rho^{\prime\prime}(\widetilde x_i^\intercal\widetilde\beta)|\frac{1}{n^2}\widetilde G^{-1}\widetilde X^\intercal D_{\widetilde\beta}\widetilde X\widetilde G^{-1} \\
& = \frac{1}{n}\sup_{i\in[n]}|\rho^{\prime\prime}(\widetilde x_i^\intercal\widetilde\beta)|\widetilde G^{-1}.
\end{aligned}
\end{equation}
By Lemma 7 in \citet{Pragya2019LikelihoodRatio}, we have $\sigma_{\min}(\widetilde G)> \lambda$ with high probability, in which $\lambda$ is some positive constant. This implies $\sigma_{\max}(\Sigma_Z) = O_p(1/n)$, which further leads to $||\widetilde G^{-1}w||_\infty \lesssim \sqrt{{ \log n}/{n}}$. Thus we complete the proof of Lemma \ref{lemma:glm-mirror-close} via
\begin{equation}
||\Delta||_\infty \lesssim n^{-1+o(1)}+ n^{-1/2+o(1)}\sqrt{{ \log n}/{n}}= O_p( n^{-1+o(1)}). 
\end{equation}

\begin{remark}
Lemma \ref{lemma:glm-mirror-close} also holds in the case where each row of the design matrix $\widetilde X$ independently follows a multivariate normal distribution with a general covariance matrix $\Sigma$ satisfying Assumption \ref{assump:GLM-data-splitting-moderate-dimension} (1).
In addition, by Lemma \ref{lemma:glm-mirror-close}, we know that the infinity norm of the absolute difference between the normalized MLEs (defined similarly as Equation \eqref{eq:glm-moderate-dimension-T-definition}) of the full model and the reduced model behave as $\sqrt{n}||\Delta||_\infty = O_p( n^{-1/2+o(1)}) = o(1)$, which is crucial in the proof of Lemma \ref{lemma:glm-moderate-mirror-covariance}.
\end{remark}

\begin{lemma}
\label{lemma:glm-mirror-conditional-variance}
As $n, p\to\infty$, we have 
\begin{equation}
\max_{j\in[p]}|c_j-\tau_j| = o_p(1),
\end{equation}
in which $c_j$ is defined in Equation \eqref{eq:Gaussian-mirror-scaling}, and $\tau_j = \text{Var}(X_j|X_{-j})$. 
\end{lemma}

\textit{Proof of Lemma \ref{lemma:glm-mirror-conditional-variance}.} Recall that $c_j = {||P_{-j}^\perp X_j||}/{||P_{-j}^\perp Z_j||}$. We note that $||P_{-j}^\perp Z_j||^2 \sim \chi^2_{n - p + 1}$ and $||P_{-j}^\perp X_j||^2\sim \tau_j^2\chi^2_{n - p + 1}$. In addition, $P_{-j}^\perp Z_j$ and $P_{-j}^\perp X_j$ are independent since $Z_j$ is independent to the design matrix $X$. The proof is thus completed by the Hoeffiding's inequality and the union bound over $j \in[p]$.

\begin{lemma}
\label{lemma:mirror-two-close}
For $j\in[p]$, we consider fitting the GLM using the response vector $y$ with respect to two augmented sets of features, $(X_{-j}, X_j+c_jZ_j, X_j-c_jZ_j)$ and $(X_{-j}, X_j+\tau_jZ_j, X_j-\tau_jZ_j)$, in which $\tau_j = \text{Var}(X_j|X_{-j})$, and $c_j$ is defined in Equation \eqref{eq:Gaussian-mirror-scaling}. Denote the MLEs associated with features $X_j+c_jZ_j, X_j-c_jZ_j$ and $X_j+\tau_jZ_j, X_j-\tau_jZ_j$ as $\widehat\beta_j^+, \widehat\beta_j^-$ and ${}^\ast\widehat\beta_j^+, {}^\ast\widehat\beta_j^-$, respectively. Further, we denote their normalized versions, defined similarly as Equation \eqref{eq:glm-moderate-dimension-T-definition}, as $T_j^+, T_j^-$ and ${}^\ast\widehat T_j^+, {}^\ast\widehat T_j^-$. Then under Assumption \ref{assump:GLM-data-splitting-moderate-dimension}, as $n, p\to\infty$ we have
\begin{equation}
\sup_{j\in[p],\ t_1, t_2\in \mathbbm{R}}\left|\mathbbm{P}(T_j^+ > t_1, T_j^- > t_2)-\mathbbm{P}({}^\ast\widehat T_j^+ > t_1, {}^\ast\widehat T_j^- > t_2)\right|\longrightarrow 0.
\end{equation}
\end{lemma}

\textit{Proof of Lemma \ref{lemma:mirror-two-close}.} We consider fitting the GLM using the response vector $y$ with respect to the augmented set of features $(X_{-j}, X_j, Z_j)$. Denote the MLEs as ${}^\dagger\widehat\beta_{-j}, {}^\dagger\widehat\beta_j, {}^\dagger\widehat\vartheta$. We have the following relationship,
\begin{equation}
\begin{aligned}
&\widehat\beta_j^++\widehat\beta_j^- = {}^\dagger\widehat\beta_j, \hspace{0.75cm}
{}^\ast\widehat\beta_j^++{}^\ast\widehat\beta_j^- = {}^\dagger\widehat\beta_j,\\
&\widehat\beta_j^+-\widehat\beta_j^- = {}^\dagger\widehat\vartheta/c_j, \ \ \  {}^\ast\widehat\beta_j^+ - {}^\ast\widehat\beta_j^- = {}^\dagger\widehat\vartheta/\tau_j.
\end{aligned}
\end{equation}
By Proposition \ref{prop:GLM-asymptotics-moderate-dimension}, $|{}^\dagger\widehat\vartheta| = O_p(1/\sqrt{n})$. By Equation \eqref{eq:bound-tau_j} and Lemma \ref{lemma:glm-moderate-tau},  \ref{lemma:glm-mirror-conditional-variance}, we have
\begin{equation}
\begin{aligned}
\max_{j\in[p]}\left|\widehat T^+_j - {}^\ast\widehat T^+_j\right| & = \max_{j\in[p]}|\widehat{\tau}_j/\sigma_\star|\left|(1/c_j - 1/\tau_j)/2\right||\sqrt{n}{}^\dagger\widehat\vartheta| = o_p(1).
\end{aligned}
\end{equation}
Similarly, we have $\max_{j\in[p]}|\widehat T^-_j - {}^\ast\widehat T^-_j| = o_p(1)$. The proof of Lemma \ref{lemma:mirror-two-close} thus completes based on similar arguments as in the proof of Lemma \ref{lemma:glm-moderate-approx-mirror-two}.

\begin{remark}
Without loss of generality, by Lemma \ref{lemma:mirror-two-close}, instead of calculating $c_j$ by Equation \eqref{eq:Gaussian-mirror-scaling}, we assume $c_j =\tau_j$ for $j\in[p]$ henceforth in order to simplify the proofs.
\end{remark}

\begin{lemma}
\label{lemma:glm-mirror-bridge}
Under Assumption \ref{assump:GLM-data-splitting-moderate-dimension}, as $n, p\to\infty$, we have
\begin{equation}
\sup_{t\in\mathbbm{R}, j\in S_0}|\mathbbm P(M_j>t)-H(t)|\to 0.
\end{equation}
\end{lemma}

\textit{Proof of Lemma \ref{lemma:glm-mirror-bridge}.} 
Recall that we fit a GLM with respect to the augmented set of features $(X_{-j}, X_j^+, X_j^-)$, of which the augmented covariance matrix and precision matrix are denoted as $\Sigma_{\text{aug}}$ and $\Theta_{\text{aug}}$, respectively. Based on $\Sigma_{\text{aug}}$ and $\Theta_{\text{aug}}$, we define $V_j^+, V_j^-$ and $\widetilde{V}_j^+, \widetilde{V}_j^-$ similarly as Equation \eqref{eq:glm-moderate-dimension-V-definition}. In particular, $V_j^+$ and $V_j^-$ are independent since $c_j = \tau_j$. Besides, the normalized MLEs 
$T_j^+, T_j^-$, as well as their approximations $\widetilde{T}_j^+, \widetilde{T}_j^-$, are defined similarly as Equation \eqref{eq:glm-moderate-dimension-T-definition}.

For $j\in S_0$, we have
\begin{equation}
\begin{aligned}
\mathbbm{P}(M_j > t) & = \mathbbm{P}\Big(\text{sign}\big(T_j^+T_j^-\big)f\big(T_j^+, T_j^-\big) > t\Big)\\
& = \mathbbm{P}\Big(\text{sign}\big(V_j^+V_j^-\big)f\big(V_j^+, V_j^-\big) > t\Big) + o(1)\\
& = H(t) + o(1).
\end{aligned}
\end{equation}
The second line follows from Corollary \ref{cor:borel-set-convergence}, in which the asymptotic vanishing term $o(1)$ is uniform over $j\in S_0$ and $t\in\mathbbm{R}$. 
The proof of Lemma \ref{lemma:glm-mirror-bridge} is thus completed.

\begin{lemma}
\label{lemma:glm-moderate-mirror-covariance}
Under Assumption \ref{assump:GLM-data-splitting-moderate-dimension}, as $n, p\to\infty$, we have 
\begin{equation}
\sup_{t\in\mathbbm{R}}\textnormal{Var}\bigg(\frac{1}{p_0}\sum_{j\in S_0}\mathbbm{1}(M_j > t)\bigg)\leq \frac{1}{4p_0}+ O(||R_{S_0}||_1) + o(1),
\end{equation}
in which $||R_{S_0}||_1 = \sum_{i,j\in S_0}R_{ij}/p_0^2$.
\end{lemma}

\textit{Proof of lemma \ref{lemma:glm-moderate-mirror-covariance}.} 
For any $j\in S_0$, let $T_j^+, T_j^-$ be the normalized MLEs, defined similarly as Equation \eqref{eq:glm-moderate-dimension-T-definition}, when we fit a GLM 
with respect to the augmented set of features $(X_{-j}, X_j^+, X_j^-)$. For any different $i, j \in S_0$, let ${}^{\ast}T_i^+, {}^{\ast}T_i^-$ and ${}^{\ast}T_j^+, {}^{\ast}T_j^-$ be the normalized MLEs when we fit a GLM with respect to the augmented set of features $(X_{-[i,j]}, X_i^+, X_i^-, X_j^+, X_j^-)$, in which $X_{-[i,j]}$ denotes the design matrix excluding the $i$-th and $j$-th columns. Correspondingly, we define ${}^{\ast}V_j^+, {}^{\ast}V_j^-$ and ${}^{\ast}\widetilde{V}_j^+, {}^{\ast}\widetilde{V}_j^-$ following Equation \eqref{eq:glm-moderate-dimension-V-definition}, and define ${}^{\ast}\widetilde{T}_j^+, {}^{\ast}\widetilde{T}_j^-$ following Equation \eqref{eq:glm-moderate-dimension-T-definition}.

For any different $i, j \in S_0$ and $t > 0$, we have
\begin{equation}
\begin{aligned}
\mathbbm{P}(M_i > t, M_j > t) & = \mathbbm{P}\Big(\text{sign}\big(T_i^+T_i^-\big)f\big(T_i^+, T_i^-\big) > t ,\text{sign}\big(T_j^+T_j^-\big)f\big(T_j^+, T_j^-\big) > t\Big)\\
& \hspace{-1.5cm} = \mathbbm{P}\Big(\text{sign}\big({}^{\ast}\widetilde{T}_i^+{}^{\ast}\widetilde{T}_i^-\big)f\big({}^{\ast}\widetilde{T}_i^+, {}^{\ast}\widetilde{T}_i^-\big) > t\ ,\text{sign}\big({}^{\ast}\widetilde{T}_j^+{}^{\ast}\widetilde{T}_j^-\big)f\big({}^{\ast}\widetilde{T}_j^+, {}^{\ast}\widetilde{T}_j^-\big) > t\Big)\hspace{0.1cm} + o(1)\\
&  \hspace{-1.5cm} = \mathbbm{P}\Big(\text{sign}\big({}^{\ast}V_i^+{}^{\ast}V_i^-\big)f\big({}^{\ast}V_i^+, {}^{\ast}V_i^-\big) > t ,\text{sign}\big({}^{\ast}V_j^+{}^{\ast}V_j^-\big)f\big({}^{\ast}V_j^+, {}^{\ast}V_j^-\big) > t\Big) + o(1),\\
\end{aligned}
\end{equation}
in which the second line follows from Lemma \ref{lemma:glm-mirror-close}, 
and the last line is based on similar arguments as the proof of Lemma \ref{lemma:glm-moderate-approx-mirror-two} and Corollary \ref{cor:borel-set-convergence}.
The rest of proof follows similarly as the proof Lemma \ref{lemma:weakly-correlated-normal}, based on a similar decomposition as Equation \eqref{eq:mirror-statistic-variance-bound-eq1} by conditioning on the signs of ${}^{\ast}V_j^+$ and ${}^{\ast}V_j^-$. The proof of Lemma \ref{lemma:glm-moderate-mirror-covariance} is thus completed.

\subsection{Proof of Proposition \ref{prop:linear-data-splitting-FDR-high-dimension}}
For the design matrix $X$, we denote $X_j$ as its $j$-th column and $x_k$ as its $k$-th row, in which $j\in[p]$ and $k\in[n]$. We introduce the normalized versions of $\Theta$ and $\Lambda$ defined as below,
\begin{equation}
\Theta^0_{ij} = \frac{\Theta_{ij}}{\sqrt{\Theta_{ii}\Theta_{jj}}},\ \ \ \Lambda^0_{ij} = \frac{\Lambda_{ij}}{\sqrt{\Lambda_{ii}\Lambda_{jj}}}.
\end{equation}

\subsubsection{Technical lemmas}

\begin{lemma}
\label{lemma:unnormalized-precision-matrix}
Under Assumption \ref{assump:linear-high-dimension}, for any $\epsilon > 0$, there exist constants $c_{1}, c_{2}, c_{3}, c_{4} > 0$ such that for large enough $n$,
\begin{equation}
\mathbbm P\left(\max_{i,j\in[p]}\left|\Lambda_{ij}-\Theta_{ij}\right|\leq \frac{1}{\sqrt{\log p}}\right) \geq 1-c_{1}p^{2}\exp\left(-c_{2}\frac{n}{\log p}\right)-c_{3}p^{2}\exp\left(-c_{4}n\right)-\epsilon.
\end{equation}
\end{lemma}

 \textit{Proof of Lemma} \ref{lemma:unnormalized-precision-matrix}. The proof follows similarly as the proof of Lemma 7.2 in \citet{javanmard2013nearly}. For any $\epsilon > 0$, by Proposition 5.1 in \citet{javanmard2013nearly} (also see \citet{van2014asymptotically}), there exists an $n_\epsilon \in \mathbbm{N}_+$ such that for $n \geq n_\epsilon$, we have
 \begin{equation}
 \mathbbm{P}\left(||\widehat{\Theta} - \Theta||_\infty \leq \frac{1}{\sqrt{\log p}}\right) \geq 1 - \epsilon.
 \end{equation}
In the following, we condition on this high probability event.
Denote $v = \Theta^{\top}e_{i}, u = \Theta^{\top}e_{j}, \delta = (\Theta-\hat\Theta)^{\top}e_{i}, \eta = (\Theta-\hat\Theta)^{\top}e_{j}$. We have the following decomposition, 
\begin{equation}
\label{eq:precision-matrix-decomposition}
\begin{aligned}
\Lambda_{ij} - \Theta_{ij} & = (v - \delta)^{\top}\hat\Sigma(u - \eta) - \Theta_{ij} \\
& = \big(v^{\top}\hat\Sigma u - \Theta_{ij}\big) - v^{\top}\hat\Sigma\eta - \delta^{\top}\hat\Sigma u + \delta^{\top}\hat\Sigma\eta.
\end{aligned}
\end{equation}

We proceed to bound each term. For the term $v^{\top}\hat\Sigma u -  \Theta_{ij}$, since $\mathbbm{E}[v^{\top}\hat\Sigma u] = v^{\top}\Sigma u = \Theta_{ij}$, it follows that
\begin{equation}
\begin{aligned}
v^{\top}\hat\Sigma u - \Theta_{ij} & = v^{\top}\hat\Sigma u - \mathbbm{E}[v^{\top}\hat\Sigma u] \\
& = \frac{1}{n}\sum_{k = 1}^{n}e_{i}^{\top}\Theta\left(x_{k}x_{k}^{\top} - \mathbbm{E}[x_{k}x^{\top}_{k}]\right)\Theta^{\top}e_{j}
\end{aligned}
\end{equation}
Denote $\xi_{k} = e_{i}^{\top}\Theta\left(x_{k}x_{k}^{\top} - \mathbbm{E}[x_{k}x_{k}^{\top}]\right)\Theta^{\top}e_{j}$ for $k\in[n]$. We note that $\xi_{k}$'s are independent, and have sub-exponential tails because
\begin{equation}
\begin{aligned}
||\xi_{k}||_{\psi_{1}} \leq 2||(e_{i}^{\top}\Theta x_{k})^2||_{\psi_{1}} \leq 4||e_{i}^{\top}\Theta x_{k}||_{\psi_{2}}^2\leq K
\end{aligned}
\end{equation}
for some constant $K > 0$, where the first inequality follows from \citet{vershynin2010introduction} (Remark 5.18), the second inequality follows from Lemma C.1 in \citet{javanmard2013nearly}, and the last inequality follows from Assumption \ref{assump:linear-high-dimension} (2). Without loss of generality, we assume $K \geq 1$.
Using a Bernstein-type inequality (Proposition 5.26 in \citet{vershynin2010introduction}), we have
\begin{equation}
\mathbbm{P}\left(\left|\frac{1}{n}\sum_{k= 1}^{n}\xi_{k}\right| \geq t\right)\leq 2\exp\left(-cn\min\left(\frac{t^{2}}{K^{2}}, \frac{t}{K}\right)\right).
\end{equation}
Plugging in $t = 1/{\sqrt{\log p}}$ and employing the union bound, we obtain
\begin{equation}
\label{eq:precision-matrix-main-term}
\mathbbm{P}\left(\max_{i,j\in[p]}\left| v^{\top}\hat\Sigma u -  \Theta_{ij} \right| \geq \frac{1}{\sqrt{\log p}}\right)\leq 2p^{2}\exp\left(-c_{2}\frac{n}{\log p}\right)
\end{equation}
for some constant $c_2 > 0$. 

Next we consider the term $\delta^{\top}\hat\Sigma\eta$. Since $\hat\Sigma\succeq 0$, 
it is sufficient for us to bound both $\delta^{\top}\hat\Sigma\delta$ and $\eta^{\top}\hat\Sigma\eta$ by the Cauchy-Schwartz inequality. Notice that 
\begin{equation}
\label{eq:precision-matrix-residual-term-bound}
\delta^{\top}\Sigma\delta = \sum_{i, j\in[p]} \hat\Sigma_{ij}\delta_{i}\delta_{j} \leq |\hat\Sigma|_{\infty}||\delta||_{1}^{2}\leq |\hat\Sigma|_{\infty}||\Theta-\hat\Theta||_{\infty}^{2} \leq  \frac{|\hat\Sigma|_{\infty}}{\log p}.
\end{equation}
We proceed to bound the tail probability $\mathbbm{P}(|\hat\Sigma|_{\infty}\geq 2)$. First, using the union bound, we have
\begin{equation}
\mathbbm{P}(|\hat\Sigma|_{\infty}\geq 2)\leq \sum_{i,j\in[p]} \mathbbm{P}(|\hat\Sigma_{ij}|\geq 2)\leq\sum_{i, j\in[p]} \mathbbm{P}(|\hat\Sigma_{ij}-\mathbbm{E}\hat\Sigma_{ij}|\geq 1).
\end{equation}
Denote $\xi_{k} = e_{i}^{\top}x_{k}x_{k}^{\top}e_{j} - \Sigma_{ij}$ for $k\in[n]$. We have
\begin{equation}
\begin{aligned}
||\xi_{k}||_{\psi_{1}}&\leq 2||(e_{i}^{\top}x_{k})^2||_{\psi_{1}} \leq 4||e_{i}^{\top}x_{k}||^2_{\psi_{2}}\leq 4 ||e_{i}^{\top}\Theta^{-1/2}\Theta^{1/2}x_{k}||^2_{\psi_{2}}\leq K 
\end{aligned}
\end{equation}
for some constant $K > 0$. Thus $\xi_{k}$‘s are independent sub-exponential random variables. By the Bernstein inequality, we have
\begin{equation}
\mathbbm{P}(|\hat\Sigma|_{\infty}\geq 2)\leq 2p^{2}\exp(-c_{4}n)
\end{equation}
for some constant $c_4 > 0$. We remark that although $\delta = (\Theta-\hat\Theta)^{\top}e_{i}$ is implicitly associated with the index $i$, the upper bound $|\hat\Sigma|_{\infty}||\Theta-\hat\Theta||_{\infty}^{2}$ in Equation \eqref{eq:precision-matrix-residual-term-bound} is irrelevant to the index $i$. Therefore, we have shown that
\begin{equation}
\label{eq:precision-matrix-residual-term}
\mathbbm{P}\left(\max_{i,j\in[p]}{|\delta^{\top}\hat\Sigma\eta|} \geq{\frac2{\log p}}\right)\leq 4p^{2}\exp(-c_{4}n) + 2\epsilon,
\end{equation}
where the maximum is taken with respect to the implicit index $i, j$ associated with $\delta$ and $\eta$.

We now consider the term $v^{\top}\hat\Sigma\eta$. We have
\begin{equation}
\begin{aligned}
\max_{i,j\in[p]}|v^{\top}\hat\Sigma\eta| & \leq \max_{i,j\in[p]}[v^{\top}\hat\Sigma v]^{1/2}[\eta^{\top}\hat\Sigma\eta]^{1/2}\\
& \leq\max_{i,j\in[p]}[|v^{\top}\hat\Sigma v-\Theta_{ii}|+|\Theta_{ii}|]^{1/2}[\eta^{\top}\hat\Sigma\eta]^{1/2} \\
&\leq [\max_{i\in[p]}|v^{\top}\hat\Sigma v-\Theta_{ii}|+\max_{i\in[p]}|\Theta_{ii}|]^{1/2}\max_{j\in[p]}[\eta^{\top}\hat\Sigma\eta]^{1/2}.
\end{aligned}
\end{equation}
Under Assumption \ref{assump:linear-high-dimension} (2), $\max_{i\in[p]}|\Theta_{ii}|$ is upper bounded. Combining the inequalities in Equation \eqref{eq:precision-matrix-main-term} and Equation \eqref{eq:precision-matrix-residual-term}, we have
\begin{equation}
\label{eq:precision-matrix-cross-term}
\mathbbm{P}\left(\max_{i,j\in[p]}|v^{\top}\hat\Sigma\eta| \geq \frac{c}{\sqrt{\log p}}\right) \leq  2p^{2}\exp\left(-c_{2}\frac{n}{\log p}\right)
 + 2p^{2}\exp(-c_{4}n) + \epsilon.
\end{equation}
The decomposition in Equation \eqref{eq:precision-matrix-decomposition}, along with the inequalities in Equation \eqref{eq:precision-matrix-main-term}, Equation \eqref{eq:precision-matrix-residual-term} and Equation \eqref{eq:precision-matrix-cross-term} together imply the claim in Lemma \ref{lemma:unnormalized-precision-matrix}.

\begin{corollary}
\label{cor:normalized-precision-matrix}
The high probability bound in Lemma \ref{lemma:unnormalized-precision-matrix} also applies to the normalized versions $\Lambda^0$ and $\Theta^0$. That is, under Assumption \ref{assump:linear-high-dimension}, for any $\epsilon > 0$, there exist constants $c_{1}, c_{2}, c_{3}, c_{4} > 0$ such that for large enough $n$,
\begin{equation}
\mathbbm P\left(\max_{i,j\in[p]}\left|\Lambda^0_{ij}-\Theta^0_{ij}\right|\leq \frac{1}{\sqrt{\log p}}\right) \geq 1-c_{1}p^{2}\exp\left(-c_{2}\frac{n}{\log p}\right)-c_{3}p^{2}\exp\left(-c_{4}n\right)-\epsilon.
\end{equation}
\end{corollary}

\textit{Proof of Corollary} \ref{cor:normalized-precision-matrix}. 
We show that $\max_{i,j\in[p]}\left|\Lambda^0_{ij}-\Theta^0_{ij}\right| \leq c/\sqrt{\log p}$ for some constant $c > 0$, conditioning on the high probability event $\left\{\max_{i,j\in[p]}\left|\Lambda_{ij}-\Theta_{ij}\right|\leq {1}/{\sqrt{\log p}}\right\}$. We have
\begin{equation}
\begin{aligned}
\max_{i,j\in[p]}\left|\Lambda^0_{ij}-\Theta^0_{ij}\right| & = \max_{i,j\in[p]}\left|\frac{\Lambda_{ij}}{\sqrt{\Lambda_{ii}\Lambda_{jj}}} - \frac{\Theta_{ij}}{\sqrt{\Theta_{ii}\Theta_{jj}}}\right|\\
&  \leq \max_{i,j\in[p]}\left|\frac{\Lambda_{ij}}{\sqrt{\Lambda_{ii}\Lambda_{jj}}} - \frac{\Lambda_{ij}}{\sqrt{\Theta_{ii}\Theta_{jj}}}\right| + \max_{i,j\in[p]}\left|\frac{\Lambda_{ij}}{\sqrt{\Theta_{ii}\Theta_{jj}}} - \frac{\Theta_{ij}}{\sqrt{\Theta_{ii}\Theta_{jj}}}\right|.
\end{aligned}
\end{equation}

Consider the first term. By Assumption \ref{assump:linear-high-dimension} (2), for large enough $p$, we have 
\begin{equation}
\frac{|\Lambda_{ij}|}{\sqrt{\Lambda_{ii}\Lambda_{jj}}} \leq \frac{|\Lambda_{ij} - \Theta_{ij}| +  \Theta_{ij}}{\sqrt{[ \Theta_{ii} - |\Lambda_{ii} - \Theta_{ii}|][\Theta_{jj} - |\Lambda_{jj} - \Theta_{jj}|]}} \leq K
\end{equation}
for some constant $K > 0$. It follows that
\begin{equation}
\label{eq:precision-matrix-corollary-1}
\begin{aligned}
\left|\frac{\Lambda_{ij}}{\sqrt{\Lambda_{ii}\Lambda_{jj}}} - \frac{\Lambda_{ij}}{\sqrt{\Theta_{ii}\Theta_{jj}}}\right| & = \frac{|\Lambda_{ij}|}{\sqrt{\Lambda_{ii}\Lambda_{jj}}} \left|\left({\frac{\Lambda_{ii}\Lambda_{jj}}{\Theta_{ii}\Theta_{jj}}}\right)^{1/2} - 1\right| \leq K  \left|\left({\frac{\Lambda_{ii}\Lambda_{jj}}{\Theta_{ii}\Theta_{jj}}}\right)^{1/2} - 1\right| \\
& \leq K  \left|\left(1 + \frac{1}{\sqrt{\log p}\Theta_{ii}}\right)^{1/2}\left(1 + \frac{1}{\sqrt{\log p}\Theta_{jj}}\right)^{1/2} - 1\right|\\
& \leq K\left|1 + \frac{2C}{\sqrt{\log p}} + \frac{C^2}{\log p} - 1\right| \leq \frac{K^\prime}{\sqrt{\log p}}
\end{aligned}
\end{equation}
for some constant $K^\prime > 0$. In the last to second inequality above, we use an elementary inequality $\sqrt{1 + x} \leq 1 + x$ for $x > 0$.

For the second term, by Assumption \ref{assump:linear-high-dimension} (2),  we have
\begin{equation}
\label{eq:precision-matrix-corollary-2}
\left|\frac{\Lambda_{ij}}{\sqrt{\Theta_{ii}\Theta_{jj}}} - \frac{\Theta_{ij}}{\sqrt{\Theta_{ii}\Theta_{jj}}}\right| \leq C{|\Lambda_{ij} - \Theta_{ij}|} \leq \frac{C}{\sqrt{\log p}}.
\end{equation}
We note that both the upper bounds in Equation \eqref{eq:precision-matrix-corollary-1} and Equation \eqref{eq:precision-matrix-corollary-2} do not depend on the indexes $i, j$. Therefore, we have shown that $\max_{i,j\in[p]}\left|\Lambda^0_{ij}-\Theta^0_{ij}\right|\leq (K^\prime + C)/\sqrt{\log p}$.
This implies the claim in Corollary \ref{cor:normalized-precision-matrix}.

\begin{lemma}
\label{lemma:bias-bound}
Under Assumption \ref{assump:linear-high-dimension}, as $n,p\to\infty$, we have 
\begin{equation}
\begin{aligned}
& \sup_{t\in\mathbbm{R},\ j\in S_0}\left|\mathbbm{P}(T_j > t) - Q(t)\right| \longrightarrow 0.
\end{aligned}
\end{equation}
\end{lemma}

\textit{Proof of Lemma} \ref{lemma:bias-bound}. Recall that 
\begin{equation}
T_j = {\sqrt{n}\widehat{\beta}_j^d}/{{\Lambda^{1/2}_{jj}}}\ \ \ \text{and}\ \ \ \widetilde{Z}_j := {Z_j}/{{\Lambda^{1/2}_{jj}}}\Big|X \sim N(0, 1).
\end{equation}
Without changing the selection result obtained via Algorithm \ref{alg:FDR-control-framework}, we multiply the normalized debiased Lasso estimator $T_j$ defined in Equation \eqref{eq:linear-normalized-debiased-Lasso-estimator} by a constant factor $\sqrt{n}$ so that it follows the standard normal distribution asymptotically. 

By Lemma \ref{lemma:unnormalized-precision-matrix}, for large enough $n$, we have $\min_{j\in[p]}\Lambda_{jj} \geq c$ for some constant $c > 0$ with high probability. Henceforth we condition on this high probability event. Recall the decomposition in Equation \eqref{eq:linear-decomposition}, thus for $j\in S_0$, i.e., $\beta^\star_j = 0$, we have $T_j = \widetilde{Z}_j + \Delta_j/{\Lambda^{1/2}_{jj}}$.
Let $\epsilon = p_1\log p/\sqrt{n} \rightarrow 0$ by Assumption \ref{assump:linear-high-dimension} (1). For the bias term $\Delta$, we have
$$\mathbbm{P}\left(||\Delta||_\infty \geq \epsilon\sqrt{c}\right) \leq \epsilon$$
by Theorem 2.3 in \citet{javanmard2013nearly}.
It follows that for any $t \in\mathbbm{R}$, 
\begin{equation}
\label{eq:bias-upper-bound}
\begin{aligned}
\mathbbm{P}\left(T_j > t\right) - Q(t) & = \mathbbm{P}\big(\widetilde{Z}_j  > t - \Delta_j/{\Lambda^{1/2}_{jj}}\big)  - Q(t) \\
& \leq  \mathbbm{P}(\widetilde{Z}_j  > t - \epsilon) +  \mathbbm{P}\left(||\Delta||_\infty \geq \epsilon\sqrt{c}\right)  - Q(t) \\
& \leq Q(t - \epsilon)  - Q(t) + \epsilon\\
&  \leq c_1\epsilon
\end{aligned}
\end{equation}
for some constant $c_1 > 0$, in which the last inequality follows from the simple fact that for any $t \in \mathbbm{R},\ \epsilon > 0$, we have
\begin{equation}
\label{eq:Q-bound}
Q(t - \epsilon) - Q(t) = \int_{t - \epsilon}^t \frac{1}{\sqrt{2\pi}}\exp\left(-\frac{x^2}{2}\right)dx \leq \frac{\epsilon}{\sqrt{2\pi}}.
\end{equation}

Similarly, we have
\begin{equation}
\label{eq:bias-lower-bound}
\begin{aligned}
\mathbbm{P}\left(T_j > t\right) - Q(t) & \geq \mathbbm{P}(\widetilde{Z}_j  > t + \epsilon)\mathbbm{P}(||\Delta||_\infty < \epsilon\sqrt{c}) - Q(t)\\&\geq Q(t + \epsilon)(1 - \epsilon) - Q(t)\\
&  \geq -c_2\epsilon.
\end{aligned}
\end{equation}
for some constant $c_2 > 0$.
Since the bounds in Equation \eqref{eq:bias-upper-bound} and Equation \eqref{eq:bias-lower-bound} are irrelevant to $t$ and the index $j$, the claim in Lemma \ref{lemma:bias-bound} follows.

\begin{lemma}
\label{lemma:mirror-statistic-variance-bound}
Under Assumption \ref{assump:linear-high-dimension}, as $n,p\to\infty$, we have
\begin{equation}
\begin{aligned}
&\sup_{t\in\mathbbm{R}}\textnormal{Var}\bigg(\frac{1}{p_0}\sum_{j\in S_0}\mathbbm{1}(M_j > t)\bigg) \longrightarrow 0. 
\end{aligned}
\end{equation}
\end{lemma}

\textit{Proof of Lemma} \ref{lemma:mirror-statistic-variance-bound}. 
Denote the correlated set as $\Gamma = \{(i,j): i,j\in S_0,\ \Theta_{ij} \neq 0\}$, and the uncorrelated set as $\Gamma^c = \{(i,j): i,j\in S_0,\ \Theta_{ij} = 0\}$. We have the following decomposition.
\begin{equation}
\begin{aligned}
\sup_{t\in\mathbbm{R}}\textnormal{Var}\bigg(\frac{1}{p_0}\sum_{j\in S_0}\mathbbm{1}(M_j > t)\bigg)
&\leq \frac{1}{p_0^2}\sup_{t\in\mathbbm{R}}\sum_{(i,j)\in\Gamma} \text{Cov}(\mathbbm{1}(M_i > t), \mathbbm{1}(M_j > t)) \\
& \hspace{0.07cm} +  \frac{1}{p_0^2}\sup_{t\in\mathbbm{R}}\sum_{(i,j)\in\Gamma^c}\text{Cov}(\mathbbm{1}(M_i > t), \mathbbm{1}(M_j > t))\\
&\hspace{-0.2cm} \leq \frac{|\Gamma|}{p_0^2} + 
\sup_{t\in\mathbbm{R}, (i,j) \in \Gamma^c}\left|\mathbbm{P}(M_i > t, M_j > t) - H^2(t)\right| \\
&\hspace{0.85cm} + 
\sup_{t\in\mathbbm{R}, (i,j) \in \Gamma^c}\left|\mathbbm{P}(M_i > t)\mathbbm{P}(M_j > t) - H^2(t)\right|.
\end{aligned}
\end{equation}
By Assumption \ref{assump:linear-high-dimension} (1), we have $|\Gamma|/p_0^2 \leq p_0 \sqrt{n} /(p_0^2 \log p) \to 0$. Further, by Lemma \ref{lemma:mirror-statistic-bias-bound}, we have 
\begin{equation}
\sup_{t\in\mathbbm{R},i,j\in S_0}\left|\mathbbm{P}(M_i > t)\mathbbm{P}(M_j > t) - H^2(t)\right| \to 0.
\end{equation} 
Therefore, it is sufficient for us to show that
\begin{equation}
\sup_{t\in\mathbbm{R}, (i,j) \in \Gamma^c}\left|\mathbbm{P}(M_i > t, M_j > t) - H^2(t)\right| \rightarrow 0,\ \ \ \text{as}\ n,p\to\infty.
\end{equation}

In the following, we condition on the high probability event $E_1\cap E_2 \cap E_3$, in which
\begin{equation}
\label{eq:linear-high-probability-event}
\begin{aligned}
E_1 & = \Big\{\min_{j\in[p]}\Lambda^{(1)}_{jj} \geq c\Big\} \cap  \Big\{\min_{j\in[p]}\Lambda^{(2)}_{jj} \geq c\Big\},\\
E_2 & = \Big\{\max_{i,j\in[p]}|\Lambda_{ij}^{0, (1)} - \Theta^0_{ij}| \leq \frac{1}{\sqrt{\log p}}\Big\} \cap \Big\{\max_{i,j\in[p]}|\Lambda_{ij}^{0, (2)} - \Theta^0_{ij}| \leq \frac{1}{\sqrt{\log p}}\Big\},\\
E_3 & = \Big\{||\Delta^{(1)}||_\infty \leq \epsilon \sqrt{c}\Big\} \cap \Big\{||\Delta^{(2)}||_\infty \leq \epsilon \sqrt{c}\Big\}.
\end{aligned}
\end{equation}
The superscripts $(1), (2)$ reflects the data splitting step, and the corresponding random variables are defined on each part of the data.
$c > 0$ is some suitable constant (see Lemma \ref{lemma:unnormalized-precision-matrix} and Assumption \ref{assump:linear-high-dimension} (2)), and $\epsilon = p_1\log p/\sqrt{n} \rightarrow 0$ by Assumption \ref{assump:linear-high-dimension} (1). 

We consider the same decomposition as in Equation \eqref{eq:mirror-statistic-variance-bound-eq1}, and proceed to upper bound $I_1$. Let $\rho = 1/\sqrt{\log p}$. Notice that for $(i,j) \in \Gamma^c$, $\Theta^0_{ij} = 0$, thus we have 
$$\max_{(i,j) \in \Gamma^c}\big|\Lambda^{0, (1)}_{i,j}\big| \leq \rho\ \ \ \text{and}\ \ \ \max_{(i,j) \in \Gamma^c}\big|\Lambda^{0, (2)}_{i,j}\big| \leq \rho$$
once we condition on the high probability event $E_2$.
Let $\widetilde{Z}^{(1)}$ and $\widetilde{Z}^{(2)}$ be the normalized version (e.g. variance 1) of $Z^{(1)}$ and $Z^{(2)}$ (see Equation \eqref{eq:linear-decomposition}), respectively, thus $\Lambda^{0, (1)}$ and $\Lambda^{0, (2)}$ are essentially the corresponding covariance matrices.
We have 
\begin{equation}
\label{eq:mirror-statistic-variance-bound-eq2}
\begin{aligned}
I_1 & \leq \mathbbm{P}\left(\widetilde{Z}_i^{(2)} > I_t\big(T_i^{(1)}\big) - \epsilon, \widetilde{Z}_j^{(2)} > I_t\big(T_j^{(1)}\big) - \epsilon, T_i^{(1)} > 0, T_j^{(1)} > 0\right)\\
& = \mathbbm{E}\left[\mathbbm{P}\big(\widetilde{Z}_i^{(2)} > I_t(x) - \epsilon, \widetilde{Z}_j^{(2)} > I_t(y) - \epsilon\big)\mid T_i^{(1)} = x > 0,\ T_j^{(1)} = y > 0\right]\\
& \leq c_1\rho + \mathbbm{E}\left[ Q\big(I_t(x) - \epsilon\big)Q\big(I_t(y) - \epsilon\big)\mid T_i^{(1)} = x > 0,\ T_j^{(1)} = y > 0\right]\\
& \leq c_1\rho + c_2\epsilon + \mathbbm{E}\left[ Q\big(I_t(x)\big)Q\big(I_t(y)\big)\mid T_i^{(1)} = x > 0,\ T_j^{(1)} = y > 0\right]
\end{aligned}
\end{equation}
for some constants $c_1, c_2 > 0$, in which the third line is based on the Mehler's identity (see Equation \eqref{eq:mehler}) and Lemma 1 in \citet{azriel2015empirical}, and the last line is based on the inequality in Equation \eqref{eq:Q-bound}.
Similarly, we can upper bound $I_2$, $I_3$ and $I_4$. Combining the four upper bounds together, we obtain an upper bound on $\mathbbm{P}(M_i > t, M_j > t)$ specified as below,
\begin{equation}
\label{eq:mirror-statistic-variance-bound-eq3}
\mathbbm{P}\left(\text{sign}(W_i^{(2)}T_i^{(1)})f(W_i^{(2)}, T_i^{(1)}) > t, \text{sign}(W_j^{(2)}T_j^{(1)})f(W_j^{(2)}, T_j^{(1)}) > t\right) + 4c_1\rho + 4c_2\epsilon, 
\end{equation}
in which $W_i^{(2)}$ and $W_j^{(2)}$ are two independent random variables (also independent to everything else) following the standard normal distribution.
We can further decompose the first term in Equation \eqref{eq:mirror-statistic-variance-bound-eq3} into four terms as Equation \eqref{eq:mirror-statistic-variance-bound-eq1} by conditioning on the signs of $W_i^{(2)}$ and $W_j^{(2)}$, and repeat the upper bound in Equation \eqref{eq:mirror-statistic-variance-bound-eq2}. This leads to
\begin{equation}
\mathbbm{P}(M_i > t, M_j > t) \leq H^2(t) + 8c_1\rho + 8c_2\epsilon.
\end{equation}
Similarly, we can establish the corresponding lower bound.
Thus we complete the proof of Lemma \ref{lemma:mirror-statistic-variance-bound}.

\subsubsection{Proof of Proposition \ref{prop:linear-data-splitting-FDR-high-dimension}}
In addition to the notations introduced in Section \ref{subsubsec:proof-glm-moderate-FDR}, we denote
\begin{equation}
\begin{aligned}
&G^0_{p}(t) = \frac{1}{p_0}\sum_{j \in S_0}\mathbbm{P}(M_j > t),\ \ \ V^0_{p}(t) = \frac{1}{p_0}\sum_{j \in S_0}\mathbbm{P}(M_j < -t).
\end{aligned}
\end{equation}
The proof of Proposition \ref{prop:linear-data-splitting-FDR-high-dimension} is essentially the same as the proof of Proposition \ref{prop:GLM-data-splitting-FDR-moderate-dimension}, with the help of the following Lemma \ref{lemma:FDR}.

\begin{lemma}
\label{lemma:FDR}
Under Assumption \ref{assump:linear-high-dimension}, as $n, p\to\infty$, we have in probability,
\begin{equation}
\begin{aligned}
&\sup_{t\in\mathbbm{R}} \left|\widehat{G}^0_{p}(t) - H(t)\right| \longrightarrow 0,\\
&\sup_{t\in\mathbbm{R}} \left|\widehat{V}^0_{p}(t) - H(t)\right| \longrightarrow 0.
\end{aligned}
\end{equation}
\end{lemma}
\textit{Proof of Lemma \ref{lemma:FDR}}. We prove the first claim. The second claim follows similarly. Notice that $H(t)$ is symmetric about 0. By Lemma \ref{lemma:mirror-statistic-bias-bound}, it is sufficient for us to show that 
\begin{equation}
\begin{aligned}
&\sup_{t\in\mathbbm{R}} \left|\widehat{G}^0_{p}(t) - G^0_{p}(t)\right| \longrightarrow 0,\\
&\sup_{t\in\mathbbm{R}} \left|\widehat{V}^0_{p}(t) - V^0_{p}(t)\right| \longrightarrow 0.
\end{aligned}
\end{equation}
The proof follows similarly as the proof of Lemma \ref{lemma:glm-moderate-supreme-convergence} using an $\epsilon$-net argument, except that we use the Chebyshev's inequality and Lemma \ref{lemma:mirror-statistic-variance-bound} in Equation \eqref{eq:supreme-bound}.

\subsection{Proof of Proposition \ref{prop:GLM-bias-high-dimension} and \ref{prop:GLM-data-splitting-FDR-high-dimension}}
Without loss of generality, we assume that the Lipschitz constant of $\ddot{\rho}(v)$ is 1 (see Assumption \ref{assump:GLM-high-dimension} (3)).
For $j\in[p]$, denote $\eta_j = X_{\beta^\star, j} - X_{\beta^\star, -j}\gamma_j$, and denote $\tau_j^2 = \mathbbm{E}[\eta_j^\top\eta_j/n]$ as the conditional variance $\text{Var}(X_{\beta^\star,j}|X_{\beta^\star, -j})$. Denote $\widehat{\Sigma} = P_n\dot{\ell}_{\widehat\beta}\dot{\ell}^\top_{\widehat\beta}$ as the sample version of $\Sigma$.

\subsubsection{Technical lemmas}
\begin{lemma}
\label{lemma:GLM-lasso-bound}
Under Assumption \ref{assump:GLM-high-dimension}, for any $\epsilon > 0$, there exists a constant $C_\epsilon > 0$, such that $\mathbbm{P}(\cap_{i = 1}^4 E^\epsilon_i) \geq 1 - \epsilon$, in which the events $E^\epsilon_1, E^\epsilon_2, E^\epsilon_3, E^\epsilon_4$ are defined as below,
\begin{equation}
\begin{aligned}
& E^\epsilon_1 = \left\{\frac{1}{n}||X(\widehat{\beta} - \beta^\star)||_2^2 \leq C_\epsilon p_1\log p/n\right\},\\
& E^\epsilon_2 = \left\{||\widehat{\beta} - \beta^\star||_1 \leq C_\epsilon p_1\sqrt{\log p/n}\right\},\\
& E^\epsilon_3 = \left\{\max_{j\in[p]}||\widehat{\gamma}_j - \gamma_j||_1 \leq C_\epsilon s \sqrt{\log p / n}\right\},\\
& E^\epsilon_4 = \left\{\max_{j\in[p]}||\widehat{\gamma}_j - \gamma_j||_2 \leq C_\epsilon  \sqrt{s\log p / n}\right\}.
\end{aligned}
\end{equation}
\end{lemma}

\noindent\textit{Proof of Lemma \ref{lemma:GLM-lasso-bound}}. Lemma \ref{lemma:GLM-lasso-bound} follows from standard arguments in \citet{bickel2009simultaneous}, \citet{raskutti2010restricted}, \citet{buhlmann2011statistics}, \citet{van2014asymptotically}.

\begin{lemma}
\label{lemma:GLM-tau-bound}
Under Assumption \ref{assump:GLM-high-dimension}, we have
\begin{equation}
\begin{aligned}
& \max_{j\in[p]}\left|\widehat{\tau}_j^2 - \tau_j^2\right| = O_p\big(\sqrt{s\log p/n}\big),\\
& \max_{j\in[p]}\left|{1}/{\widehat{\tau}_j^2} - {1}/{\tau_j^2}\right| = O_p\big(\sqrt{s\log p/n}\big).
\end{aligned}
\end{equation}
\end{lemma}

\noindent\textit{Proof of Lemma \ref{lemma:GLM-tau-bound}}. The proof follows similarly as the proof of Theorem 3.2 in \citet{van2014asymptotically}, combined with some enriched arguments that we detail below. For any $\epsilon > 0$, we condition on the high probability event $\cap_{i = 1}^4 E^\epsilon_i$. 
Since $X_{\widehat{\beta}} = W_{\widehat{\beta}}W_{\beta^\star}^{-1}X_{\beta^\star}$, we have the following decomposition, 
\begin{equation}
\widehat{\tau}^2_j - \tau_j^2  = \frac{1}{n}X_{\widehat{\beta}, j}^\top\left(X_{\widehat{\beta}, j} - X_{\widehat{\beta}, -j}\widehat{\gamma}_j\right) - \tau_j^2 :=I_1 + I_2,
\end{equation}
in which 
\begin{equation}
\begin{aligned}
I_1 & = \frac{1}{n}X_{\beta^\star, j}^\top\left(X_{\beta^\star, j} - X_{\beta^\star, -j}\widehat{\gamma}_j\right) - \tau_j^2,\\
I_2 & = \frac{1}{n}X_{\beta^\star, j}^\top\left(W^2_{\widehat{\beta}}W_{\beta^\star}^{-2} - I\right)\left(X_{\beta^\star, j} - X_{\beta^\star, -j}\widehat{\gamma}_j\right).
\end{aligned}
\end{equation}
We proceed to upper bound $I_1$ and $I_2$. 

We further decompose $I_1$ into four terms, $I_1 = I _{11} +  I _{12} +  I _{13} +  I _{14}$, 
in which
\begin{equation}
\begin{aligned}
 I _{11} & = \frac{1}{n}\eta_i^\top\eta_j - \tau_j^2,\\
   I _{12} & = \frac{1}{n}\eta_j^\top X_{\beta^\star, -j}(\gamma_j - \widehat{\gamma}_j),\\
  I _{13} & = \frac{1}{n}\gamma_j^\top X_{\widehat{\beta}, -j}^\top\left(X_{\widehat{\beta}, j} - X_{\widehat{\beta}, -j}\widehat{\gamma}_j\right),\\
 I _{14} & = \frac{1}{n}\gamma_j^\top X_{\beta^\star, -j}^\top\left(I - W^2_{\widehat{\beta}}W_{\beta^\star}^{-2}\right)\left(X_{\beta^\star, j} - X_{\beta^\star, -j}\widehat{\gamma}_j\right).
\end{aligned}
\end{equation}
For $ I _{11}$, by Assumption \ref{assump:GLM-high-dimension} (2), we have 
\begin{equation}
\label{eq:eta-infinity}
||\eta_j||_\infty \leq ||X_{\beta^\star, j}||_\infty + ||X_{\beta^\star, -j}\gamma_j||_\infty \leq 2C_1.
\end{equation}
Since $\mathbbm{E}[\eta_j^\top\eta_j/n] = \tau_j^2$, we have 
\begin{equation}
\max_{j\in[p]} I _{11} = O_p(\sqrt{\log p/n}),
\end{equation}
based on the union bound and the Hoeffding's inequality. 

For $I _{12}$, since $\mathbbm{E}[\eta_j^\top X_{\beta^\star, -j}] = 0$, we have
\begin{equation}
\max_{j\in[p]}||\eta_j^\top X_{\beta^\star, -j}/n||_\infty = O_p(\sqrt{\log p/n}),    
\end{equation}
by the union bound and the Hoeffding's inequality. It follows that
\begin{equation}
\begin{aligned}
\max_{j\in[p]} I _{12} & \leq \max_{j\in[p]}||\eta_j^\top X_{\widehat{\beta}, -j}/n||_\infty\max_{j\in[p]}||\widehat{\gamma}_j - \gamma_j||_1  \\
& = O_p(s \log p/n).
\end{aligned}
\end{equation}

For $I_{13}$, since $\gamma_j$ is $s-$sparse, by Assumption \ref{assump:GLM-high-dimension} (2), we have
\begin{equation}
\label{eq:gamma-l1}
\begin{aligned}
\max_{j\in[p]}||\gamma_j||_1  & \leq  \sqrt{s}\max_{j\in[p]}||\gamma_j||_2\\
& = \sqrt{s}\max_{j\in[p]}[\Sigma_{j, -j}\Sigma_{-j,-j}^{-2}\Sigma_{-j, j}]^{1/2}\\ 
& \leq \sqrt{s}C_1^2/C_2.
\end{aligned}
\end{equation} 
In addition, by the KKT condition of the $j$-th nodewise Lasso regression, we have
\begin{equation}
\max_{j\in[p]} ||X_{\widehat{\beta}, -j}^\top\left(X_{\widehat{\beta}, j} - X_{\widehat{\beta}, -j}\widehat{\gamma}_j\right)/n||_\infty \leq \lambda_j.
\end{equation}
 It follows that
\begin{equation}
\begin{aligned}
\max_{j\in[p]} I _{13} & \leq \max_{j\in[p]}||\gamma_j||_1 \max_{j\in[p]} ||X_{\widehat{\beta}, -j}^\top\left(X_{\widehat{\beta}, j} - X_{\widehat{\beta}, -j}\widehat{\gamma}_j\right)/n||_\infty \\
& = O_p(\sqrt{s\log p/n}).
\end{aligned}
\end{equation}

For $I_{14}$, we first notice that 
\begin{equation}
\begin{aligned}
||X_{\beta^\star, j} - X_{\beta^\star, -j}\widehat{\gamma}_j||_\infty & \leq ||\eta_j||_\infty + ||X_{\beta^\star, -j}(\gamma_j - \widehat{\gamma}_j)||_\infty\\
& \leq 2C_1 + C_1||\gamma_j - \widehat{\gamma}_j||_1 \\
& \leq 3C_1 
\end{aligned}
\end{equation}
by Assumption \ref{assump:GLM-high-dimension} (2) and Equation \eqref{eq:eta-infinity}. By Assumption \ref{assump:GLM-high-dimension} (3), it follows that 
\begin{equation}
\begin{aligned}
\max_{j\in[p]} I _{14} & \leq \frac{3C_1^2}{n}\sum_{i = 1}^n\left|\frac{\ddot{\rho}(x_i^\intercal \widehat{\beta}) - \ddot{\rho}( x_i^\intercal \beta^\star)}{\ddot{\rho}(x_i^\intercal\beta^\star)}\right| \\
& = O_p(||X(\widehat{\beta} - \beta^\star)||_2/\sqrt{n})\\
& = O_p(\sqrt{p_1\log p/n}),
\end{aligned}
\end{equation}
in which the second inequality follows from Cauchy-Schwarz inequality. Combining the upper bound on $\max_{j\in[p]} I _{11}$, $\max_{j\in[p]} I _{12}$, $\max_{j\in[p]} I _{13}$ and $\max_{j\in[p]} I _{14}$, we show that 
\begin{equation}
\max_{j\in[p]} I _1 = O_p(\sqrt{s\log p/n}).
\end{equation}

Finally, $\max_{j\in[p]}I_2$ can be upper bounded similarly as $\max_{j\in[p]} I _{14}$. This completes the proof of the first claim in Lemma \ref{lemma:GLM-tau-bound}. The second claim follows by noticing that $1/\tau_j^2$ is uniformly upper bounded because
\begin{equation}
\label{eq:inverse-tau-square-lower-bound}
1/\tau_j^2 = \Theta_{j,j} \leq \sigma_{\max}(\Theta) = 1/\sigma_{\min}(\Sigma) \leq C_2.
\end{equation}
This completes the proof of Lemma \ref{lemma:GLM-tau-bound}

\begin{lemma}
\label{lemma:GLM-Theta-bound}
Under Assumption \ref{assump:GLM-high-dimension}, we have
\begin{equation}
\begin{aligned}
& \max_{j\in[p]}||\widehat{\Theta}_{j,\cdot} - \Theta_{j,\cdot}||_1  = O_p(s\sqrt{\log p/n}),\\
& \max_{j\in[p]}||\widehat{\Theta}_{j,\cdot} - \Theta_{j,\cdot}||_2  = O_p(\sqrt{s\log p/n}),\\
& \max_{j\in[p]}|\widehat{\Theta}_{j,\cdot} \Sigma \widehat{\Theta}_{j,\cdot}^\top - \Theta_{j,j}|  = O_p(\sqrt{s\log p/n}).
\end{aligned}
\end{equation}
\end{lemma}

\noindent\textit{Proof of Lemma \ref{lemma:GLM-Theta-bound}}. By Lemma \ref{lemma:GLM-lasso-bound} and \ref{lemma:GLM-tau-bound}, for any $\epsilon > 0$, there exists a constant $C_\epsilon > 0$, such that $\mathbbm{P}(\cap_{i = 1}^6 E^\epsilon_i) \geq 1 - \epsilon$, in which $E_i^\epsilon$ for $i\in[4]$ are specified in Lemma \ref{lemma:GLM-lasso-bound}, and
\begin{equation}
\begin{aligned}
& E_5^\epsilon = \left\{\max_{j\in[p]}\left|\widehat{\tau}_j^2 - \tau_j^2\right| \leq C_\epsilon \sqrt{s\log p/n}\right\},\\
& E_6^\epsilon = \left\{\max_{j\in[p]}\left|1/\widehat{\tau}_j^2 - 1/\tau_j^2\right| \leq C_\epsilon \sqrt{s\log p/n}\right\}.
\end{aligned}
\end{equation}
In the following, we condition on this high probability event $\cap_{i = 1}^6 E^\epsilon_i$. By Assumption \ref{assump:GLM-high-dimension} (1), we assume that $n$ is large enough so that $C_\epsilon \sqrt{s\log p/n} \leq C_2$.

We first note that by Equation \eqref{eq:inverse-tau-square-lower-bound},
\begin{equation}
1/\widehat{\tau}_j^2 \leq |1/\widehat{\tau}_j^2 - 1/\tau_j^2| + 1/\tau_j^2 \leq 2C_2.
\end{equation}
Thus, by Equation \eqref{eq:gamma-l1}, we have
\begin{equation}
\begin{aligned}
 \max_{j\in[p]}||\widehat{\Theta}_{j,\cdot} - \Theta_{j,\cdot}||_1 & = \max_{j\in[p]}||\widehat{C}_{j,\cdot}/\widehat{\tau}_j^2 - C_{j,\cdot}/\tau_j^2||_1\\
 & \leq \max_{j\in[p]}||\widehat{\gamma}_j - \gamma_j||_1\max_{j\in[p]}1/\widehat{\tau}_j^2 + \max_{j\in[p]}||\gamma_j||_1\max_{j\in[p]}\left|1/\widehat{\tau}_j^2 - 1/\tau_j^2\right|\\
 & \leq 2C_2C_\epsilon s \sqrt{\log p / n} + C_1^2/C_2C_\epsilon\sqrt{s}\sqrt{s\log p/n}\\
 & = O_p(s\sqrt{\log p /n}).
\end{aligned}
\end{equation}

Similarly, we have
\begin{equation}
\begin{aligned}
\max_{j\in[p]}||\widehat{\Theta}_{j,\cdot} - \Theta_{j,\cdot}||_2 & = \max_{j\in[p]}||\widehat{C}_{j,\cdot}/\widehat{\tau}_j^2 - C_{j,\cdot}/\tau_j^2||_2\\
& \leq \max_{j\in[p]}||\widehat{\gamma}_j - \gamma_j||_2\max_{j\in[p]}1/\widehat{\tau}_j^2 + \max_{j\in[p]}||\gamma_j||_2\max_{j\in[p]}\left|1/\widehat{\tau}_j^2 - 1/\tau_j^2\right|\\
& \leq 2C_2C_\epsilon \sqrt{s\log p / n} + C_1^2/C_2C_\epsilon\sqrt{s\log p/n}\\
 & = O_p(\sqrt{s\log p /n}).
\end{aligned}
\end{equation}

For the last claim in Lemma \ref{lemma:GLM-Theta-bound}, we employ the following decomposition. 
\begin{equation}
\begin{aligned}
\widehat{\Theta}_{j,\cdot} \Sigma \widehat{\Theta}_{j,\cdot}^\top - \Theta_{j,j} & = (\widehat{\Theta}_{j,\cdot} - \Theta_{j,\cdot})\Sigma(\widehat{\Theta}_{j,\cdot} - \Theta_{j,\cdot})^\top + 2\Theta_{j,\cdot}\Sigma(\widehat{\Theta}_{j,\cdot} - \Theta_{j,\cdot})^\top \\
& := I_1 + I_2.
\end{aligned}
\end{equation}
For $I_1$, by Assumption \ref{assump:GLM-high-dimension} (2), we have
\begin{equation}
\begin{aligned}
\max_{j\in[p]}I_1 & \leq \sigma_{\max}(\Sigma)\max_{j\in[p]}||\widehat{\Theta}_{j,\cdot} - \Theta_{j,\cdot}||^2_2 \\
& = O_p(s\log p /n).
\end{aligned}
\end{equation}
For $I_2$, we have
\begin{equation}
\begin{aligned}
\max_{j\in[p]}I_2 & = 2\max_{j\in[p]} e_j^\top (\widehat{\Theta}_{j,\cdot} - \Theta_{j,\cdot})^\top \\
& = 2\max_{j\in[p]}\left|1/\widehat{\tau}_j^2 - 1/\tau_j^2\right| \\
& =  O_p(\sqrt{s\log p /n}).
\end{aligned}
\end{equation}
This completes the proof of Lemma \ref{lemma:GLM-Theta-bound}.

\begin{lemma}
\label{lemma:GLM-sigma-bound}
Under Assumption \ref{assump:GLM-high-dimension}, we have
\begin{equation}
\begin{aligned}
& \max_{j\in[p]}|\widehat{\sigma}_j - \sigma_j| = O_p(s\sqrt{\log p /n}),\\
& \max_{j\in[p]}|1/\widehat{\sigma}_j - 1/\sigma_j| = O_p(s\sqrt{\log p /n}).
\end{aligned}
\end{equation}
\end{lemma}

\noindent\textit{Proof of Lemma \ref{lemma:GLM-sigma-bound}}. By Assumption \ref{assump:GLM-high-dimension} (2), $\sigma^2 = \Theta_{j,j} \geq \sigma_{\min}(\Theta) \geq 1/C_2 > 0$. Therefore, we only need to consider the first claim in the lemma. In addition, since
\begin{equation}
\max_{j\in[p]}|\widehat{\sigma}_j - \sigma_j| \leq \max_{j\in[p]}|\widehat{\sigma}^2_j - \sigma^2_j|/\sigma_j \leq \sqrt{C_2}\max_{j\in[p]}|\widehat{\sigma}^2_j - \sigma^2_j|,
\end{equation}
it is sufficient to show that 
\begin{equation}
\max_{j\in[p]}|\widehat{\sigma}^2_j - \sigma^2_j| = O_p(s\sqrt{\log p /n}).
\end{equation}
We note that
\begin{equation}
\begin{aligned}
\max_{j\in[p]}|\widehat{\sigma}^2_j - \sigma^2_j| & \leq \max_{j\in[p]}|\widehat{\Theta}_{j,\cdot} \Sigma \widehat{\Theta}_{j,\cdot}^\top - \Theta_{j,j}| + \max_{j\in[p]}|\widehat{\Theta}_{j,\cdot} \widehat{\Sigma} \widehat{\Theta}_{j,\cdot}^\top - \widehat{\Theta}_{j,\cdot} \Sigma \widehat{\Theta}_{j,\cdot}^\top |.
\end{aligned}
\end{equation}
By Lemma \ref{lemma:GLM-Theta-bound}, the first term is $O_p(\sqrt{s\log p /n})$. Using the same arguments as in the proof of Theorem 3.1 in \citet{van2014asymptotically} (page 1198), we can show that the second term is $O_p(s\sqrt{\log p /n})$. This completes the proof of Lemma \ref{lemma:GLM-sigma-bound}.

\subsubsection{Proof of Proposition \ref{prop:GLM-bias-high-dimension}}
We first note that for $j\in[p]$, the bias term $\Delta_j$ can be decomposed into the following three terms,
\begin{equation}
\begin{aligned}
R_{1, j} & = \frac{\sqrt{n}}{\sigma_j}
\Theta_{j,\cdot}P_n\dot{\ell}_{\beta^\star} -  \frac{\sqrt{n}}{\widehat{\sigma}_j}
\widehat{\Theta}_{j,\cdot}P_n\dot{\ell}_{\beta^\star},\\
R_{2, j} & = -\frac{\sqrt{n}}{\widehat{\sigma}_j}\left(\widehat{\Theta}_{j,\cdot}P_n\ddot{\ell}_{\widehat{\beta}} - e_j^\top\right)(\widehat{\beta} - \beta^\star),\\
R_{3, j} & = -\frac{\sqrt{n}}{\widehat{\sigma}_j}\widehat{\Theta}_{j,\cdot}P_n(\dot{\ell}_{\widehat{\beta}} - \dot{\ell}_{\beta^\star}) + \frac{1}{\sqrt{n}\widehat{\sigma}_j}\widehat{\Theta}_{j,\cdot}X^\top W^2_{\widehat{\beta}}X(\widehat{\beta} - \beta^\star).
\end{aligned}
\end{equation}
We proceed to bound each term. For any $\epsilon > 0$, there exists a constant $C_\epsilon > 0$ such that $\mathbbm{P}(\cap_{i = 1}^{10}E_i^\epsilon) \geq 1 - \epsilon$, in which $E_i^\epsilon$ for $i\in[4]$ are defined in Lemma \ref{lemma:GLM-lasso-bound}, $E_5^\epsilon$ and $E_6^\epsilon$ are defined in Lemma \ref{lemma:GLM-Theta-bound}, and $E_7^\epsilon, E_8^\epsilon, E_9^\epsilon, E_{10}^\epsilon$ are defined as follows,
\begin{equation}
\begin{aligned}
& E^\epsilon_7 \hspace{0.1cm} = \left\{\max_{j\in[p]}||\widehat{\Theta}_{j,\cdot} - \Theta_{j,\cdot}||_1 \leq C_\epsilon s \sqrt{\log p / n}\right\},\\ 
& E^\epsilon_8 \hspace{0.1cm} = \left\{\max_{j\in[p]}||\widehat{\Theta}_{j,\cdot} - \Theta_{j,\cdot}||_2 \leq C_\epsilon  \sqrt{s\log p / n}\right\},\\
& E^\epsilon_9 \hspace{0.1cm} = \left\{\max_{j\in[p]}|\widehat{\sigma}_j - \sigma_j|  \leq C_\epsilon s\sqrt{\log p/n}\right\},\\
& E^\epsilon_{10} = \left\{\max_{j\in[p]}|1/\widehat{\sigma}_j - 1/\sigma_j| \leq C_\epsilon s\sqrt{\log p/n}\right\}.
\end{aligned}
\end{equation}
In the following, we condition on this high probability event $\cap_{i = 1}^{10}E_i^\epsilon$. By Assumption \ref{assump:GLM-high-dimension} (1), we further assume that $n, p$ are large enough so that $C_\epsilon s\sqrt{\log p/n} \vee C_\epsilon \sqrt{s\log p/n} \leq 1$.

For $R_{1, j}$, we have
\begin{equation}
\begin{aligned}
\max_{j\in[p]}R_{1, j} & \leq \max_{j\in[p]}\left|\frac{\sqrt{n}}{\sigma_j}
\left(\Theta_{j,\cdot} - \widehat{\Theta}_{j,\cdot}\right)P_n\dot{\ell}_{\beta^\star} \right| + \max_{j\in[p]}\left|\sqrt{n}\widehat{\Theta}_{j,\cdot}P_n\dot{\ell}_{\beta^\star}\left(\frac{1}{\widehat{\sigma}_j} - \frac{1}{\sigma_j}\right)\right|\\
& := I_1 + I_2.
\end{aligned}
\end{equation}
For $I_1$, since $\mathbbm{E}[P_n\dot{\ell}_{\beta^\star}] = 0$ and $||\dot{\ell}_{\beta^\star}||_\infty$ is upper bounded by Assumption \ref{assump:GLM-high-dimension} (2)(3), we have 
\begin{equation}
||P_n\dot{\ell}_{\beta^\star}||_\infty = O_p(\sqrt{\log p/n})
\end{equation}
using the union bound and the Hoeffding's inequality. Since $\sigma^2 \geq 1/C_2$ (see the proof of Lemma \ref{lemma:GLM-sigma-bound}), it follows that
\begin{equation}
\begin{aligned}
I_1 & \leq \sqrt{C_2}\sqrt{n}||P_n\dot{\ell}_{\beta^\star}||_\infty\max_{j\in[p]}||\widehat{\Theta}_{j,\cdot} - \Theta_{j,\cdot}||_1\\
& = \sqrt{n}O_p(\sqrt{\log p/n})O_p(s \sqrt{\log p / n})\\
& = o_p(1)
\end{aligned}
\end{equation}
by Assumption \ref{assump:GLM-high-dimension} (1). For $I_{2}$, using the same argument, we have
\begin{equation}
\begin{aligned}
I_{2} & \leq \sqrt{n} \max_{j\in[p]}||\widehat\Theta_{j}||_{\infty}||P_n\dot{\ell}_{\beta^\star}||_\infty\max_{j \in [p]}\left|\frac{1}{\widehat{\sigma}_j}-\frac{1}{\sigma_{j}}\right|\\
& \leq \sqrt{n}\left(||\widehat{\Theta}-\Theta||_{\infty}+ ||\Theta||_{\infty}\right)O_{p}(\sqrt{\log p/n})O_{p}(s\sqrt{\log p/n})\\
&= o_{p}(1).
\end{aligned}
\end{equation}

For $R_{2,j}$, using the KKT condition of the nodewise Lasso regression, we have
\begin{equation}
\left|\left|\widehat{\Theta}_{j,\cdot}P_n\ddot{\ell}_{\widehat{\beta}} - e_j^\top\right|\right|_\infty \leq \lambda_j/\widehat{\tau}_j^2.
\end{equation}
Since 
\begin{equation}
\begin{aligned}
\max_{j\in[p]}1/\widehat{\tau}_j^2  & \leq \max_{j\in[p]}|1/\widehat{\tau}_j^2 - 1/\tau_j^2| + \max_{j\in[p]}1/\tau_j^2\\
& \leq C_\epsilon \sqrt{s\log p/n} + \max_{j\in[p]}1/\Theta_{j,j}\\
& \leq 1 + C_2
\end{aligned}
\end{equation}
and 
\begin{equation}
\begin{aligned}
\max_{j\in[p]}1/\widehat{\sigma}_j  & \leq \max_{j\in[p]}|1/\widehat{\sigma}_j - 1/\sigma_j| + \max_{j\in[p]}1/\sigma_j\\
& \leq C_\epsilon s\sqrt{\log p/n} + 1/\sigma^{1/2}_{\min}(\Sigma)\\
& \leq 1 + \sqrt{C_2}
\end{aligned}
\end{equation}
by Assumption \ref{assump:GLM-high-dimension} (3), we have
\begin{equation}
\begin{aligned}
R_{2,j} \leq (1 + C_2)\sqrt{n}\left|\left|\widehat{\Theta}_{j,\cdot}P_n\ddot{\ell}_{\widehat{\beta}} - e_j^\top\right|\right|_\infty||\widehat{\beta} - \beta^\star||_1 = o_p(1)
\end{aligned}
\end{equation}
by Assumption \ref{assump:GLM-high-dimension} (1).

For $R_{3,j}$, the first term is $o_{p}(1)$ following the proof of Theorem 3.1 in \citet{van2014asymptotically}. For the second term, we note that
\begin{equation}
\begin{aligned}
||X\widehat{\Theta}_{j,\cdot}^\top||_\infty & \leq ||X\Theta_{j,\cdot}^\top||_\infty + ||X(\widehat{\Theta}_{j,\cdot} - \Theta_{j,\cdot})^\top||_\infty\\
& \leq ||X||_\infty + ||X_{-j}\gamma_j||_\infty +  ||X||_\infty||(\widehat{\Theta}_{j,\cdot} - \Theta_{j,\cdot})^\top||_1\\
& \leq ||X||_\infty + ||X_{\beta^\star, -j}\gamma_j||_\infty/\inf|\ddot{\rho}(x^\intercal\beta^\star)| +  ||X||_\infty||(\widehat{\Theta}_{j,\cdot} - \Theta_{j,\cdot})^\top||_1\\
& < +\infty
\end{aligned}
\end{equation}
by Assumption \ref{assump:GLM-high-dimension} (2)(3). By the mean value Theorem and Assumption \ref{assump:GLM-high-dimension}, it follows that
\begin{equation}
R_{3,j} \leq \frac{1}{\sqrt{n}\widehat{\sigma}_j}||X\widehat{\Theta}_{j,\cdot}^\top||_\infty||X(\widehat{\beta} - \beta^\star)||_2^2 = o_p(1).
\end{equation}
This completes the proof of Proposition \ref{prop:GLM-bias-high-dimension}.

\subsubsection{Proof of Proposition \ref{prop:GLM-data-splitting-FDR-high-dimension}}
The proof of Proposition  \ref{prop:GLM-data-splitting-FDR-high-dimension} essentially follows the same as the proof of Proposition \ref{prop:linear-data-splitting-FDR-high-dimension}. The only change is that $Z_j = -{\sqrt{n}\Theta_{j, \cdot}P_n\dot{\ell}_{\beta^\star}}/{\sigma_j}$ is not exactly normal, but asymptotically normal. The discrepancy between the joint law of $(Z_1, \cdots, Z_p)$ and the corresponding multivariate normal distribution (with the covariance matrix $\Sigma$) can be quantified using the Berry-Esseen Theorem, which is in the order of $o(1/\sqrt{n})$. Thus, we can establish the GLM version of Lemma \ref{lemma:bias-bound} similarly.

\section{Additional Numerical Results}
\subsection{The moderate-dimensional setting}
\subsubsection{Logistic regression}
\label{subsub:logistic-moderate-dimension-additional-result}
\begin{figure*}
\begin{center}
\includegraphics[width=0.45\columnwidth]{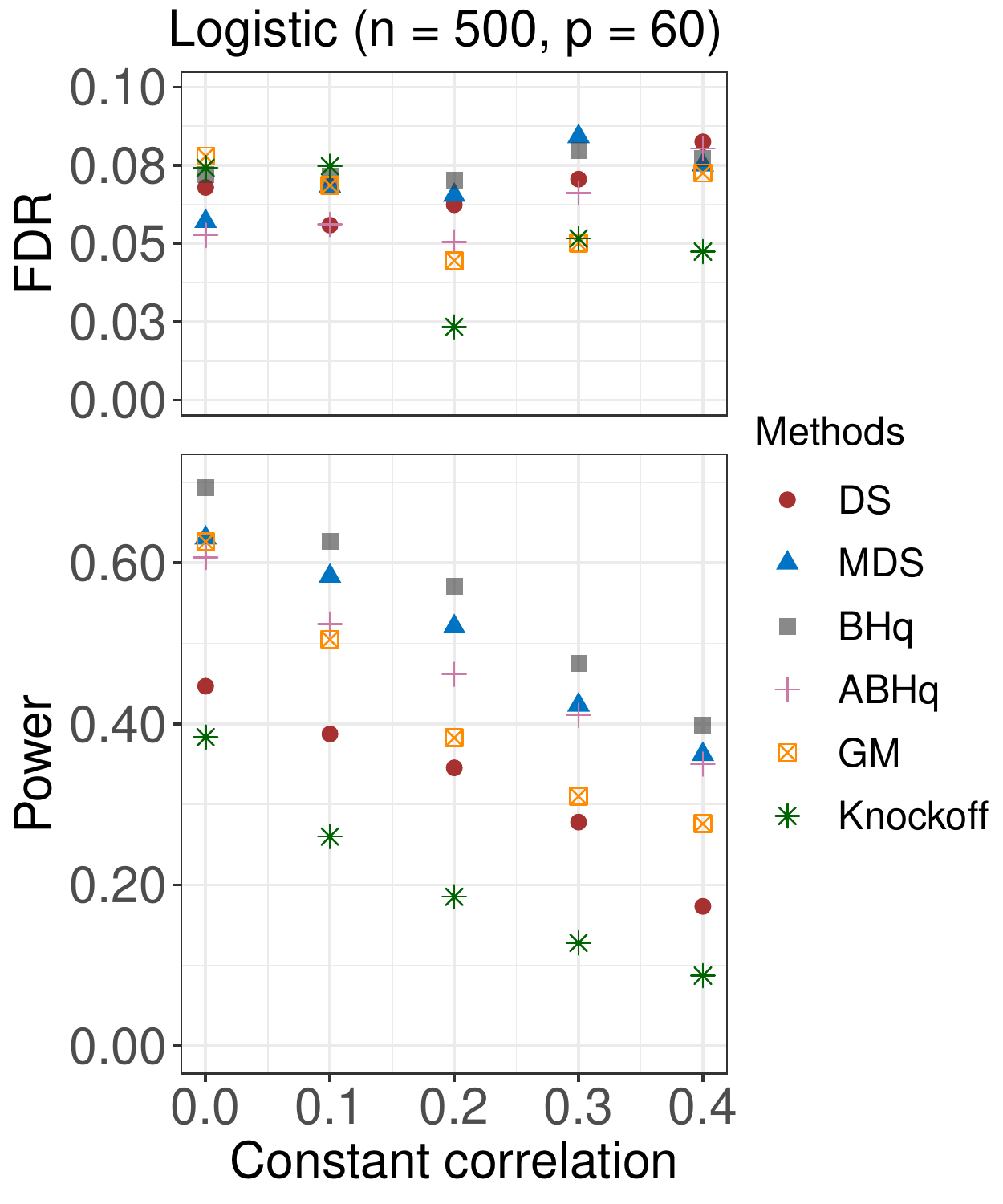}
\includegraphics[width=0.45\columnwidth]{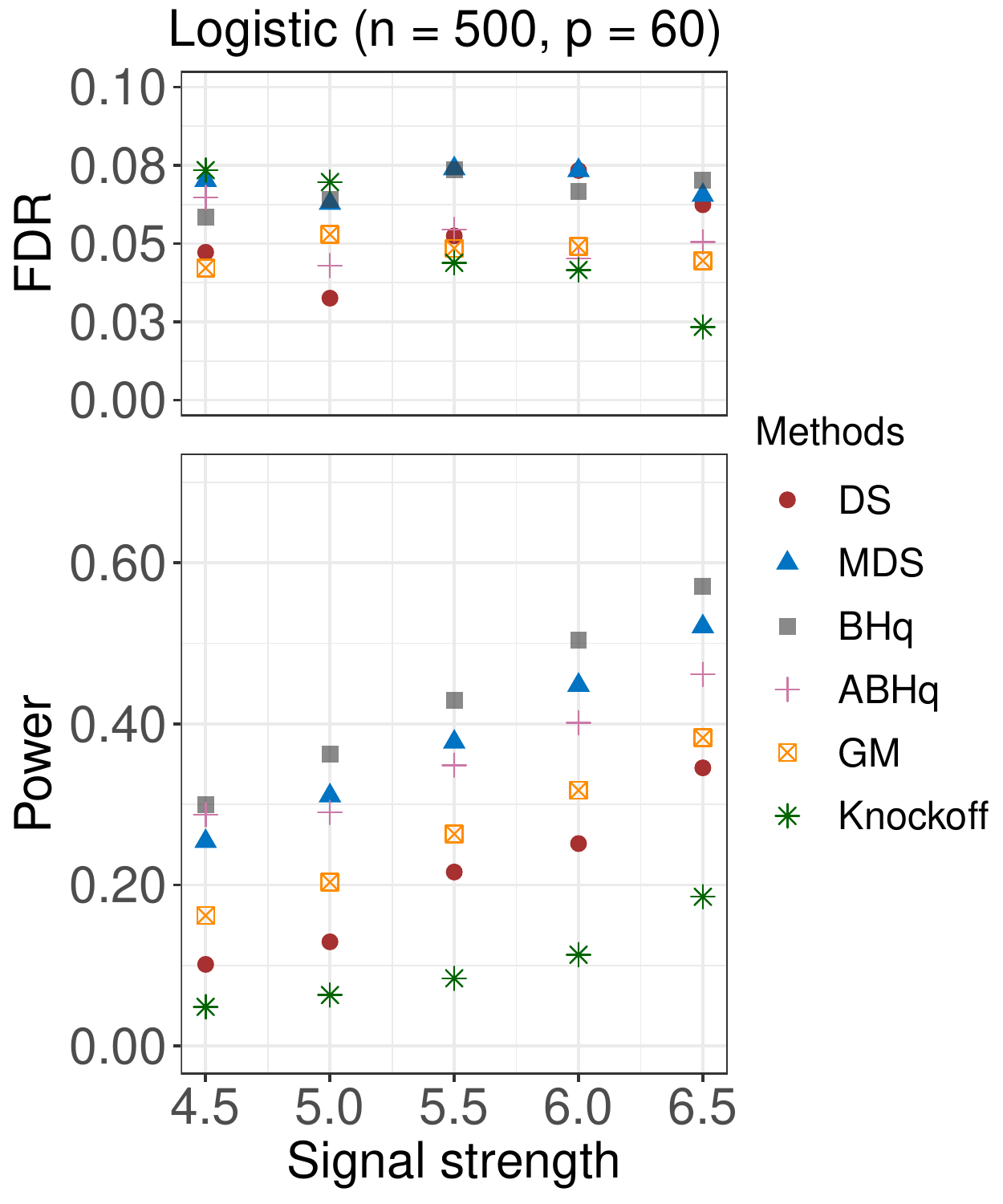}
\end{center}
\caption{Empirical FDRs and powers for the logistic regression model in the small-$n$-and-$p$ setting. 
The algorithmic settings are as per Figure \ref{fig:moderate-dimensional-logistic-small-n-and-p-toeplitz-correlation}, except that features have constant pairwise correlation, i.e.,
$\Sigma_{ij} = r^{\mathbbm{1}(i\neq j)}$.}
\label{fig:moderate-dimensional-logistic-small-n-and-p-constant-correlation}
\end{figure*}

\begin{figure*}
\begin{center}
\includegraphics[width=0.45\columnwidth]{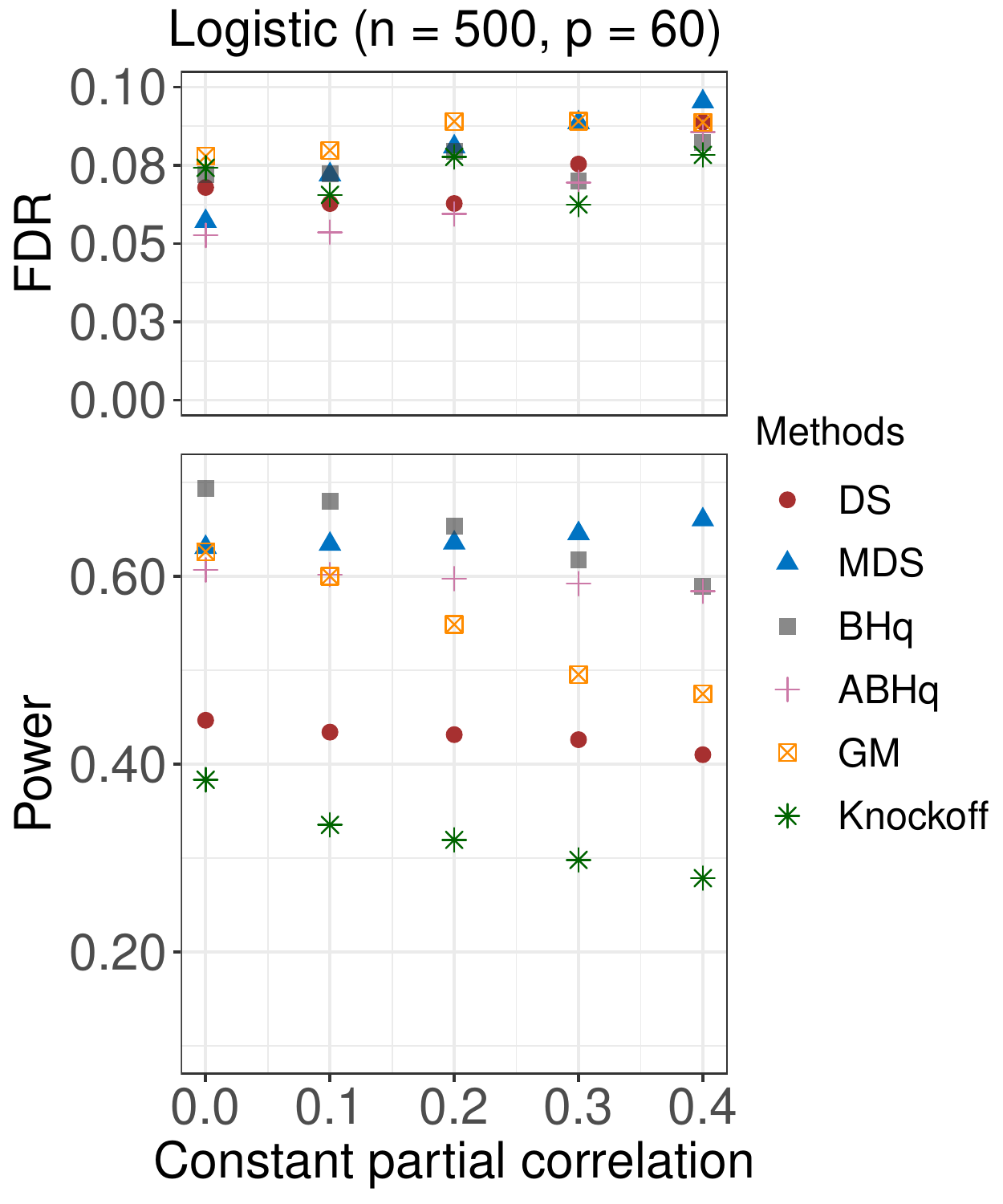}
\includegraphics[width=0.45\columnwidth]{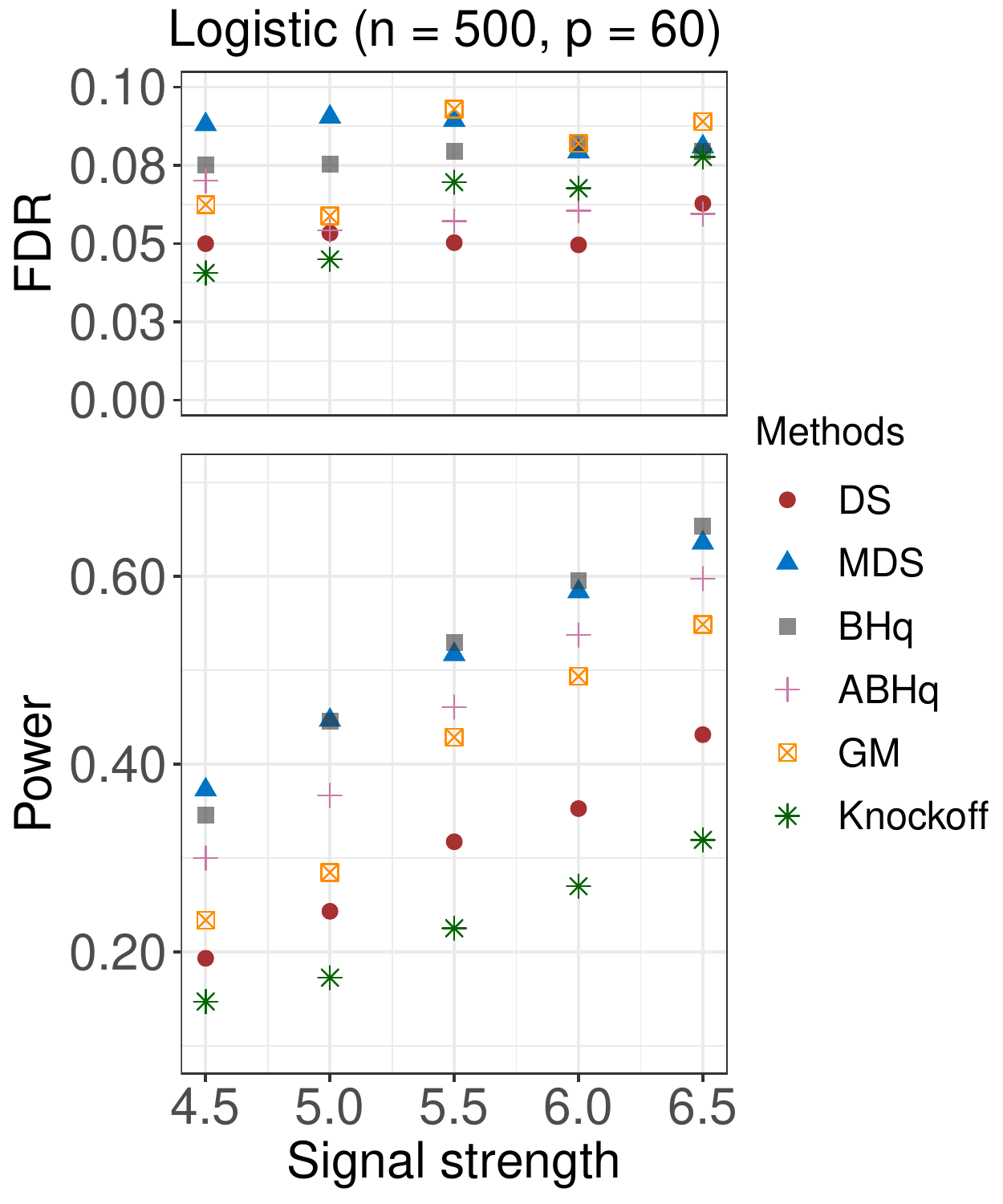}
\end{center}
\caption{Empirical FDRs and powers for the logistic regression model in the small-$n$-and-$p$ setting. 
The algorithmic settings are as per Figure \ref{fig:moderate-dimensional-logistic-small-n-and-p-toeplitz-correlation}, except that features have constant pairwise partial correlation, i.e.,
$\Sigma^{-1}_{ij} = r^{\mathbbm{1}(i\neq j)}$.}
\label{fig:moderate-dimensional-logistic-small-n-and-p-constant-partial-correlation}
\end{figure*}

\begin{figure*}
\begin{center}
\includegraphics[width=0.45\columnwidth]{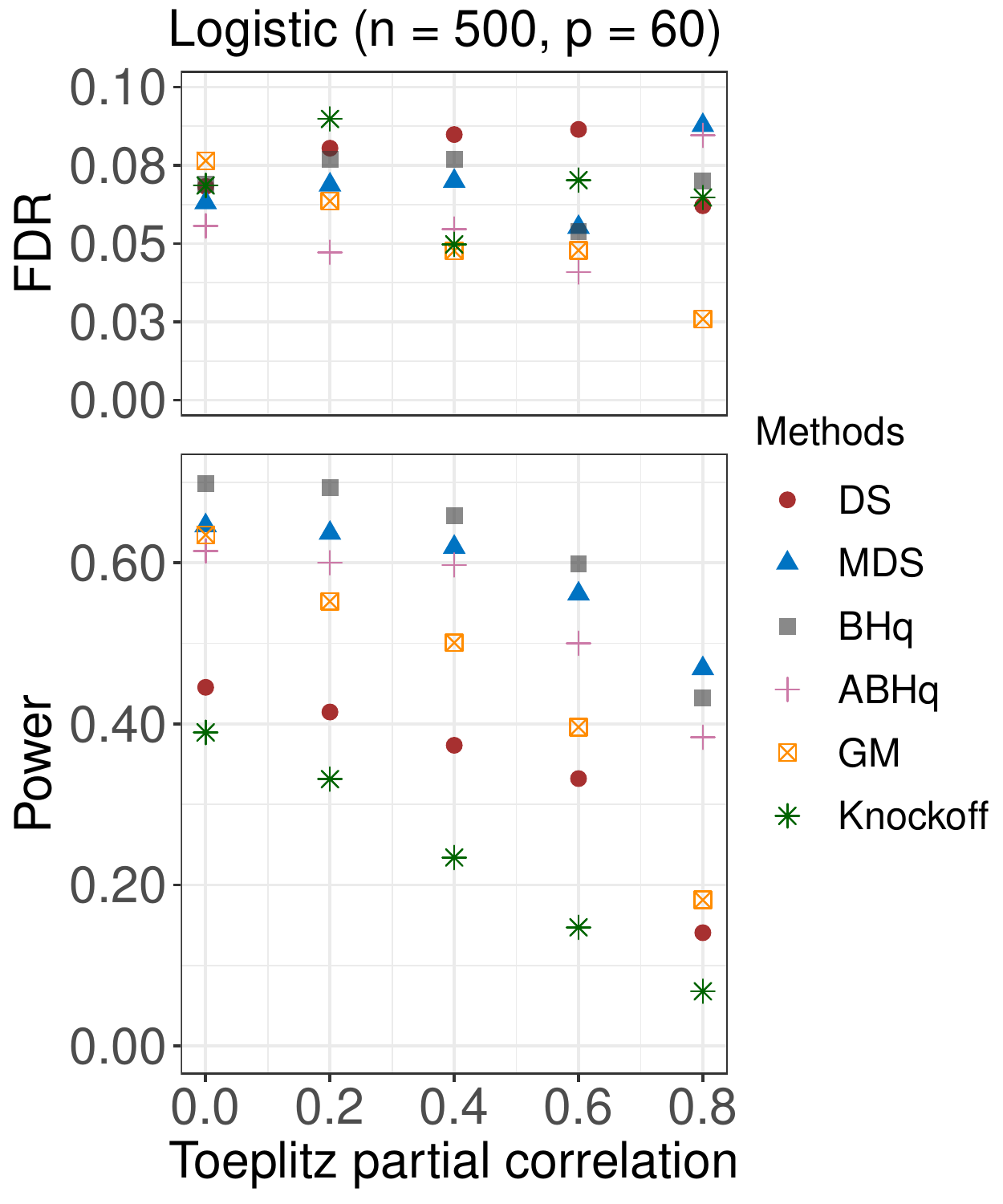}
\includegraphics[width=0.45\columnwidth]{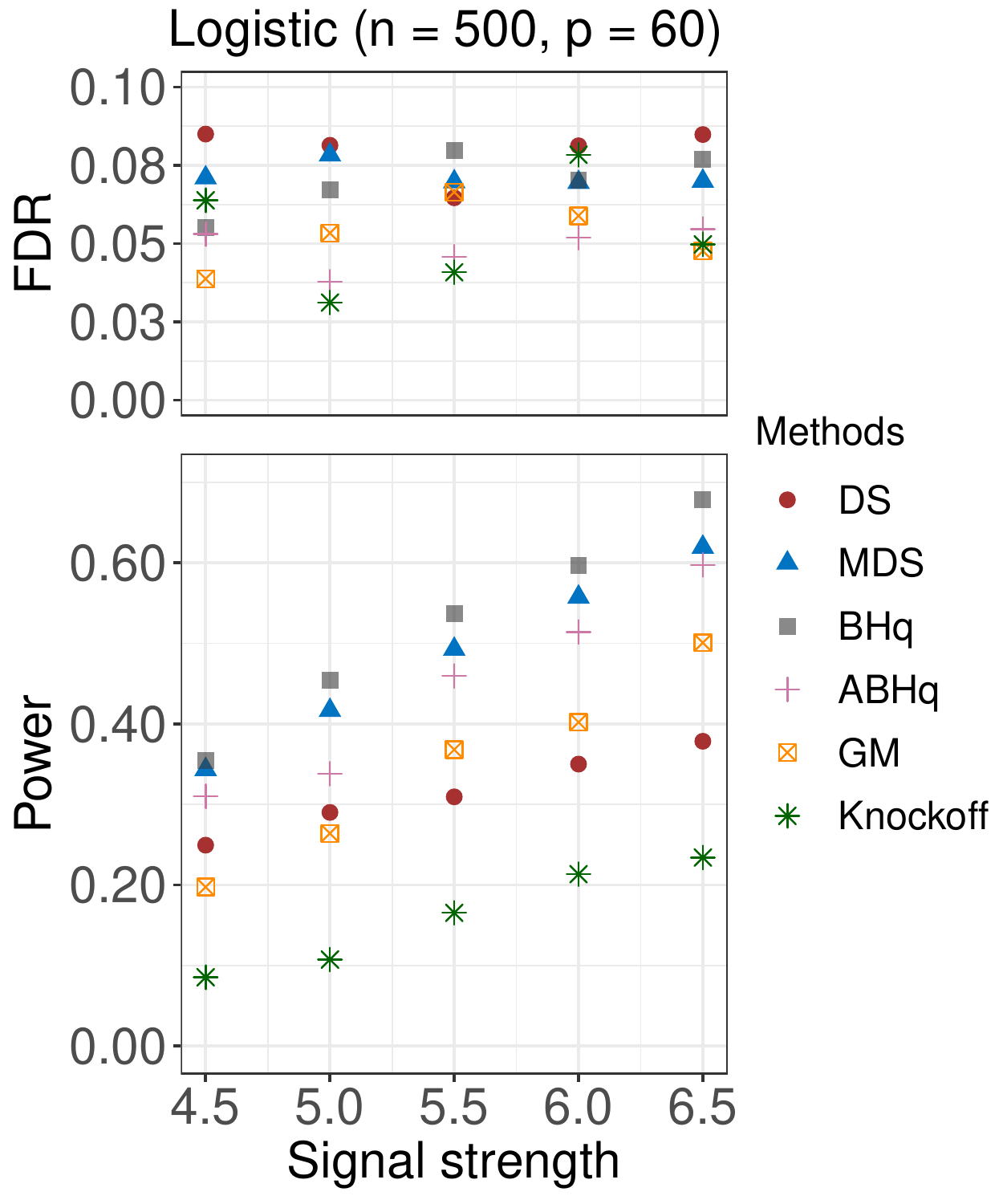}
\end{center}
\caption{Empirical FDRs and powers for the logistic regression model in the small-$n$-and-$p$ setting. 
The algorithmic settings are as per Figure \ref{fig:moderate-dimensional-logistic-small-n-and-p-toeplitz-correlation}, except that features have Toeplitz partial correlation, i.e.,
$\Sigma^{-1}_{ij} = r^{|i - j|}$.}
\label{fig:moderate-dimensional-logistic-small-n-and-p-toeplitz-partial-correlation}
\end{figure*}

\begin{figure*}
\begin{center}
\includegraphics[width=0.45\columnwidth]{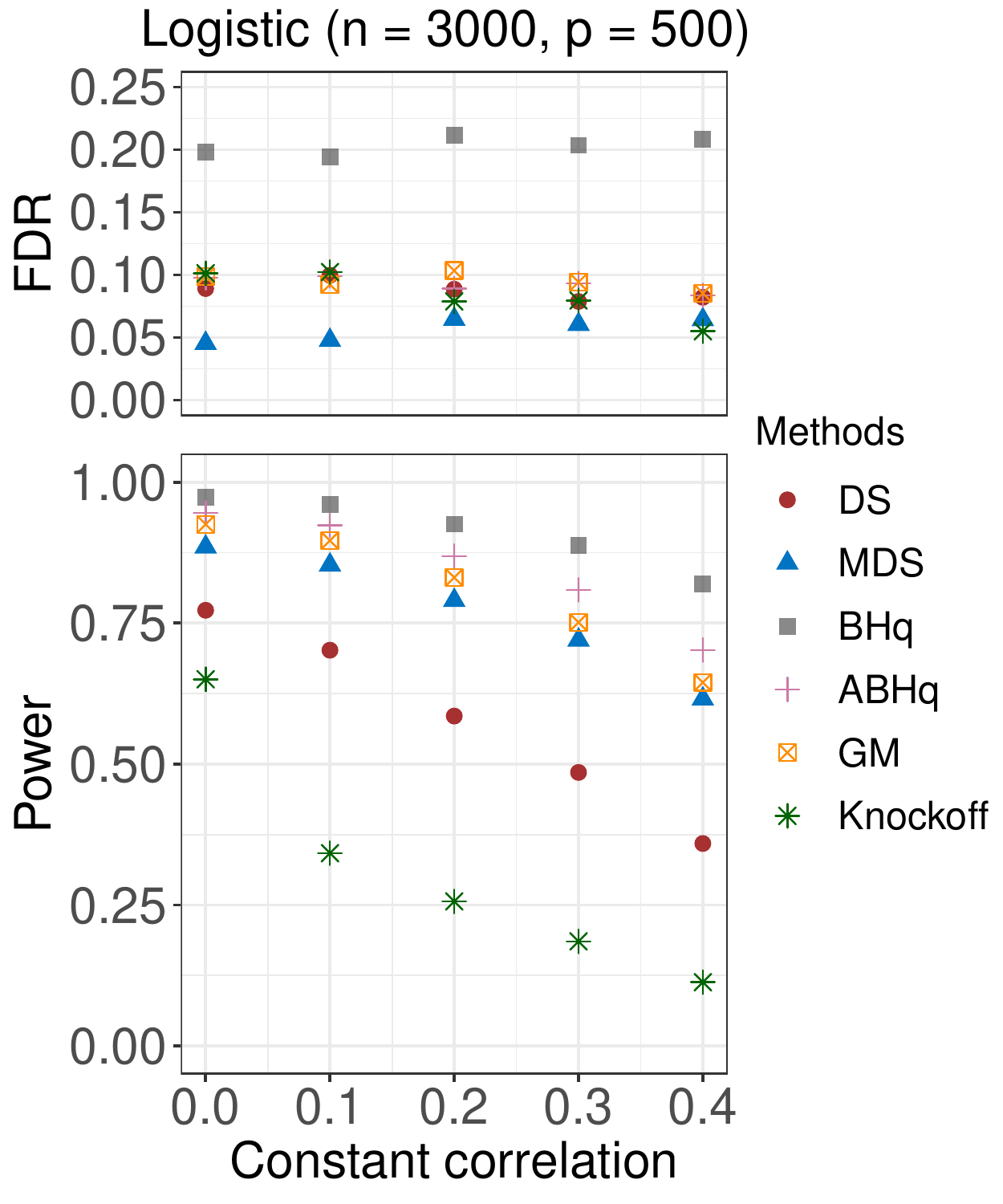}
\includegraphics[width=0.45\columnwidth]{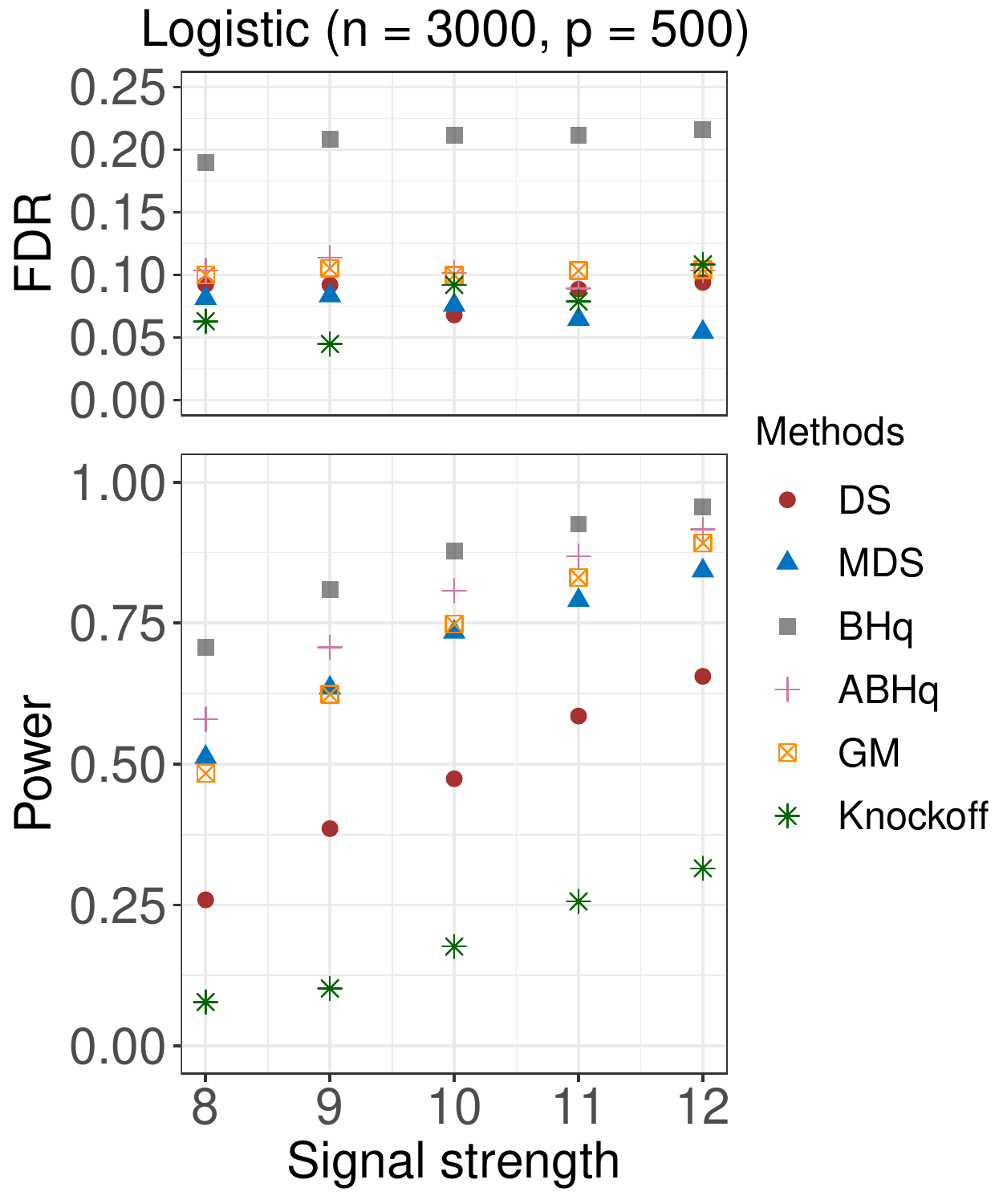}
\end{center}
\caption{Empirical FDRs and powers for the logistic regression model in the large-$n$-and-$p$ setting. 
The algorithmic settings are as per Figure \ref{fig:moderate-dimensional-logistic-large-n-and-p-toeplitz-correlation}, except that features have constant pairwise correlation, i.e.,
$\Sigma_{ij} = r^{\mathbbm{1}(i\neq j)}$.}
\label{fig:moderate-dimensional-logistic-large-n-and-p-constant-correlation}
\end{figure*}

We provide additional numerical results to complement Section \ref{subsub:moderate-dimensional-logistic} by considering the following types of the covariance matrix $\Sigma$.
\begin{enumerate}
\item The small-$n$-and-$p$ setting.
\begin{enumerate}
\item Features have constant pairwise correlation, i.e., $\Sigma_{ij} = r^{\mathbbm{1}(i\neq j)}$.
\item Features have constant pairwise partial correlation, i.e., $\Sigma^{-1}_{ij} = r^{\mathbbm{1}(i\neq j)}$.
\item Features have Toeplitz partial correlation, i.e., $\Sigma^{-1}_{ij} = r^{|i - j|}$.
\end{enumerate}
\item The large-$n$-and-$p$ setting.
\begin{enumerate}
\item Features have constant pairwise correlation, i.e., $\Sigma_{ij} = r^{\mathbbm{1}(i\neq j)}$.
\end{enumerate}
\end{enumerate}
For the small-$n$-and-$p$ setting, the empirical FDRs and powers of different methods in scenarios (a), (b) and (c) are summarized in Figures \ref{fig:moderate-dimensional-logistic-small-n-and-p-constant-correlation}, \ref{fig:moderate-dimensional-logistic-small-n-and-p-constant-partial-correlation} and \ref{fig:moderate-dimensional-logistic-small-n-and-p-toeplitz-partial-correlation}, respectively. For the large-$n$-and-$p$ setting, the results in scenario (a) are shown in Figure \ref{fig:moderate-dimensional-logistic-large-n-and-p-constant-correlation}.

\subsubsection{Negative binomial regression}
\label{subsub:negative-binomial-moderate-dimension-additional-result}
\begin{figure*}
\begin{center}
\includegraphics[width=0.45\columnwidth]{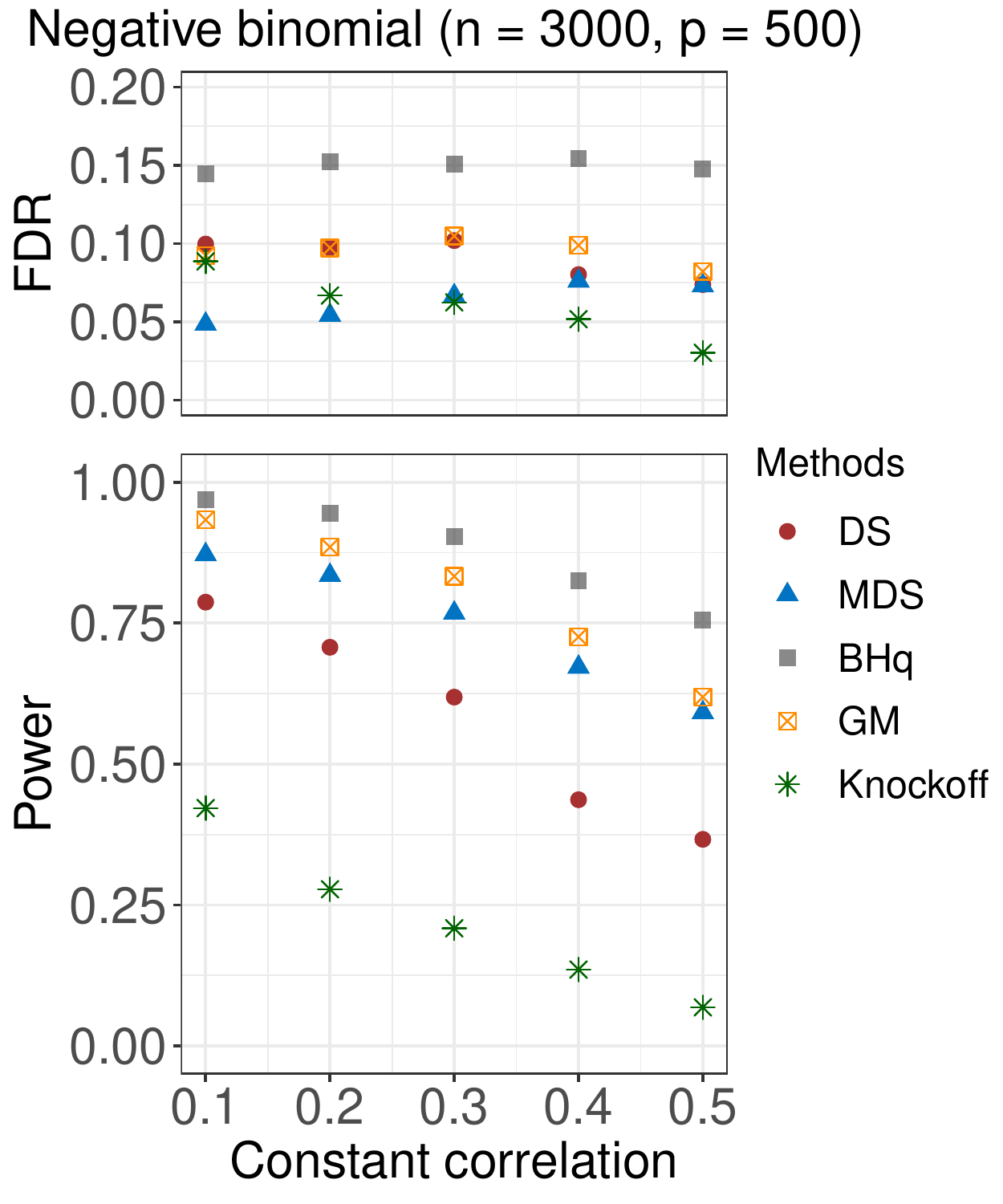}
\includegraphics[width=0.45\columnwidth]{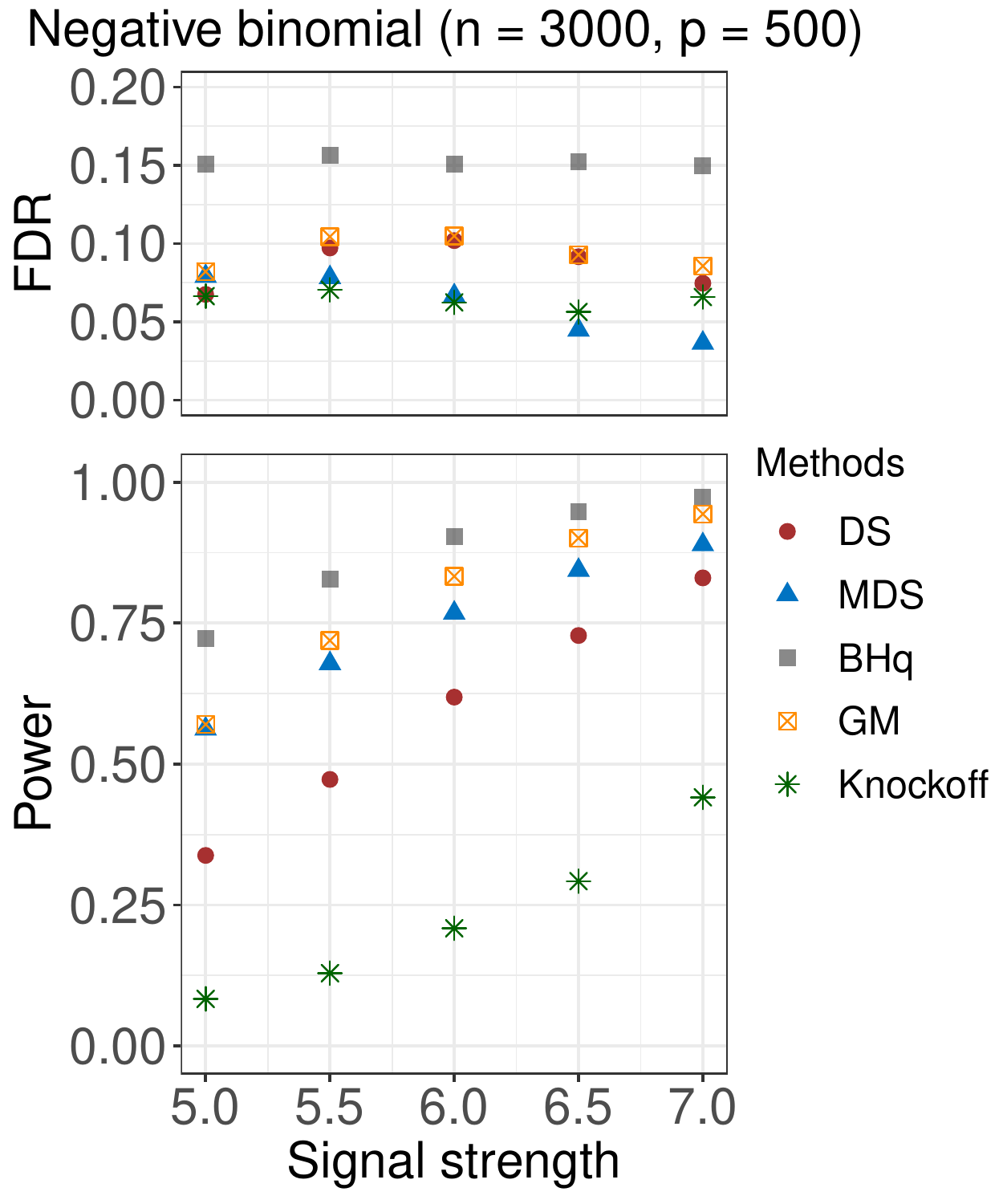}
\end{center}
\caption{Empirical FDRs and powers for the negative binomial regression model. The algorithmic settings are as per Figure \ref{fig:moderate-dimensional-negative-binomial-toeplitz-correlation}, except that features have constant pairwise correlation, i.e.,
$\Sigma_{ij} = r^{\mathbbm{1}(i\neq j)}$.}
\label{fig:moderate-dimensional-negative-binomial-constant-correlation}
\end{figure*}

To complement Section \ref{subsub:moderate-dimensional-negative-binomial}, we provide additional numerical results for the case where features have constant pairwise correlation, i.e., $\Sigma_{ij} = r^{\mathbbm{1}(i\neq j)}$. The empirical FDRs and powers of different methods are summarized in Figure \ref{fig:moderate-dimensional-negative-binomial-constant-correlation}.

\subsection{The high-dimensional setting}
\subsubsection{Linear regression}
\label{subsub:linear-high-dimension-additional-result}
To complement Section \ref{subsub:numeric-linear-high-dimension}, we first detail the Toeplitz matrix aligned along the diagonal of the covariance matrix $\Sigma$ as below,
\begin{equation}
\label{eq:blockwise-diagonal-formula}
\begin{bmatrix}
1 & \frac{(p^\prime-2)r}{p^\prime-1} & \frac{(p^\prime-3)r}{p^\prime-1} & \ldots & \frac{r}{p^\prime-1} & 0\\
& & & & & \\
\vspace{0.5cm}
\frac{(p^\prime-2)r}{p^\prime-1} & 1 & \frac{(p^\prime-2)r}{p^\prime-1} & \ldots & \frac{2r}{p^\prime-1} & \frac{r}{p^\prime-1} \\
\vspace{0.5cm}
\vdots & & \ddots & & & \vdots\\
0  & \frac{r}{p^\prime-1} & \frac{2r}{p^\prime-1} & \ldots & \frac{(p^\prime-2)r}{p^\prime-1} & 1
\end{bmatrix},
\end{equation}
where $p^\prime = p/10$. Throughout, we refer $r\in (0, 1)$ as the correlation factor. 

\begin{figure*}
\begin{center}
\includegraphics[width=0.45\columnwidth]{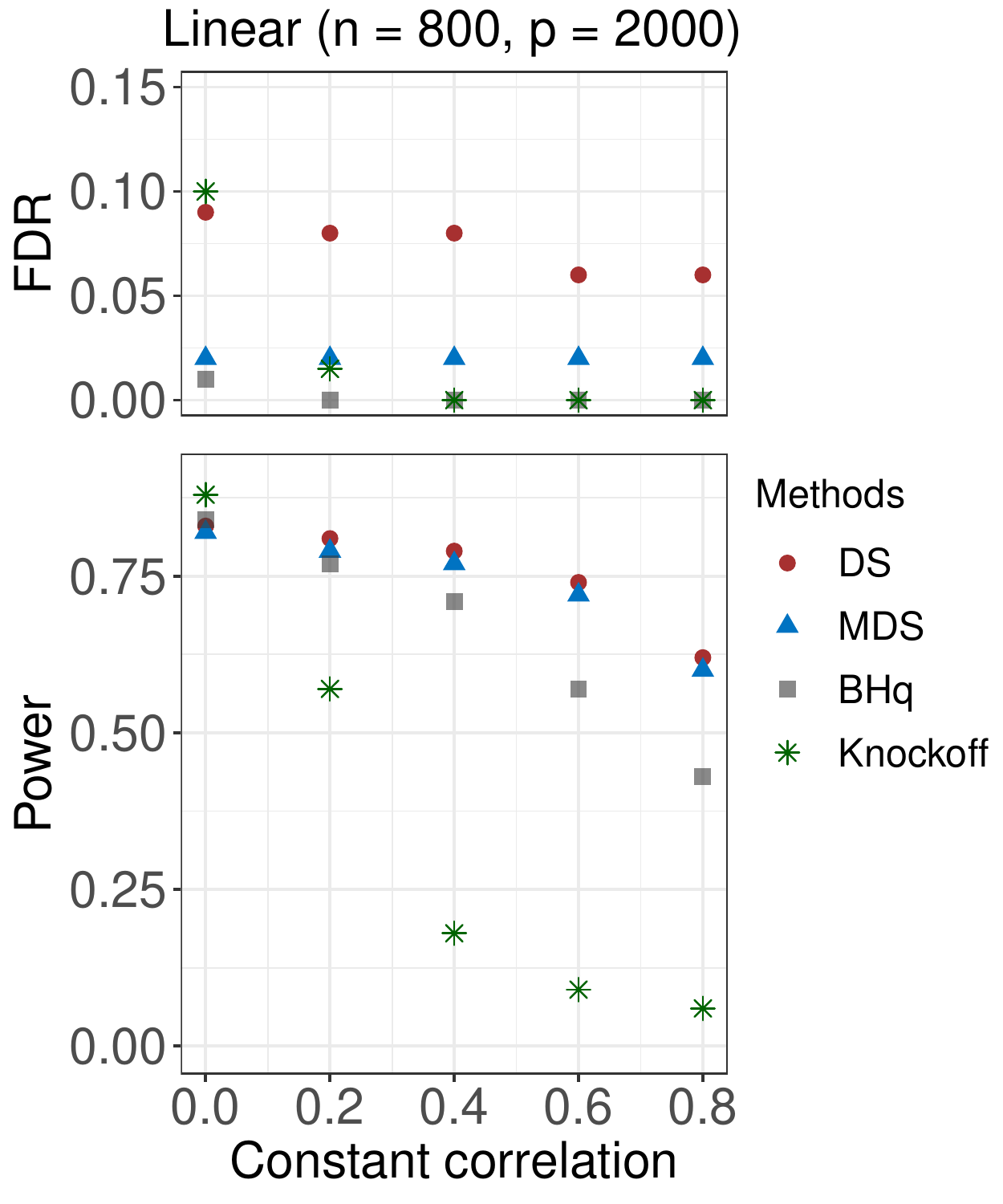}
\includegraphics[width=0.45\columnwidth]{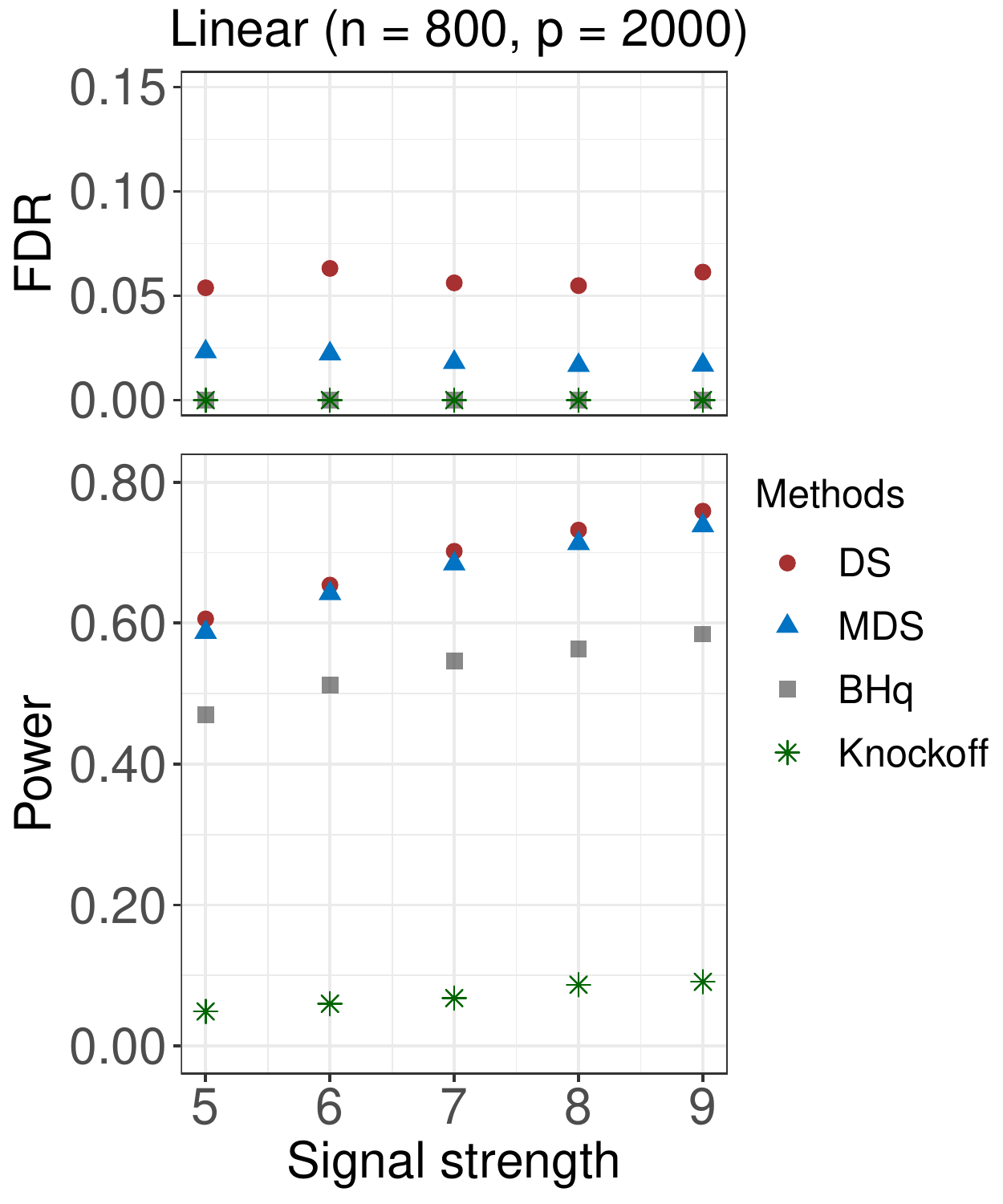}
\end{center}
\caption{Empirical FDRs and powers for the linear model. Each row of the design matrix is independently drawn from $N(0, \Sigma)$, with $\Sigma_{ij} = r^{\mathbbm{1}(i\neq j)}$, i.e., features have constant pairwise correlation.
$\beta_j^\star$ for $j\in S_1$ are i.i.d. samples from $N(0,s^2)$, where
$s$ 
is referred to as the signal strength. 
The signal strength along the x-axis of the right panel shows multiples of 
$\sqrt{\log p/n}$.
In the left panel, we fix the signal strength at $8\sqrt{\log p/n}$ and vary the correlation $r$. 
In the right panel, we fix the ccorrelation at $r = 0.6$ and vary the signal strength.
The number of relevant features is 50 across all settings, and
the designated FDR control level is $q = 0.1$.
Each dot in the figure represents the average from 50 independent runs.}
\label{fig:linear-high-dimension-constant-correlation}
\end{figure*}

We then provide additional numerical results for the case where features have constant pairwise correlation, i.e., $\Sigma_{ij} = r^{\mathbbm{1}(i\neq j)}$.
All the algorithmic settings are the same as discussed in Section \ref{subsub:numeric-linear-high-dimension}, except that the number of relevant features is $50$ across all settings, i.e. $p_1 = 50$.
The empirical FDRs and powers of different methods are summarized in Figure \ref{fig:linear-high-dimension-constant-correlation}. 

\end{document}